\pdfoutput=1
\documentclass[%
 sor4-1,
 reprint,
 amsmath,amssymb,
]{revtex4-1}

\usepackage{graphicx}% Include figure files
\usepackage{dcolumn}% Align table columns on decimal point
\usepackage{bm}% bold math
\usepackage{xcolor}
\usepackage{enumerate}
\usepackage{gensymb}

\begin{document}
	
\preprint{APS/123-QED}

\title{Medium Amplitude Parallel Superposition (MAPS) Rheology \\ Part 2: Experimental Protocols and Data Analysis}
\author{Kyle R. Lennon$^1$}
\author{Michela Geri$^2$}
\author{Gareth H. McKinley$^2$}
\author{James W. Swan$^1$}
\thanks{Corresponding author; Electronic mail: jswan@mit.edu}
\affiliation{1. Department of Chemical Engineering, \\ 2. Department of Mechanical Engineering, Massachusetts Institute of Technology, Cambridge, MA 02139 \looseness=-1\\}
\date{\today}

\begin{abstract}

An experimental protocol is developed to directly measure the new material functions revealed by medium amplitude parallel superposition (MAPS) rheology. This experimental protocol measures the medium amplitude response of a material to a simple shear deformation composed of three sine waves at different frequencies. Imposing this deformation and measuring the mechanical response reveals a rich data set consisting of up to 19 measurements of the third order complex modulus at distinct three-frequency coordinates. We discuss how the choice of the input frequencies influences the features of the MAPS domain studied by the experiment. A polynomial interpolation method for reducing the bias of measured values from spectral leakage and variance due to noise is discussed, including a derivation of the optimal range of amplitudes for the input signal. This leads to the conclusion that conducting the experiment in a stress-controlled fashion possesses a distinct advantage to the strain-controlled mode. The experimental protocol is demonstrated through measurements of the MAPS response of a model complex fluid: a surfactant solution of wormlike micelles. The resulting data set is indeed large and feature-rich, while still being acquired in a time comparable to similar medium amplitude oscillatory shear (MAOS) experiments. We demonstrate that the data represents measurements of an intrinsic material function by studying its internal consistency, its compatibility with low-frequency predictions for Coleman-Noll simple fluids, and its agreement with data obtained via MAOS amplitude sweeps. Finally, the data is compared to predictions from the corotational Maxwell model to demonstrate the power of MAPS rheology in determining whether a constitutive model is consistent with a material's time-dependent response.
 
\end{abstract}


\maketitle

\section{Introduction}

The experimental protocols with which we study the simple shear rheology of viscoelastic materials vary widely. They include methods that measure time-domain functions, such as stress relaxation upon the imposition of a step strain or stress growth upon the imposition of a steady shear rate \cite{bird-1987}, as well as methods that measure frequency-domain functions, such as large amplitude oscillatory shear (LAOS) \cite{giacomin-1993,mckinley-2008,hyun-2011,cho-2016} and parallel superposition (PS) \cite{tanner-1968,vermant-1998,yamamoto-1971}. Though these protocols are typically discussed in the context of controlling the shear strain and measuring the shear stress, each can in principle be conducted under stress control as well. The data taken from each of these experiments in general looks quite different, and represents distinct aspects of a material's full viscoelastic response space. Except in the limit of linear viscoelasticity, none of these experimental protocols, nor the mathematical frameworks for interpreting their data, give a result that can be used to directly predict the data that will be obtained by another of those protocols, however. This leaves experimental rheologists with the choice between different measurement protocols that probe different aspects of a material's nonlinear viscoelastic response under specific and difficult-to-generalize conditions.

In Part 1 of this work, we introduced a remedy to address the issue of disparate experimental data in the weakly nonlinear regime of simple shear flows through a framework called Medium Amplitude Parallel Superposition (MAPS) rheology \cite{lennon-2020}. MAPS rheology describes material functions that span the entire weakly nonlinear response space of a viscoelastic material to an imposed shear flow. Full knowledge of a MAPS material function allows one to predict the response of that material under an arbitrary, weakly nonlinear simple shear deformation, including all of the protocols listed previously. This presents a new, comprehensive choice to the experimental rheologist: a single experimental framework capable of providing data that generically describes the weakly nonlinear response of an unknown material.

Part 1 of this work developed the mathematics of MAPS rheology, including relationships between MAPS and common experimental protocols such as medium amplitude oscillatory shear (MAOS) \cite{ewoldt-2013,davis-1978} and PS. We also presented theoretical studies of different constitutive viscoelastic models. In this part, we develop a general framework for experimental protocols that can access the entire MAPS domain -- that is, probe the entire weakly nonlinear simple shear response space of viscoelastic materials. This experimental framework has the distinct benefit that it produces data sampled from a high-dimensional domain, and does so with high throughput. We will show how parameters in the experimental design can be varied to probe different regions of the domain covered by each response function, to yield data that map closely to MAOS, PS, and steady shear flow experiments, respectively.

Another distinct benefit of the experimental framework for MAPS rheology developed herein is that it generates information-rich experimental data sets capable of representing orders of magnitude more data when compared to MAOS or PS tests that require comparable data-acquisition time. In oscillatory rheological probes such as MAOS or PS, the data-acquisition time is limited by the time scale on which the material is studied, which in turn is set by the fundamental oscillatory frequency $\omega_0$. MAPS rheology takes advantage of the fact that we are free to study the material response simultaneously at faster time scales without increasing the data-acquisition time. This can be easily accomplished by simultaneously imposing multiple oscillatory tones at integer multiples of $\omega_0$. With a modest number of input tones, the data throughput of experiments can be increased tenfold or more.

The development of the experimental framework for MAPS rheology in this part proceeds as follows. After a brief review of the key mathematical details of MAPS rheology, the third order expansion for the shear stress is written for an imposed strain signal consisting of three sine waves of equal amplitude but different frequencies imposed in parallel. This expansion demonstrates how the output stress response can be directly translated to discrete values of the third order complex modulus, $G^*_3(\omega_1,\omega_2,\omega_3)$. We then consider how the specific choice of the input tones influences the regions of the MAPS domain that are probed by the experiment. To increase the robustness of experimental measurements to noise, a protocol for obtaining MAPS data via polynomial interpolation is presented, followed by a discussion of how to best balance the effects of variance from noise with bias from higher-order responses. Finally, the developed MAPS experimental framework is applied to measurements of a model solution of wormlike micelles using the stress-controlled version of the protocol. Though the discussion in this work is limited to three-tone input signals, the experimental framework can be readily extended to general multi-tone inputs. Discussion of the complexities and additional experimental design suited for general multi-tone inputs is left to future studies.

Although we develop in detail the mathematics behind the analysis of MAPS data in this work, a principal goal of the development of the MAPS experimental framework is to design experimental protocols that are simple and easily accessible to the experimental rheologist. There are two primary efforts that we have taken to this effect. The first is to simplify the experimental design by reducing the number of experimental design variables to a minimal set, and to clearly identify the consequences of varying each parameter. In the case of a three-tone MAPS experiment, this set of design variables is the fundamental frequency of the input signal, $\omega_0$, the set of integers specifying the three input tones, $\{n_1,n_2,n_3\}$, and the maximal amplitude at which the experiment is run. The first variable controls the longest time-scale $t_{\mathrm{exp}} \sim 2\pi/\omega_0$ on which a material is studied, and can be freely chosen by the user within the physical constraints of their experimental equipment and sample. The additional input variables are subject to some constraints, and we discuss examples and best practices for selecting these variables in the present paper.

The second effort that we have taken to increase the accessibility of MAPS experiments is to develop an open-source software package, MITMAPS, to enable the analysis of raw data output from commercial rheometers. The responsibilities of an experimental rheologist are therefore focused on selecting appropriate experimental design variables, entering the resulting input signals into the commercial software for their rheometer, and feeding the output data into our software package for analysis. The software directly outputs all obtainable values of the MAPS material functions ($G^*_3(\omega_1,\omega_2,\omega_3)$, $\eta^*_3(\omega_1,\omega_2,\omega_3)$, $J^*_3(\omega_1,\omega_2,\omega_3)$, and $\phi^*_3(\omega_1,\omega_2,\omega_3)$) from the experimental data, either in tabular form or as a collection of Bode and Nyquist plots. A version of this software package is included in the Supplementary Material.

\section{A very brief review of the MAPS framework}
\label{sec:volterra}

In simple shear rheometry, the shear stress is a nonlinear functional of the time-dependent shear strain \cite{noll-1958,rivlin-1959,coleman-1961}.  A Volterra series expansion of this functional relationship is compactly expressed in the frequency domain with the Fourier transform of the stress indicated by a caret:
\begin{equation*}
    \hat \sigma( \omega ) = \int_{-\infty}^\infty e^{-i \omega t} \sigma(t) \, dt,
\end{equation*}
and correspondingly for transforms of the the shear strain, $ \hat \gamma( \omega ) $, and the strain rate, $ \hat{ \dot \gamma}( \omega ) $.  To cubic order in the shear strain, the shear stress can be written as:
\begin{align}
    \hat{\sigma}(\omega) &= G_1^*( \omega ) \hat \gamma( \omega ) \label{eq:strain_controlled3} \\ & + \frac{1}{(2\pi)^2} \iiint_{-\infty}^{\infty}G^{*}_{3}(\omega_1,\omega_2,\omega_3)\delta(\omega - \sum_{j=1}^3 \omega_j) \nonumber \\
    & \quad \quad \quad \times \hat{\gamma}(\omega_1) \hat{\gamma}(\omega_2) \hat{\gamma}(\omega_3) \, d\omega_1 d\omega_2 d\omega_3 + O( \hat \gamma( \omega )^5)\nonumber,
\end{align}
where $ G_1^*( \omega ) $ is the complex viscosity familiar from linear response theory and $ G_3^*( \omega_1, \omega_2, \omega_3 ) $ is called the third order complex modulus.  The third order modulus possesses two key symmetries that simplify manipulations and measurement of this response function.  First, the third order modulus is permutation-symmetric:
 \begin{align*}
    &G^*_3(\omega_1,\omega_2,\omega_3) = G^*_3(\omega_2,\omega_3,\omega_1) = G^*_3(\omega_3,\omega_1,\omega_2) \nonumber \\
    & \,\, = G^*_3(\omega_3,\omega_2,\omega_1) = G^*_3(\omega_2,\omega_1,\omega_3) = G^*_3(\omega_1,\omega_3,\omega_2),
\end{align*}   
which allows for the interchange of arguments.  Second, it is Hermitian-symmetric:
\begin{align*}
        G^{\prime}_3(-\omega_1,-\omega_2,-\omega_3) = G^{\prime}_3(\omega_1,\omega_2,\omega_3),\\
        G^{\prime\prime}_3(-\omega_1,-\omega_2,-\omega_3) = -G^{\prime\prime}_3(\omega_1,\omega_2,\omega_3),
    \end{align*}
which indicates that the real part of the third order modulus, $ G_3^\prime( \omega_1, \omega_2, \omega_3 ) $, describes an elastic response while the imaginary part, $ G_3^{\prime\prime}( \omega_1, \omega_2, \omega_3 ) $, describes a viscous response. A stress-controlled analog of equation \ref{eq:strain_controlled3} also exists, in which case the relevant response functions are the first and third order complex compliances, $J^*_1(\omega)$ and $J^*_3(\omega_1,\omega_2,\omega_3)$. The stress-controlled description of MAPS rheology is presented in Section IIB of Part 1 of this work.

\begin{figure}[t]
    \centering
    \includegraphics[width=1.0\columnwidth]{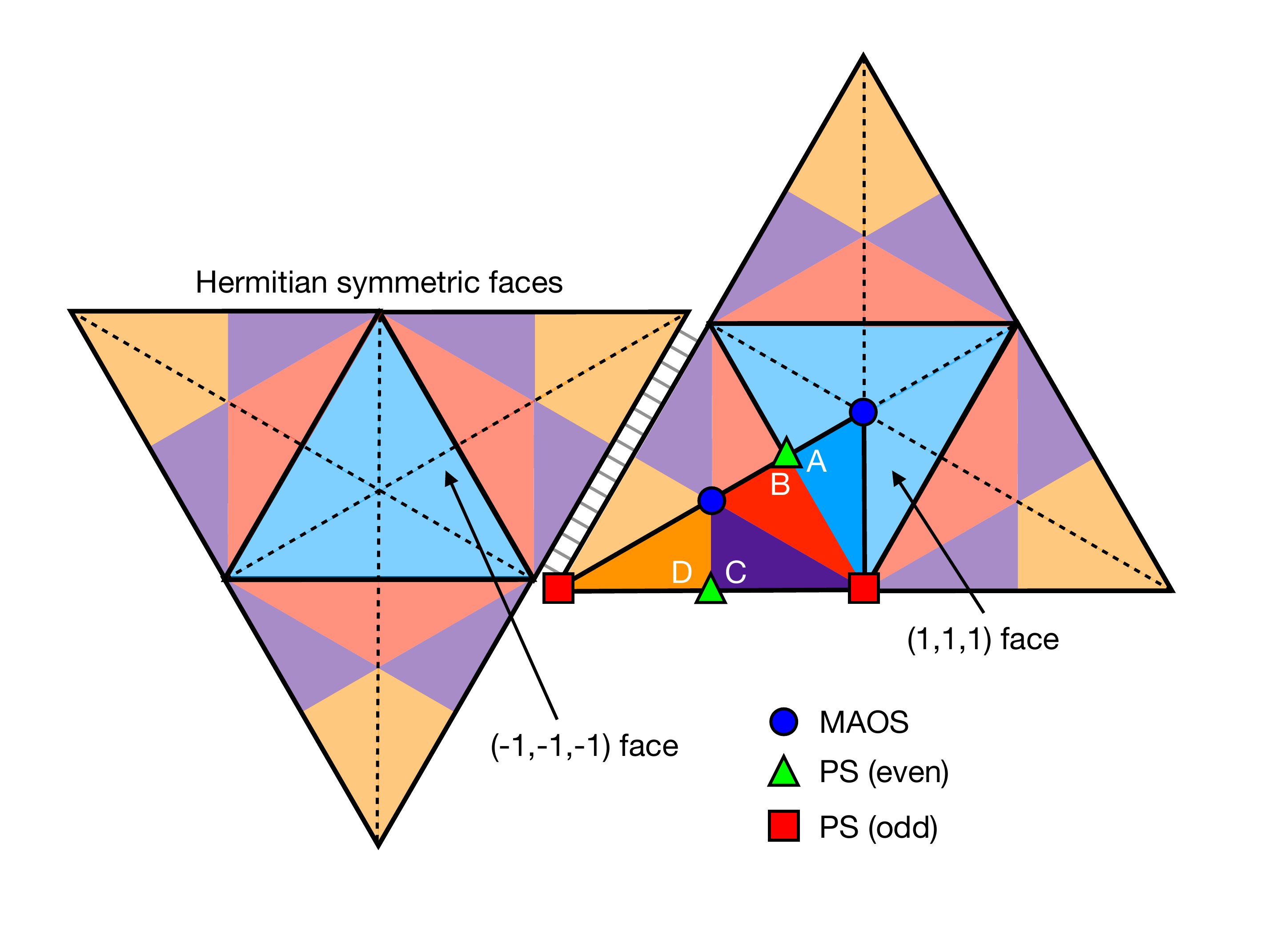}
    \caption{The faces of the surface on which $ | \omega_1 | + | \omega_2 | + | \omega_3 | $ is constant -- an octahedron -- laid flat in the plane.  Dashed lines distinguish the permutation symmetry of $ G_3^*( \omega_1, \omega_2, \omega_3 ) $.  The four triangular sub-regions: A, B, C, and D, pattern the faces of the octahedron in ways that respect that permutation symmetry.  Hermitian symmetry is enforced by substituting the complex conjugate of $ G_3^*( \omega_1, \omega_2, \omega_3 )$ on the Hermitian symmetric faces.}
    \label{fig:octahedron}
\end{figure}

The nature of the symmetries in the third order complex modulus is easily understood by examining the structure of this response function on a constant $L^1$-norm surface:
\begin{equation}
    | \omega_1 | + | \omega_2 | + | \omega_3 | = |\boldsymbol{\omega}|_1.
\end{equation}
Such a surface is given by the faces of an octahedron having vertices aligned with the Cartesian axes in three dimensional frequency space $(\omega_1,\omega_2,\omega_3)$.  Figure \ref{fig:octahedron} depicts the net of an octahedron by laying its faces flat in the plane while preserving the connections among many of the edges.  On the $(-1,-1,-1)$ face of the octahedron, $ \omega_1, \omega_2, \omega_3 \le 0 $.  The dashed lines on this face indicate the lines of permutation symmetry across which swapping any two arguments cannot change the value of $ G_3^*( \omega_1, \omega_2, \omega_3 ) $.  While the $(-1,-1,-1)$ and $(1,1,1)$ faces have three lines of symmetry passing through them, the other faces have only one.  The triangles labeled A, B, C, D together describe $ 1 / 12 $th of the octahedral surface, but can be reflected across the net using permutation symmetry.  Values of $ G^*_3( \omega_1, \omega_2, \omega_3 ) $ on the $(1,1,1)$ face and its directly appended partners are related to those values on the $(-1,-1,-1)$ face and its partners through the Hermitian symmetry of the function.

The consequence of these symmetries means that $ G_3^*( \omega_1, \omega_2, \omega_3 ) $ need only be measured in the identical triangular subspaces labeled A, B, C, D and described by the inequalities:  
\begin{subequations}
\begin{align}
    &\text{A:} \quad \omega_1 \geq \omega_3 \geq \omega_2, \quad \omega_1,\omega_2,\omega_3 \geq 0, \\
    &\text{B:} \quad \omega_1 \geq \omega_3 \geq -\omega_2, \quad \omega_1,\omega_3 \geq 0 \geq \omega_2, \\
    &\text{C:} \quad \omega_1 \geq -\omega_2 \geq \omega_3, \quad \omega_1,\omega_3 \geq 0 \geq \omega_2, \\
    &\text{D:} \quad -\omega_2 \geq \omega_1 \geq \omega_3, \quad \omega_1,\omega_3 \geq 0 \geq \omega_2.
\end{align}
\label{eq:subspaces}
\end{subequations}
These subspaces are hemiequilateral triangles in which the third order complex modulus takes on distinct values.  As discussed in Part 1, $ G_3^*( \omega_1, \omega_2, \omega_3 ) $ can be measured directly at the vertices of these triangular subspaces using the MAOS and PS techniques.  Each subspace has two vertices associated with the PS protocol and one associated with the MAOS protocol.  The experimental procedure discussed in this work enables measurement of the third order modulus at other points on the perimeter or in the interior of these subspaces.

\begin{figure}[t]
    \centering
    \includegraphics[width=0.75\columnwidth]{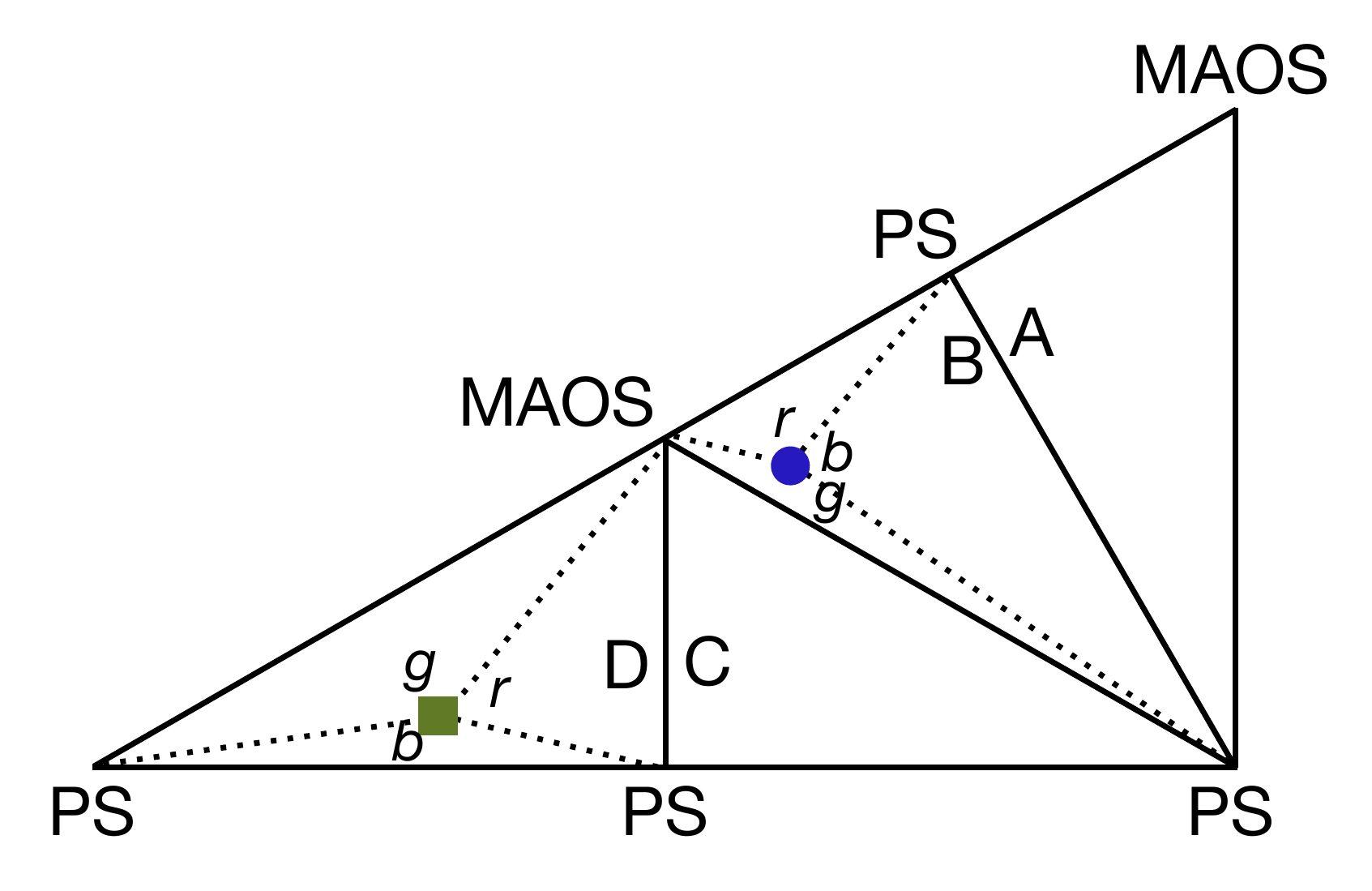}
    \caption{Depiction of a barycentric visualization scheme for each triangular subspace. Within one of the four unique triangular subspaces (here denoted A, B, C, D) a set of barycentric coordinates $(r,g,b)$ can be constructed.  The position of data within this triangle can be associated with a color using the $ (r,g,b) $ coordinates as color channels or with a unique symbol.}
    \label{fig:barycentric}
\end{figure}

To encode the coordinates within each triangular subspace, we use a barycentric coordinate system (Figure \ref{fig:barycentric}).  A point within each triangle is described by a coordinate $(r,g,b)$ with $ r + g + b = 1 $, where  $ r,g,b \in [0,1] $ are the area fraction of the triangles formed by connecting that interior point to each of the vertices.  We use the convention that $ b $ describes the relative area of the triangle opposing the MAOS vertex, while the coordinate $ g $ opposes the right angle.  With this scheme, the third order modulus, $ G_3^*( \omega_1, \omega_2, \omega_3 ) $ can be specified in terms of an alternate set of coordinates: $\{ \mathcal{S}, r,g,b, |\boldsymbol{\omega}|_1 \}$, where $ \mathcal{S} $ is the specific triangular subspace associated with the point $ ( \omega_1, \omega_2, \omega_3 ) $.  While this may appear to be a more complicated representation, it helps break the data into physically distinct parts that are readily measurable and then easily visualized.  For example, the experimental procedure discussed in Section \ref{measurements} acquires values of the third order modulus at certain barycentric coordinates within the triangular subspaces regardless of the specified value of the frequency $L^1$-norm, $ |\boldsymbol{\omega}|_1 $.  Thus, multiple MAPS experiments might acquire data at different values of $|\boldsymbol{\omega}|_1$ on the same barycentric coordinate, and the value of the modulus at each barycentric coordinate can be plotted as a function of $ |\boldsymbol{\omega}|_1 $ much as with traditional Bode and Nyquist plots of linear response data.  The data at different barycentric coordinates can be demarcated by different symbols or colors and measurements of the modulus in different triangular subspaces plotted separately to aid visualization.  Extensive examples of these visualization schemes are given in Part 1 and the same methods are employed in this work.

Finally, many of the mathematical manipulations needed to relate experimentally measured quantities to the response functions in the Volterra series expansions involve summations over three indices for each of the components in frequency space.  Throughout this work, we will employ the shorthand notations:
\begin{equation*}
\sum_j \Longleftrightarrow \sum_{j=1}^3, \quad  \sum_{j \ne k} \Longleftrightarrow \sum_{j=1}^3 \sum_{\substack{k=1\\j\ne k}}^3,  \quad  \sum_{j,k,l} \Longleftrightarrow \sum_{j=1}^3 \sum_{k=1}^3 \sum_{l=1}^3 ,
\end{equation*}
to make these calculations more compact.

\section{Experimental Protocol for MAPS Rheology} 
\label{measurements}

\subsection{Medium Amplitude Stress Response to a Three-tone Strain}
\label{sec:three_tones}

As we have seen, the Volterra series is a useful representation of the higher order nonlinearities in a material's viscoelastic stress or strain response. The form of the third order complex modulus, viscosity, or their stress-controlled equivalents can be derived for a wide range of models of various complexity. Another distinct benefit of the Volterra series representation is that the third order response functions can be directly measured using widely available tools. We have seen in Part 1 that subsets of the complete third order response functions are currently measured in MAOS and PS experiments. These techniques, however, sample only from a few one-dimensional manifolds embedded in the full three-dimensional domain of the third order response functions. In other words, current techniques do not exploit the data-rich nature of nonlinear viscoelasticity, a richness that might prove extremely valuable in data science driven tasks such as material classification, model selection, and performance evaluation and optimization.

In this section, our interest turns to developing a procedure for using commercial rheometers to collect data sets that reflect the richness of a material's full, weakly nonlinear viscoelastic response space in simple shear. To motivate such a procedure, we can consider the stress response to a relatively simple, three-tone signal,
\begin{equation}
    \gamma(t) = \gamma_0 \sum_j \sin( \omega_j^* t + \alpha_j ),
    \label{eq:three_tones}
\end{equation}
where $ \omega_j^* $ and $\alpha_j$ are the frequency and phase of the $j$th tone, and $\gamma_0$ is the amplitude of the signal. Note that the input tones $\omega^*_j$ are distinct from the arguments to the third order complex modulus, $(\omega_1,\omega_2,\omega_3)$. In particular, though we have chosen here to explore signals with three input tones, we are free to explore the weakly nonlinear response to signals with more than three input tones using MAPS rheology; however, the third order complex modulus will always have three arguments by definition, regardless of the input signal under study. In the following analysis, we will see that a multi-tone input signal with input tones $\omega^*_j$ probes values of the third order complex modulus for which the arguments $(\omega_1,\omega_2,\omega_3)$ each take on a value from the set $\{-\omega^*_j, \omega^*_j\}$.

The Fourier transform of the strain signal in equation \ref{eq:three_tones} is:
\begin{align}
    \hat \gamma( \omega ) &= -i \pi \gamma_0 \sum_{j} \left[ e^{i \alpha_j } \delta( \omega - \omega_j^* ) - e^{-i \alpha_j } \delta( \omega + \omega_j^* ) \right] \nonumber \\
    &= -i \pi \gamma_0
    \sum_j \sum_{p=1}^2 s_p e^{i s_p \alpha_j } \delta( \omega - s_p \omega_j^* ),
    \label{eq:three_tone_FT}
\end{align}
with $ s_p = (-1)^p $.  The linear response to this signal is given by substituting the above expression into equation \ref{eq:strain_controlled3} and considering only the first term in the expansion:
\begin{widetext}
\vspace{-6mm}
\begin{align}
    \hat \sigma( \omega ) &= -i \pi \gamma_0
    \sum_j \sum_{p=1}^2 G_1^*( s_p \omega_j^* ) s_p e^{i s_p \alpha_j } \delta( \omega - s_p \omega_j^* ) \label{eq:threetonesLR} \\
    &= -i\pi\gamma_0\left[G^*_1(\omega^*_1)e^{i\alpha_1}\delta(\omega - \omega^*_1) - G^*_1(-\omega^*_1)e^{-i\alpha_1}\delta(\omega + \omega^*_1) + G^*_1(\omega^*_2)e^{i\alpha_2}\delta(\omega - \omega^*_2) - G^*_1(-\omega^*_2)e^{-i\alpha_2}\delta(\omega + \omega^*_2)\right. \nonumber \\
    &\quad\quad\quad\quad \left. + \, G^*_1(\omega^*_3)e^{i\alpha_3}\delta(\omega - \omega^*_3) - G^*_1(-\omega^*_3)e^{-i\alpha_3}\delta(\omega + \omega^*_3)\right] \nonumber
\end{align}
\end{widetext}
The delta functions in the above expression are like discrete channels along which data about the complex modulus is transmitted. For example, the term containing $\delta(\omega - \omega^*_1)$ in equation \ref{eq:threetonesLR} multiplies $G^*_1(\omega_1)$, thus the stress response at frequency $\omega = \omega^*_1$ carries information about this particular element of the linear response. If the three input tones are distinct, then:
\begin{equation*}
\int_{\omega_1^* -\epsilon}^{\omega_1^* +\epsilon} \hat \sigma( \omega ) d \omega = -i \pi \gamma_0 e^{ i \alpha_1 } G_1^*( \omega_1^* ).
\end{equation*}
Just as in a single-tone small amplitude oscillatory shear (SAOS) experiment, the stress in the immediate vicinity ($\pm \epsilon$) of the first harmonic of an input tone measures an element of the linear response. A three-tone input therefore produces three distinct measurements of $G^*_1(\omega)$, one at each input frequency \cite{winter-1988}. Information about $G^*_1(\omega)$ also transmits along channels at negative frequency; however, the information on these channels is related to the information transmitted on the positive channels by Hermitian symmetry. For this reason, information content at negative frequencies is typically not considered in linear viscoelasticity. 

Note that here, we refer to the channels that correspond to the frequencies of each input tone as `first harmonic' channels. In a single-tone experiment, the `first harmonic' of the input signal is synonymous to the `fundamental frequency', thus these terms are often used interchangeably. For multi-tone signals, however, we will shortly demonstrate that the fundamental frequency $\omega_0$ is defined separately from the frequencies of each input tone, thus the terms `first harmonic' and `fundamental frequency' are no longer synonymous. In particular, there exists a first harmonic channel for each input tone in a multi-tone signal, corresponding exactly to the frequency of that input tone, while there exists only one fundamental frequency for the multi-tone signal.

Now, considering both the first- and third order terms upon substitution of equation \ref{eq:three_tone_FT} into equation \ref{eq:strain_controlled3}, integration with respect to the primed frequencies yields the Fourier transform of the weakly nonlinear shear stress response:
\begin{align}
    &\hat \sigma( \omega ) = -i \pi \gamma_0
    \sum_j \sum_{p=1}^2 G_1^*( s_p \omega_j^* ) s_p e^{i s_p \alpha_j } \delta( \omega - s_p \omega_j^* )  \label{eq:threetonestress} \\
    & \quad + \frac{i \pi \gamma_0^3}{4} \sum_{j,k,l} \sum_{p,q,r=1}^2 s_p s_q s_r e^{i s_p \alpha_j + i s_q \alpha_k + i s_r \alpha_l }   \nonumber \\
    & \quad \times G_3^*\left( s_p \omega_j^*, s_q \omega_k^*, s_r \omega_l^* \right) \delta( \omega - s_p \omega_j^* - s_q \omega_k^* - s_r \omega_l^* ) . \nonumber
\end{align}
The delta functions in the second term of the above expression reveal the channels on which data about the third order modulus are transmitted.  The term containing $ \delta( \omega - 3 \omega_1^* ) $ coming from the sextuple sum in equation \ref{eq:threetonestress} multiplies $ G_3^*( \omega_1^*, \omega_1^*, \omega_1^* ) $, for example.  This indicates that the stress response at frequency $\omega = 3 \omega_1^* $ carries information about this particular element of the third order response.  Assuming that no other terms in the sextuple sum transmit on this same channel, then:
\begin{equation*}
\int_{3 \omega_1^* -\epsilon}^{3 \omega_1^* +\epsilon} \hat \sigma( \omega ) d \omega = \frac{i \pi^3 \gamma_0^3}{4} e^{3 i \alpha_1 } G_3^*( \omega_1^*, \omega_1^*, \omega_1^* ),
\vspace{2mm}
\end{equation*}
and, as in MAOS, the stress in the immediate vicinity of the third harmonic of an input tone measures an intrinsic nonlinearity in the material response. Shortly, we will describe the conditions that the input tones $\omega^*_j$ must obey for this assumption to hold.

\begin{figure}[t]
    \centering
    \includegraphics[width = \columnwidth]{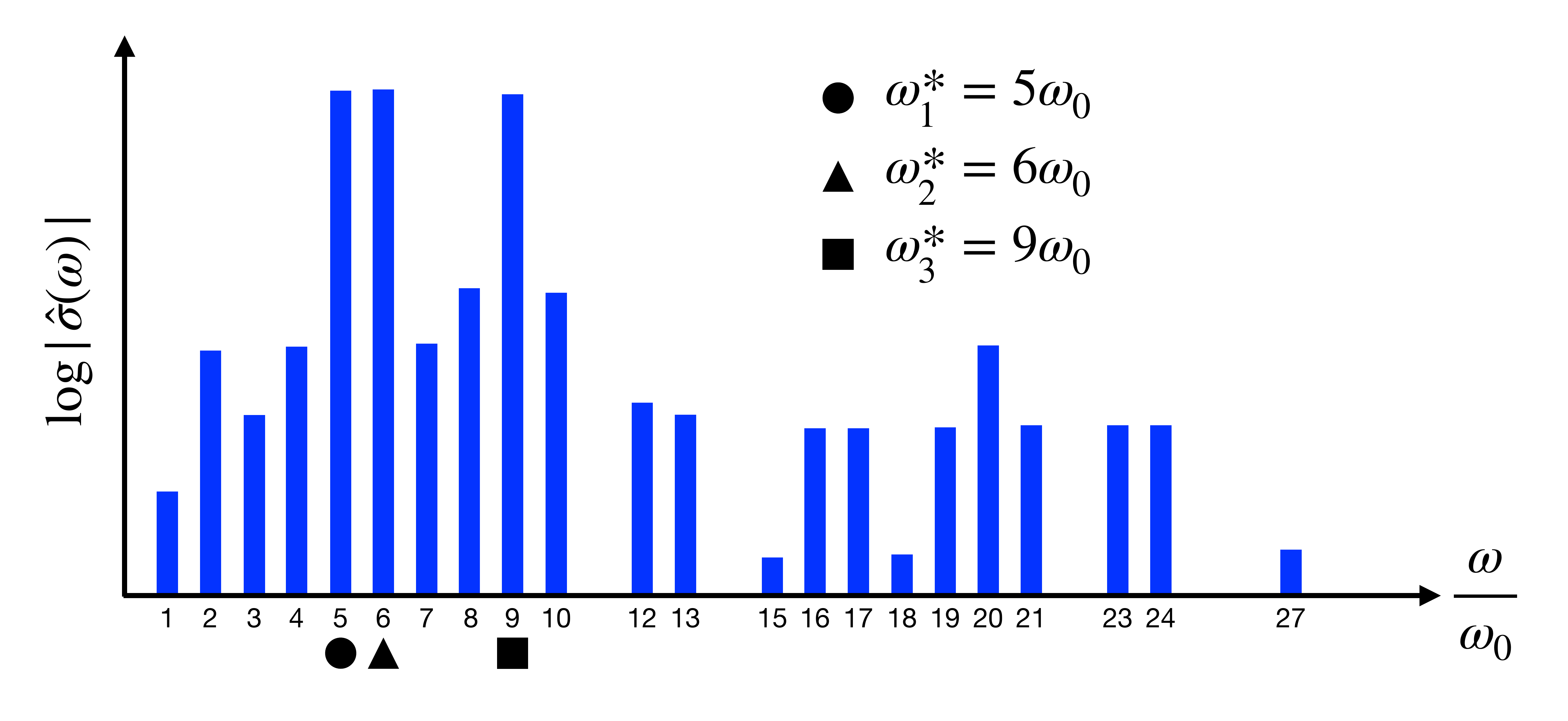}
    \caption{An example MAPS response of a material to a three-tone deformation of the form of equation \ref{eq:three_tones} with $\{\omega^*_1,\omega^*_2,\omega^*_3\} = \{5\omega_0,6\omega_0,9\omega_0\}$. The magnitude of the stress response on each of the 22 channels given by equation \ref{eq:channelterms} is represented by a blue bar, and the response at the first harmonic of each input tone is labelled. Because the magnitude of the stress response is an even function of the frequency, the Hermitian symmetric terms in equation \ref{eq:channelterms} can be found by reflecting each bar about the y-axis.}
    \label{fig:example_response}
\end{figure}

Given that the three input tones are distinct, the sextuple sum in equation \ref{eq:threetonestress} expands to a sum over 44 terms each containing a distinct delta function of the form $ \delta( \omega \pm \omega_i^* \pm \omega_j^* \pm \omega_k^* ) $ for $ i, j, k \in \{1, 2, 3\} $.  Because of Hermitian symmetry, only 22 of those terms transmit unique information about the third order response function.  Application of permutation symmetry to the response function can be used to write the Fourier transform of the stress compactly in terms of the 22 unique channels: 
\begin{widetext}
\vspace{-5mm}
\begin{align}
    \hat \sigma( \omega ) &= -i \pi \gamma_0 \sum_{j}  G_1^*( \omega^*_j ) e^{i\alpha_j} \delta( \omega - \omega^*_j ) + \frac{i \pi \gamma_0^3}{4} \left\{\sum_{j} e^{3i\alpha_j} G_3^*( \omega^*_j, \omega^*_j, \omega^*_j ) \delta( \omega - 3 \omega^*_j ) \right. \nonumber \\
    & + 6 e^{i\sum_j\alpha_j} G_3^*( \omega^*_1, \omega^*_2, \omega^*_3 ) \delta( \omega - \omega^*_1 - \omega^*_2 - \omega^*_3 ) - 6 e^{i(\alpha_1 + \alpha_2 - \alpha_3)} G_3^*( \omega^*_1, \omega^*_2, -\omega^*_3 ) \delta( \omega - \omega^*_1 - \omega^*_2 + \omega^*_3 ) \nonumber \\
    & - 6 e^{i(\alpha_1 - \alpha_2 + \alpha_3)} G_3^*( \omega^*_1, -\omega^*_2, \omega^*_3 ) \delta( \omega - \omega^*_1 + \omega^*_2 - \omega^*_3 ) + 6 e^{i(\alpha_1 - \alpha_2 - \alpha_3)} G_3^*( \omega^*_1, -\omega^*_2, -\omega^*_3 ) \delta( \omega - \omega^*_1 + \omega^*_2 + \omega^*_3 ) \nonumber \\
    & + 3 \sum_{j \ne k} \left[ e^{i(\alpha_j - 2\alpha_k)} G_3^*( \omega^*_j, -\omega^*_k, -\omega^*_k ) \delta( \omega - \omega^*_j + 2 \omega^*_k ) + e^{i(\alpha_j + 2\alpha_k)} G_3^*( \omega^*_j, \omega^*_k, \omega^*_k ) \delta( \omega - \omega^*_j - 2 \omega^*_k ) \right] \nonumber \\
    & \left. - 3  \left[ \sum_{j} e^{i\alpha_j} G_3^*( \omega^*_j, \omega^*_j, - \omega^*_j ) \delta( \omega - \omega^*_j ) + 2 \sum_{j\ne k} e^{i\alpha_j} G_3^*( \omega^*_j, \omega^*_k, -\omega^*_k ) \delta( \omega - \omega^*_j ) \right] \right\} + \textrm{HSTs}, \label{eq:channelterms}
\end{align}
\end{widetext}
which contains one half of the frequency response spectrum.  The other half of the channels are denoted as HSTs (Hermitian symmetric terms) and can be derived by negating the frequencies in the delta functions and applying the principle of Hermitian symmetry to the corresponding response functions. Even considering only the 22 terms that are unique by Hermitian symmetry, many terms multiply values of the third order complex modulus with negative frequency arguments. Therefore, considering both positive and negative frequencies is critical in nonlinear rheology. Figure \ref{fig:example_response} depicts a specific example of the 22 channels that are excited in a MAPS experiment with $\{\omega^*_1,\omega^*_2,\omega^*_3\} = \{5\omega_0,6\omega_0,9\omega_0\}$, and is a graphical depiction of equation \ref{eq:channelterms} with delta functions represented by blue bars.

The channels $ \omega \in \{ \omega^*_1, \omega^*_2, \omega^*_3 \} $ carry information about the linear response and a superposition of different values of the third order complex modulus.  This is made most clear by examining the second set of square brackets in equation \ref{eq:channelterms}, which contains terms like:
\begin{align*}
& \left[ G_3^*( \omega^*_1, \omega^*_1, -\omega^*_1 ) \right. \\
& \left. + 2 G_3^*( \omega^*_1, \omega^*_2, -\omega^*_2 ) + 2 G_3^*( \omega^*_1, \omega^*_3, -\omega^*_3 ) \right] \delta( \omega- \omega^*_1 ).
\end{align*}
This linear combination of values of the third order complex modulus can be determined as a single composite group through regression using stress measurements at different strain amplitudes as will be discussed in more detail in the next sections.  However, these particular three superposed values cannot be separated using the methods discussed in this work. In future work, we will present a more sophisticated experimental design that is able to segregate multiple values of the third order complex modulus residing on the same frequency channel.

It is possible that other channels in the stress response transmit information about commingled elements of third order modulus. In any such cases, the experimental protocol discussed in this work is unable to segregate the individual elements of the third order modulus that reside on these channels, as is the case for those residing on the first harmonic channels. The channels containing information about only a single element of the modulus are those whose frequency is a unique sum taken over three values drawn from the set $ \{\pm \omega^*_1, \pm \omega^*_2, \pm \omega^*_3\}$ \cite{boyd-1983,chua-1989}. As discussed in the next section, it is possible to choose the three input tones as integer multiples of a fundamental frequency $\omega_0$ such that the remaining 19 channels besides the first harmonic channels at $\omega_1^*$, $\omega_2^*$, and $\omega_3^*$, respectively, satisfy this condition, and thus transmit information about a single element of the third order modulus. In principle, the 19 values of the third order complex modulus that reside on these channels can be directly inferred from a single measurement. However, the measured values in this case would be highly prone to error from noise and spectral leakage from the first harmonic channels. A method for regressing the values of the third order complex modulus at these channels from discrete amplitude sweeps in order to minimize error is presented in Section \ref{sec:vandermonde}. 

Out of the 19 elements of the third order modulus that can be obtained from a three-tone experiment, three correspond to points measurable in MAOS: $G^*_3(\omega^*_j,\omega^*_j,\omega^*_j)$, $j = 1,2,3$. The presence of higher harmonics of the input tones in the output of a nonlinear system is a well-known phenomenon, commonly called \textit{harmonic distortion}. The remaining 16 elements of the third order modulus measured from a three-tone experiment correspond to points not measurable by either MAOS or PS. These points reside on either the edge or interior of the MAPS subspaces labelled in Figure \ref{fig:octahedron}, and appear in the output at frequencies that are triplet sums and differences of the input tones. The phenomenon by which nonlinear systems produce responses at these frequencies is called \textit{intermodulation distortion}, or simply \textit{intermodulation} \cite{volterra-1959}. We see from this simple three-tone example that intermodulation reveals a much richer set of data than harmonic distortion alone. Like harmonic distortion, however, the time-scale at which intermodulation effects occur is always shorter than that set by the fundamental frequency of the input signal, $\omega_0$. Though it may be unclear at this point what we mean by the ``fundamental'' frequency of a multi-tone input signal, we will see shortly that this frequency can be easily defined and controlled. For now, it suffices to say that $\omega_0$ is the minimum frequency resolution necessary to resolve all output channels, equal to the greatest common denominator of the input tones. Therefore, in principle the size of the measured data set can be freely increased by adding higher frequency tones to the input signal, thus increasing the number of intermodulation modes, without increasing the acquisition time, which is set by the fundamental frequency $\omega_0$ or the corresponding time scale $t_{\mathrm{exp}} \sim 2\pi/\omega_0$. 

From the discussion in this section, it is clear why the multi-tone input signal set forth in equation \ref{eq:three_tones} is appropriate for measurements in MAPS rheology. With only three tones, this protocol can unambiguously access up to 19 distinct points in the domain of the MAPS response functions. At the same time, constructing such input signals is feasible in many commercial rheometers, and the resulting data output by the rheometer requires only a modest amount of data processing. The remainder of this paper is dedicated to developing the experimental protocol for MAPS rheology centered around the three-tone input signal of equation \ref{eq:three_tones}. Though we focus on the three-tone signal here as an introduction to MAPS rheological experimentation, the discussion in this section can be amended to suit any number of input tones, which can further increase the data-throughput of MAPS experiments. The discussion of these more complex protocols is left to future work.

\subsection{Design of MAPS Experiments With Three-tone Input Signals}

\begin{figure*}[ht!]
    \centering
    \includegraphics[width=\textwidth]{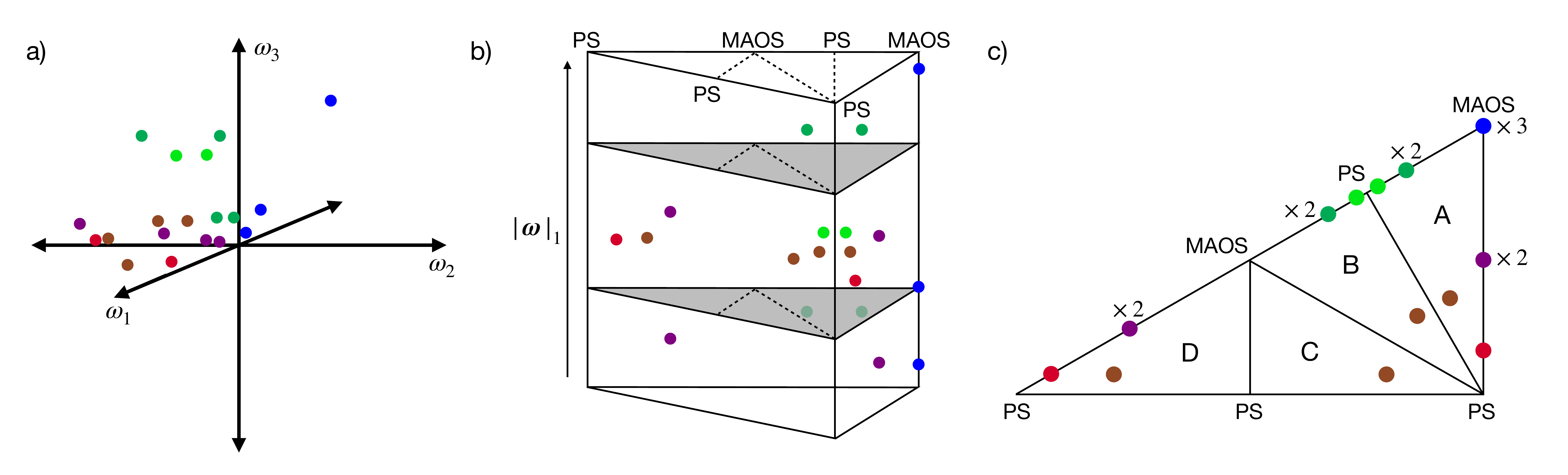}
    \caption{Different qualitative depictions of the points in the MAPS domain probed by an input signal given by equations \ref{eq:three_tones} and \ref{eq:w_to_n} with $\{n_1,n_2,n_3\} = \{1,4,16\}$. a) Depiction of points probed in the domain $(\omega_1,\omega_2,\omega_3)$. Only points that fall into the MAPS subspaces described in equation \ref{eq:subspaces} are shown. b) The spread of points in the domain $\{ \mathcal{S}, r,g,b, |\boldsymbol{\omega}|_1 \}$, depicted as a triangular prism with the triangular faces representing constant $L^1$-norm MAPS surfaces laid flat as in Figure \ref{fig:barycentric}, scaled to the same area, and the vertical dimension representing the frequency $L^1$-norm: $|\boldsymbol{\omega}|_1$. c) Projection of each measured point onto the four triangular MAPS subspaces, equivalent to looking down from above the triangular prism shown in (b).}
    \label{fig:projections}
\end{figure*}

Equation \ref{eq:three_tones} is perhaps the simplest example of a multi-tone input signal that can be used in MAPS rheology.  This simplicity allows for such signals to be easily implemented in many commercial rheometers using either built-in multi-tone functionality, arbitrary waveform tools, or a signal generator. Still, there is considerable flexibility in selecting where in the three-dimensional frequency space (Figure \ref{fig:octahedron}) a particular set of three tones probes the third order response functions, and in choosing which response function to probe. The flexibility of the input signal is manifested in the design variables: $\gamma_0$, $\omega^*_j$, and $\alpha_j$, as well as the choice of imposing a controlled-strain or controlled-stress deformation. Each of these design variables controls a distinct feature of the experimental study. In particular, $\gamma_0$ modulates the measurement error, the set of $\omega_j$ controls the regions of three-frequency space probed by the deformation, and the set of $\alpha_j$ set the peak values of the input signal and its derivative, which are important for ensuring that the deformation does not exceed the physical limitations of the rheometer. Controlled-strain experiments will directly measure the third order complex modulus (or complex viscosity), while controlled-stress experiments directly measure the third order complex compliance (or complex fluidity). In this section, we explore the consequences of changing each of these design variables, as well as the choice of stress vs. strain control, and outline best practices for selecting their values.

\subsubsection{Selecting the Input Tones}
\label{sec:input_tones}

For the three-tone input signal, a principal constraint is that the set of tones $\left\{ \pm \omega^*_1, \pm \omega^*_2, \pm \omega^*_3\right\}$ contains only unique frequency triplet sums so that the 19 elements of the third order modulus not residing on first harmonic channels can be independently calculated. In practice, it is useful to choose the tones as integer multiples of a fundamental frequency, $ \omega_0 $ so that:
\begin{equation}
    \omega^*_j = n_i\omega_0 \quad \textrm{for} \quad i = 1, 2, 3, \quad n_1 < n_2 < n_3.
    \label{eq:w_to_n}
\end{equation} 
Ensuring unique frequency triplet sums becomes equivalent to ensuring unique triplet sums of the set of integers: $\{-n_3, -n_2, -n_1, n_1, n_2, n_3 \}$. If this is the case, then the 19 channels listed in the first column of Table \ref{tab:channels} are distinct, and the corresponding values of the third order modulus listed in the second column can be determined directly from the material's stress response on the corresponding channel.

\begin{table}[t]
    \centering
    \begin{tabular}{c|c}
        Channel & Measured stress response $ / \left( i \pi \gamma_0^3/4 \right)$ \\  \hline
         $ \omega_0 ( n_1 + n_2 + n_3 )$ & $ 6 e^{i(\alpha_1 + \alpha_2 + \alpha_3)} \times G_3^*( \omega_0 n_1, \omega_0 n_2, \omega_0 n_3 )$ \\ 
         $ \omega_0 ( n_1 + n_2 - n_3 )$ & $ -6 e^{i(\alpha_1 + \alpha_2 - \alpha_3)} \times G_3^*( \omega_0 n_1, \omega_0 n_2, -\omega_0 n_3 )$ \\
         $ \omega_0 ( n_1 - n_2 + n_3 )$ & $ -6 e^{i(\alpha_1 - \alpha_2 + \alpha_3)} \times G_3^*( \omega_0 n_1, -\omega_0 n_2, \omega_0 n_3 )$ \\
         $ \omega_0 ( n_1 - n_2 - n_3 )$ & $ 6 e^{i(\alpha_1 - \alpha_2 - \alpha_3)} \times G_3^*( \omega_0 n_1, -\omega_0 n_2, -\omega_0 n_3 )$ \\
         $ \omega_0 ( 2 n_1 + n_2 )$ & $ 3 e^{i(2\alpha_1 + \alpha_2)} \times G_3^*( \omega_0 n_1, \omega_0 n_1, \omega_0 n_2 )$ \\
         $ \omega_0 ( 2 n_1 - n_2 )$ & $ -3 e^{i(2\alpha_1 - \alpha_2)} \times G_3^*( \omega_0 n_1, \omega_0 n_1, -\omega_0 n_2 )$ \\
         $ \omega_0 ( 2 n_1 + n_3 )$ & $ 3 e^{i(2\alpha_1 + \alpha_3)} \times G_3^*( \omega_0 n_1, \omega_0 n_1, \omega_0 n_3 )$ \\
         $ \omega_0 ( 2 n_1 - n_3 )$ & $ -3 e^{i(2\alpha_1 - \alpha_3)} \times G_3^*( \omega_0 n_1, \omega_0 n_1, -\omega_0 n_3 )$ \\
         $ \omega_0 ( 2 n_2 + n_1 )$ & $ 3 e^{i(2\alpha_2 + \alpha_1)} \times G_3^*( \omega_0 n_2, \omega_0 n_2, \omega_0 n_1 )$ \\
         $ \omega_0 ( 2 n_2 - n_1 )$ & $ -3 e^{i(2\alpha_2 - \alpha_1)} \times G_3^*( \omega_0 n_2, \omega_0 n_2, -\omega_0 n_1 )$ \\
         $ \omega_0 ( 2 n_2 + n_3 )$ & $ 3 e^{i(2\alpha_2 + \alpha_3)} \times G_3^*( \omega_0 n_2, \omega_0 n_2, \omega_0 n_3 )$ \\
         $ \omega_0 ( 2 n_2 - n_3 )$ & $ -3 e^{i(2\alpha_2 - \alpha_3)} \times G_3^*( \omega_0 n_2, \omega_0 n_2, -\omega_0 n_3 )$ \\
         $ \omega_0 ( 2 n_3 + n_1 )$ & $ 3 e^{i(2\alpha_3 + \alpha_1)} \times G_3^*( \omega_0 n_3, \omega_0 n_3, \omega_0 n_1 )$ \\
         $ \omega_0 ( 2 n_3 - n_1 )$ & $ -3 e^{i(2\alpha_3 - \alpha_1)} \times G_3^*( \omega_0 n_3, \omega_0 n_3, -\omega_0 n_1 )$ \\
         $ \omega_0 ( 2 n_3 + n_2 )$ & $ 3 e^{i(2\alpha_3 + \alpha_2)} \times G_3^*( \omega_0 n_3, \omega_0 n_3, \omega_0 n_2 )$ \\
         $ \omega_0 ( 2 n_3 - n_2 )$ & $ -3 e^{i(2\alpha_3 - \alpha_2)} \times G_3^*( \omega_0 n_3, \omega_0 n_3, -\omega_0 n_2 )$ \\
         $3 \omega_0 n_1$ & $ e^{3i\alpha_1} \times G_3^*( \omega_0 n_1, \omega_0 n_1, \omega_0 n_1 )$ \\
         $3 \omega_0 n_2$ & $ e^{3i\alpha_2} \times G_3^*( \omega_0 n_2, \omega_0 n_2, \omega_0 n_2 )$ \\
         $3 \omega_0 n_3$ & $ e^{3i\alpha_3} \times G_3^*( \omega_0 n_3, \omega_0 n_3, \omega_0 n_3 )$ \\
    \end{tabular}
    \caption{Stress response channels and the corresponding measured values of the third order complex modulus for a three-tone strain signal. In the right column, the stress is normalized by a constant prefactor and by the cube of the strain amplitude.}
    \label{tab:channels}
\end{table} 

Based on this constraint alone, there are an infinite number of integer triplets admissible for a MAPS experiment. Not all of these combinations are practical, however. In particular, very high harmonics of the fundamental frequency will cause issues in both implementation and data collection.  For example, commercial rheometers contain control software that will constrain the highest harmonic ($n_3$) below a certain threshold, through a low-pass filter, in order to avoid damaging the instrument as it drives the oscillating tool. Because the highest frequency channel in the stress response is at three times the highest frequency input tone, there are additional complications for the measurement.  A large disparity between the fundamental frequency ($\omega_0$) and the highest frequency component ($3n_3\omega_0$) will require both a very high sampling rate and a long measurement duration \cite{boyd-1983}.  This can drastically expand the amount of data needed to accurately evaluate the Fourier transform of the stress response. To avoid all of these complications, one should consider sets of positive integers for which $ n_3 $ is sufficiently small.

Even if the maximum input tone is constrained to a modest value, for example the 16th harmonic, there are many options for the three-tone frequency set that meet the uniqueness criterion. A brute force search finds that there are 146 such positive integer triplets with a maximum member smaller than 16. We will not present an exhaustive list of all possible combinations.  Instead, we consider a few examples that demonstrate important features of these sets as they relate to the triangular subspaces of the MAPS domain, and we provide guidelines for selecting sets of tones for MAPS measurements.

\begin{figure}[t]
    \centering
    \includegraphics[scale=0.35]{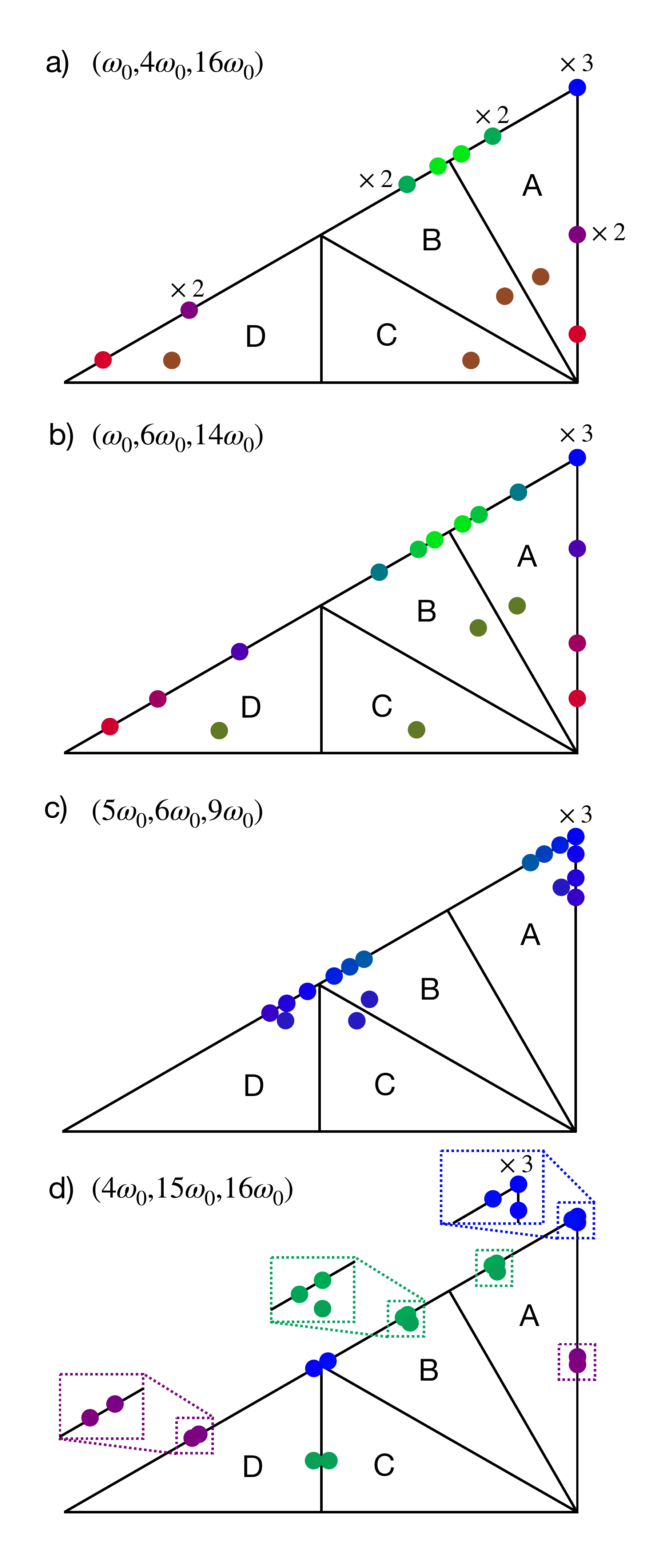}
    \caption{Projections of the 19 points of $G^*_3(\omega_1,\omega_2,\omega_3)$ measured by a three-tone strain-controlled experiment onto the constant $L^1$-norm triangular subspaces, for four different examples of integer triplet sets of input tones. Points are colored based on the barycentric coloring scheme from Figure \ref{fig:barycentric}.}
    \label{fig:frequencies}
\end{figure}

Though each of these 146 three-tone input sets leads to a stress response that samples unambiguously from 19 different points of the third order modulus, the specific values of $n_1$, $n_2$, and $n_3$ significantly affect where in the three-dimensional frequency space those 19 points reside. Choosing $\{n_1,n_2,n_3\} = \{1,4,16\}$, for example, produces the distribution of points in $\mathbb{R}^3$ shown in Figure \ref{fig:projections}a. To more conveniently visualize the spread of the points in the MAPS domain, we can project each point onto the planar depiction of the MAPS subspaces along different constant $L^1$-norm surface as shown in Figure \ref{fig:barycentric}, re-scale these surfaces to a constant size, then arrange increasing $L^1$-norm surfaces vertically to create a triangular prism as shown in Figure \ref{fig:projections}b. This is an equivalent representation of the data points, now in terms of the coordinates $\{ \mathcal{S}, r,g,b, |\boldsymbol{\omega}|_1 \}$. The value of the frequency $L^1$-norm at which the points reside, represented by the vertical scale in Figure \ref{fig:projections}b, can be adjusted freely through the choice of $\omega_0$ (though the relative vertical spacing between points is fixed for a given choice of $\{n_1,n_2,n_3\}$). The effect of choosing different integers $\{n_1,n_2,n_3\}$ on the distribution of the sampled data can be most clearly viewed by projecting the points in the prism downward onto a single triangular surface, as shown in Figure \ref{fig:projections}c. These projections are determined directly by the choice of $\{n_1,n_2,n_3\}$, and are independent of the fundamental frequency $\omega_0$ chosen for the input signal. These projections are quite different for different sets of integers, and are a convenient visualization tool for assessing how broadly and how densely these integer triplet sets probe certain regions in the MAPS domain. Figure \ref{fig:frequencies} depicts similar projections for four different sets of integers.

These projections illuminate some similarities between all integer triplet sets. First, each frequency set produces only ten points in subspace A, four points in each of subspaces B and D, and one point in subspace C. Second, we see that the three measurements corresponding to the third harmonics of the input tones always reside at the same projected point in subspace A: the MAOS vertex. Because these three measurements appear on different $L^1$-norm surfaces (as shown in Figure \ref{fig:projections}b), they represent in essence a MAOS frequency sweep over three frequencies. Third, there is always just one point per subspace that is not located on an edge or vertex. This point resides at the same barycentric coordinate within each triangle. These same observations are general for any set of three integers with unique triplet sums and can be deduced from the structure of the channels depicted in the Fourier transform of the stress (equation \ref{eq:channelterms}), but they are especially clear when considering examples graphically.  Besides these three general observations, different sets of $\left\{n_1,n_2,n_3\right\}$ produce quite different spreads of points within the triangles. The important characteristics of the four example sets shown in Figure \ref{fig:frequencies} are considered below.

\begin{enumerate}[a.]
    \item $\left\{1,4,16\right\}$
    
    The case where $\left\{n_1,n_2,n_3\right\} = \left\{1,4,16\right\}$ is interesting because the three input tones are evenly spaced on a logarithmic scale. Therefore, the 19 MAPS points tend to distribute themselves rather uniformly throughout the triangles (particularly in triangle A, which contains the most independent measurement points). This is a good candidate signal for obtaining measurements near the PS vertices as well as at a MAOS vertex. Due to the even logarithmic spacing of the tones, there are overlapping points along the edges of the subspaces in barycentric space. The pair of three-frequency coordinates $(4,1,1)\omega_0$ and $(16,4,4)\omega_0$ and the pair $(4,1,4)\omega_0$ and $(16,4,16)\omega_0$ are each located at the same barycentric coordinates on edges of subspace A, as are the pair $(4,-1,4)\omega_0$ and $(16,-4,16)\omega_0$ in subspace B and the pair $(1,-4,1)\omega_0$ and $(4,-16,4)\omega_0$ in subspace D. The multiplicity of these projected points are indicated in Figure \ref{fig:frequencies}a. Because these points reside at different $|\boldsymbol{\omega}|_1$, they do provide unique information about the underlying MAPS response function. In a sense, these points represent a truncated ``frequency sweep'' at the respective barycentric coordinate. However, having such overlap decreases the density of sampled points in barycentric space.
    
    \item $\left\{1,6,14\right\}$
    
    The integer set $\left\{1,6,14\right\}$ is similar to the set $\left\{1,4,16\right\}$, in that the three input tones are spaced rather widely, but the spacing is no longer uniform on a logarithmic scale. Therefore, this set is still able to sample the triangular spaces quite uniformly, with points near the MAOS and PS vertices, while having no additional overlapped points along the edges of the subspaces in barycentric space. Thus this set is able to sample the barycentric space more densely than did the triplet $\left\{1,4,16\right\}$, but does not provide the same ``frequency sweep''-like data at certain barycentric coordinates.
    
    \item $\left\{5,6,9\right\}$
    
    In many cases, the integers in the admissible set $\left\{n_1,n_2,n_3\right\}$ are not as spread out as in the above examples. One example of a more closely spaced set is $\left\{5,6,9\right\}$. Because the maximum integer in this set is only 9, this set might be easier to implement in a rheometer than the two widely-distributed sets above. However, sets in which the three harmonics are relatively close in value tend to produce points that are closely clustered around the MAOS vertices. Therefore, the resulting data sets of the material response will not capture the full spectrum of weakly nonlinear behavior, especially pertaining to near-PS points.
    
    \item $\left\{4,15,16\right\}$
    
    The set $\left\{4,15,16\right\}$ shows traits from both the widely and closely spaced sets. Here, the highest two input tones are close in value, but are much greater in frequency than the lowest tone. Because of this, the projected points tend to cluster closely in groups of two or three, but the clusters are spread across the subspaces.  Therefore, sets such as this can densely sample certain areas near both the MAOS vertices and the PS vertices, but cannot broadly sample the entire triangular spaces.
    
\end{enumerate}

The four example input tone sets considered above illustrate the importance of choosing the proper input tones to suit specific experimental goals. If the primary goal is to sample a material's third order response as broadly as possible, one should use input tones that most uniformly span the triangular subspaces. The uniformity of a set in these spaces can be assessed visually, or by some appropriate numerical measure, such as the average nearest-neighbor distance of points within a given triangular subspace. If, however, the goal is to sample particular sections of the response surfaces as densely as possible, such as the regions near the MAOS vertices, the most suitable input tones will not produce the most uniform sampling but rather the densest sampling in the desired region.

\subsubsection{The Vandermonde Method for Channel Estimation}
\label{sec:vandermonde}

Given a three tone input, let $ \omega^* $ be one of the 19 channels of interest in the Fourier transform of the shear stress containing unique information about third order complex modulus.  Then in the limit of vanishingly small strain amplitudes, the value of the stress measured in this channel can, in principle, be used directly. There are experimental complications, however,  that make directly inferring values of the third order modulus from a single experiment difficult. For one, rheometers have finite sensitivity. Thus the strain amplitude must be small enough such that effects above third order are negligible, but large enough that the third order response is measurable above the noise floor. Another complication is that the present experimental design requires measurement of third order features in the stress on channels very near to others that carry the linear response. The linear response has much greater magnitude than the third order features, and spectral leakage from the linear response can influence measured values of the third order complex modulus on the nearby channels. For these reasons, directly inferring values of the third order modulus from a single experiment is not recommended.

In other measurements of intrinsic nonlinearities in viscoelastic materials, such as MAOS, it is common to apply an amplitude sweep at a constant frequency, and fit the response to a cubic polynomial to infer values of the intrinsic material functions \cite{ewoldt-2013}. One could apply this same methodology to MAPS to obtain estimates for the third order modulus that are robust to noise and spectral leakage. Though a thorough amplitude sweep will improve the estimates of the third order modulus, this procedure is quite time-consuming, and will drastically decrease the data-throughput of the experimental technique. However, it has recently been shown that third order measurements can be accurately segregated from first order effects in MAOS with only two experiments at different amplitudes, as long as both measurements are taken within an amplitude range known to excite third order nonlinearities without substantial higher order effects \cite{singh-2018}. Here, we will develop a similar procedure for isolating third order effects in MAPS from first order effects such as spectral leakage.

To do so, two strain amplitudes are selected, $ \gamma_a $ and $\gamma_b $, to replace $\gamma_0$ in the strain input expression in equation \ref{eq:three_tones}. In the weakly nonlinear limit, the Fourier transform of the stress in the experiment with strain amplitude $ \gamma_a $ on a channel of interest, $ \omega^* $, can be expressed as a third order polynomial:
\begin{equation*}
\hat \sigma_a( \omega^* ) = \gamma_a \hat \sigma^{(1)}(\omega^*) + \gamma_a^3 \hat \sigma^{(3)}( \omega^* ),
\end{equation*}
and likewise for the stress resulting from the experiments with strain amplitude $ \gamma_b $.  The terms $ \hat \sigma^{(i)}( \omega^* ) $ are the coefficients of the order $ \gamma_a^i $ contributions to the Fourier transform of the stress.  By convention, we choose the following relationship for the strain amplitudes in the two experiments: $ \gamma_a = r \gamma_b $, with $ 0 < r < 1 $. In the small amplitude limit, the coefficients of the stress, $ \hat \sigma^{(i)}( \omega^* ) $, are common across experiments at different strain amplitude and can be determined from the solution to a system of linear equations:
\begin{equation}
\left(\begin{array}{c} \hat \sigma_a( \omega^* ) \\ \hat \sigma_b( \omega^* ) \end{array}\right) = \left(\begin{array}{ccc}  r & r^3 \\ 1 & 1 \end{array}\right) \left(\begin{array}{c} \gamma_b \hat \sigma^{(1)}( \omega^* ) \\ \gamma_b^3 \hat \sigma^{(3)}( \omega^* ) \end{array}\right).  \label{eq:vandermonde}
\end{equation}
The coefficient matrix in equation \ref{eq:vandermonde}, denoted $ \mathbf{V} $, is termed the Vandermonde matrix \cite{boyd-1983,klinger-1967}. The $L^\infty$-norm condition number of the Vandermonde matrix is:
\begin{equation}
    \| \mathbf{V} \|_\infty \| \mathbf{V}^{-1} \|_\infty = \frac{2 + 2r}{r - r^3}.
\end{equation}
The value of $r$ that minimizes the condition number of the Vandermonde matrix is: $ r = \frac{1}{2} $, with minimal condition number: 8.  This means that relative uncertainties in the stress coefficients are guaranteed to be smaller than eight times the relative uncertainties in the measured stresses. With this value of $ r $, the experimental design is optimal for estimating $ \hat \sigma^{(i)}( \omega^* ) $ from solution of equation \ref{eq:vandermonde}.  The value of $ \hat \sigma^{(1)}( \omega^* ) $ found from solution of this equation can be used to determine the linear response if $ \omega^* $ is one of the three tones in the input strain.  The value of $ \hat \sigma^{(3)}( \omega^* ) $ determined from solution of equation \ref{eq:vandermonde} on all the other channels of interest can be used to determine the third order modulus by reading off the appropriate row of Table \ref{tab:channels}. For example, if $\omega^* = \omega_0(2n_1 + n_2)$, then:
\begin{equation}
    G^*_3(\omega_0 n_1, \omega_0 n_1, \omega_0 n_2) = \frac{4}{3 i \pi} e^{-i(2\alpha_1 + \alpha_2)}\hat{\sigma}^{(3)}(\omega_0(2n_1 + n_2)).
\end{equation}

It is important to note that while this cubic regression mitigates bias in third order measurements from linear effects such as spectral leakage, it does slightly increase the variance in the regressed result, and does not correct for bias from $O(\gamma_b^n)$, $n > 3$ responses. This bias, which is the systematic error resulting from approximating the stress response as only a third order polynomial, can be quantified by computing the difference between solutions of equation \ref{eq:vandermonde} using third order and fifth order polynomials to generate the data for $ \hat \sigma_a( \omega^* ) $ and $ \hat  \sigma_b( \omega^* ) $.  The magnitude of the difference between $\hat \sigma^{(3)}(\omega^*)$ determined from the third and fifth order approximations is: $ (5/4) \, \gamma_b^2 | \hat \sigma^{(5)}( \omega^* ) | $, where $ \hat \sigma^{(5)}( \omega^* ) $ is the coefficient of the fifth order response that is neglected in the third order expansion of the measured stress.  As expected, the bias grows with increasing imposed strain amplitude.

The variance in the inferred value of $ \hat \sigma^{(3)}( \omega^* ) $ arises from the finite noise amplitude present in each channel, which we denote $ \epsilon( \omega^* ) $.  The mean-squared value of the stress on channels not carrying a first or third order nonlinearity can be used to estimate the noise in channels of interest, or the noise amplitude may be measured \emph{a priori} as part of a detailed characterization of the rheometer.  Assuming the noise is normally distributed, the variance in $ \hat \sigma^{(3)}( \omega^* ) $ using just two three-tone experiments is set by the value of $ \sqrt{\mathbf{e}_2^T ( \mathbf{V}^T \mathbf{V} )^{-1} \mathbf{e}_2} $, with $ \mathbf{e}_2 = \epsilon(\omega^*) \times ( 0, 1) $. With the optimal ratio of strain amplitudes, $r = \frac{1}{2}$, the variance in $\hat \sigma^{(3)}(\omega^*)$ is  $ ( 4 \sqrt{5} / 3 )  \, \epsilon( \omega^* ) \gamma_b^{-3} $.  The variance grows cubically with decreasing strain amplitude in the MAPS experiments as noise will dominate the response at small enough strain amplitudes. In contrast, the bias determined above grows with increasing strain amplitude. This results in a trade off between bias and variance that defines the range of amplitudes appropriate for measuring third order response in MAPS experiments using two different input strain amplitudes.

\begin{figure}[t]
    \centering
    \includegraphics[scale=0.6]{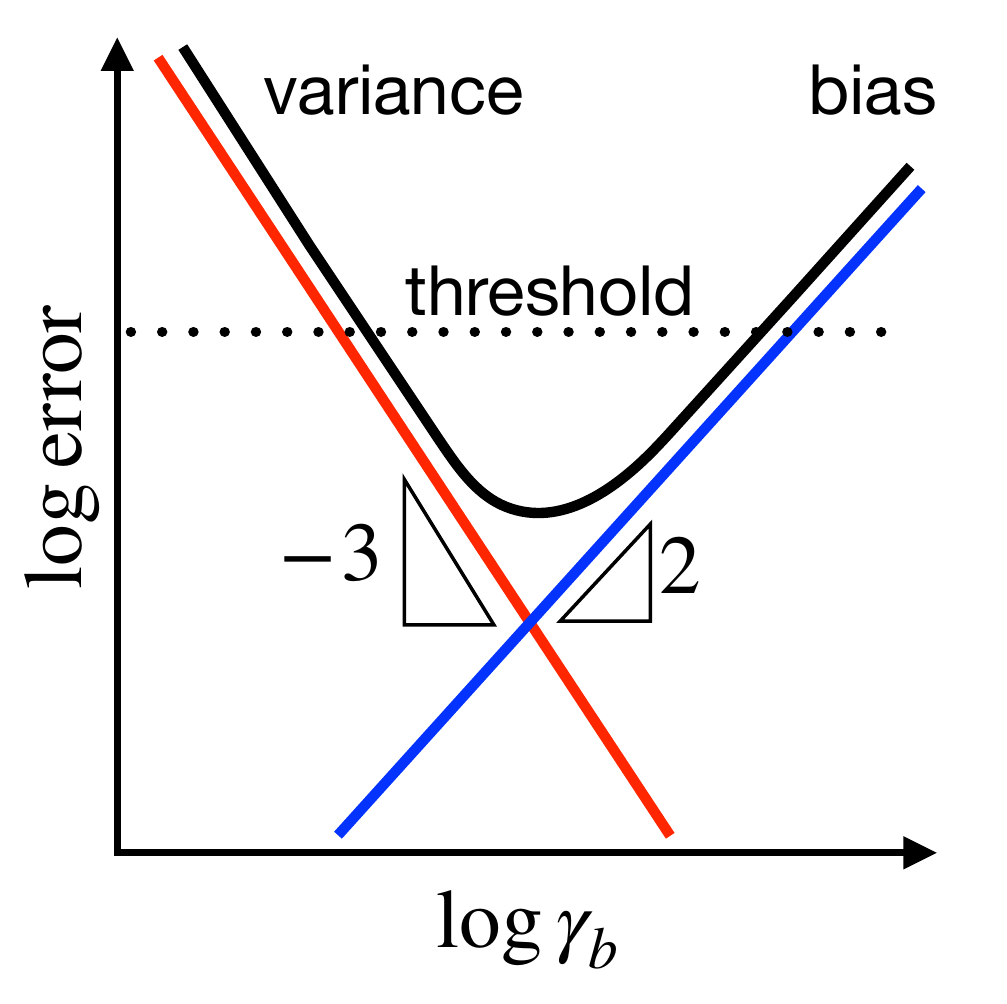}
    \caption{The bias-variance trade-off in estimating the third order stress response with a third order polynomial model.  The variance is an instrumental error and its power law decay with strain amplitude does not depend on the fitting model.  The amplitude of the bias depends on the material under study and its power law growth with strain is set by the third order fit to the measured data.  Any strain amplitude giving a total bias and variance below a specified threshold can be used to measure the third order response accurately.}
    \label{fig:biasvar}
\end{figure}

The bias-variance trade off involved in inferring $ \hat \sigma^{(3)}( \omega^* ) $ is illustrated graphically in Figure \ref{fig:biasvar}.  From this analysis it is clear that the strain amplitudes of the experiments must be chosen carefully to introduce neither systematic deviations from the third order approximation nor to magnify random errors in the measurement.  The bias-variance trade-off can be managed by selecting a value of $ \gamma_b $ for which the summed magnitudes of the bias and the variance are below a threshold tolerance.  The optimal strain amplitude can be found by minimizing this total error function. With just two three-tone experiments, this optimal balance of bias and variance suggests that the optimal strain amplitude is:
\begin{equation}
    \gamma_b \approx 1.3 \left( \frac{\epsilon( \omega^* )}{ |\hat \sigma^{(5)}( \omega^* ) |} \right)^{1/5}. \label{eq:biasvar}
\end{equation}
The minimal error in the inferred value of third order response function resulting from equal bias and variance is then approximately:
\begin{equation}
    3.5 \, \epsilon( \omega^* )^{2/5} | \hat \sigma^{(5)}( \omega^* ) |^{3/5}.
    \label{eq:minimum_error}
\end{equation}  
This value can serve as a useful \emph{a posteriori} check on whether the response function has been accurately measured in the sense that its inferred magnitude exceeds significantly the expected error in the optimal case.  Here, the minimal error is slightly more sensitive to the magnitude of the material response $|\hat \sigma^{(5)}(\omega^*)|$ than to the instrumental noise characteristics $\epsilon(\omega^*)$.  If one were to repeat this analysis for experiments with three distinct strain amplitudes, instead using a fifth order polynomial to infer the third order modulus, the power law exponents in the minimal error would change: $ \epsilon( \omega^* )^{2/5} | \hat \sigma^{(7)}( \omega^* ) |^{3/5} \rightarrow \epsilon( \omega^* )^{4/7} | \hat \sigma^{(7)}( \omega^* ) |^{3/7}$.  In the case of a fifth order fit, the error would be more sensitive to the instrumental response than the material response. This analysis is conducted in detail in Appendix \ref{app:fifth_order}.

% As discussed, the linear response can be measured during a MAPS experiment on the channels matching the three input tones: $ \omega^* = \omega_1 $ or  $\omega_2 $ or $ \omega_3 $.  Of course the stress on these channels also has a contribution from a linear combination of the third order responses.  Appendix \ref{app:MAPSlinear} discusses the bias and variance in the estimates of the linear response taken from the three tone experiment.  The measurement of linear response in a MAPS experiment can serve as a check on the consistency of the measurement method by comparison with linear response data collected via alternative methodologies \cite{geri-2017}.

\subsubsection{Amplitude Selection in Strain vs. Stress Control}
\label{sec:amplitude_region}

The Vandermonde method and analysis makes clear the importance of choosing an appropriate input amplitude in medium-amplitude experiments. The range of appropriate amplitudes here has been defined as the amplitudes that produce a total error below a certain threshold, as illustrated in Figure \ref{fig:biasvar}. In practice, the threshold will be chosen relative to the desired measured quantity: the third order modulus. Thus the range of appropriate amplitudes depends on the instrument response and both the third and fifth order features of the material response. These material response features in general depend on the measured frequency channel, and therefore the appropriate range of amplitudes will be frequency-dependent. The frequency dependence of an appropriate range for the strain amplitude has recently been demonstrated experimentally in strain-controlled MAOS \cite{singh-2018}, which has shown that this range of strain amplitudes can vary substantially with frequency, particularly in the low-frequency regime.

Though the above analysis focuses on strain-controlled experimentation, it can also be identically repeated in the stress-controlled MAPS framework. In this case, the results would be identical upon the substitutions $ \hat{\sigma}_m(\omega^*) \rightarrow \hat{\gamma}_m(\omega^*) $, $ \hat{\sigma}^{(i)}(\omega^*) \rightarrow \hat{\gamma}^{(i)}(\omega^*) $, and $ \gamma_m \rightarrow \sigma_m $. In the stress-controlled case, the relevant MAPS response function probed by experiments is the third order complex compliance, $J^*_3(\omega_1,\omega_2,\omega_3)$, rather than the third order complex modulus. In a nonlinear viscoelastic system, the frequency dependence of the strain response in stress control is distinct from that of the stress response in strain control. As a consequence, the frequency dependence of the appropriate range of stress amplitudes $ \sigma_b $ will differ from that of $ \gamma_b $.

\begin{figure*}[t]
    \centering
    \includegraphics[width = \textwidth]{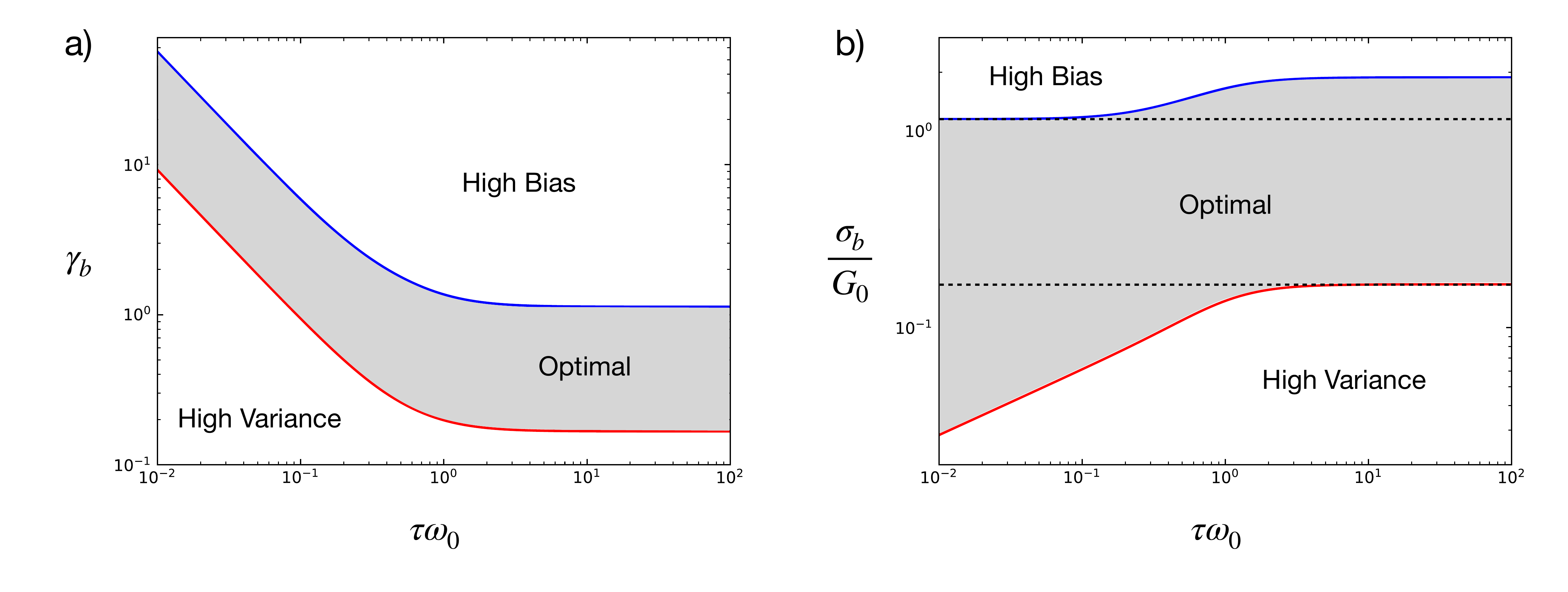}
    \caption{The bounds of the inequalities in equations \ref{eq:thresholds} on the third harmonic in a MAOS experiment as a function of the imposed frequency, in both (a) strain control and (b) stress control for the corotational Maxwell model, with an instrument response function $\epsilon(\omega_0) = G_0 \times 10^{-4}$ in (a) and $\epsilon(\omega_0) = 10^{-4}$ s in (b). The blue lines represent the upper limit for either $\gamma_b$ or $\sigma_b$, above which the total error exceeds the threshold value due to bias. The red lines represent the lower limit, below which the total error exceeds the threshold due to variance. The gray region represents the range of either $\gamma_b$ or $\sigma_b$, respectively, for which the total error is beneath the threshold value.}
    \label{fig:bounds_white}
\end{figure*}

In strain-controlled MAOS, the optimal range for strain amplitudes is used to inform MAOS frequency sweeps, in which only two experiments at different strain amplitude are performed at each frequency to infer the third order intrinsic material functions akin to the procedure outlined above \cite{singh-2018}. In strain control, however, the strong dependence of the optimal range of strain amplitudes on the imposed frequency, particularly at low frequency, introduces an experimental complication, in that this range must be determined \emph{a priori} at each frequency to ensure an accurate frequency sweep. Given this observation, it is reasonable to ask whether the same difficulty arises in stress control. Below, we provide insight to this question with a brief theoretical study of the bias-variance trade-off in stress and strain control using the corotational Maxwell model.

If we define the total error threshold to be half of the magnitude of the measured third order stress response on a channel $\omega^*$, then the inequalities defining the region of appropriate values of the stress or strain amplitudes in stress- or strain-controlled experiments are, respectively:
\begin{subequations}
  \begin{equation}
    \frac{4\sqrt{5}}{3}\frac{\epsilon(\omega^*)}{|\hat{\gamma}^{(3)}(\omega^*)|}\sigma_b^{-3} + \frac{5}{4}\frac{|\hat{\gamma}^{(5)}(\omega^*)|}{|\hat{\gamma}^{(3)}(\omega^*)|}\sigma_b^2 \leq 0.5,
  \end{equation}
\begin{equation}
  \frac{4\sqrt{5}}{3}\frac{\epsilon(\omega^*)}{|\hat{\sigma}^{(3)}(\omega^*)|}\gamma_b^{-3} + \frac{5}{4}\frac{|\hat{\sigma}^{(5)}(\omega^*)|}{|\hat{\sigma}^{(3)}(\omega^*)|}\gamma_b^2 \leq 0.5.
\end{equation}
\label{eq:thresholds}
\end{subequations}
In general, the third and fifth order elements of the material response (either stress or strain) on a specific channel $\omega^*$ depend on the details of the deformation protocol. For simplicity and the purpose of comparison with the results of Ewoldt and co-workers \cite{singh-2018}, we consider measurements of the third harmonic in a single-tone MAOS experiment, where $\omega^* = 3\omega_0$ represents the third harmonic of the input frequency $\omega_0$. In this case, we can obtain analytic expressions for the third and fifth order elements of the stress and strain response of the corotational Maxwell model. The mathematical details of this solution are left to Appendix \ref{app:crm}.

To find the roots $\gamma_b^*(3\omega_0)$ and $\sigma_b^*(3\omega_0)$ that result in the equality in equations \ref{eq:thresholds}, we must specify a functional form for the instrumental noise response, $\epsilon(\omega^*)$. A simple form of this function is the case of white noise, for which the average instrumental response is constant, $\epsilon(\omega^*) = \epsilon_0 $. The roots of the equalities in equations \ref{eq:thresholds} can be found numerically, and are shown as a function of the imposed frequency $\omega_0$ in Figure \ref{fig:bounds_white}.

The behavior depicted in Figure \ref{fig:bounds_white}a is in qualitative agreement with that observed by Ewoldt and co-workers \cite{singh-2018}. At high frequency, the bias and variance reach a plateau, so the region of optimal strain input $\gamma_b$ becomes constant. This corresponds to the onset of predominantly elastic behavior, for which the stress responds proportionally to the strain. At low frequency, we observe that the bias and variance both decrease with frequency, corresponding to a viscous regime in which the stress responds proportionally to the strain rate. As a result, the optimal region for $\gamma_b$ is shifted progressively to higher strains for lower frequencies, with $\gamma_b\omega = \dot\gamma_b \approx \mathrm{constant}$.

The behavior depicted in Figure \ref{fig:bounds_white}b for stress control is quite different than that for strain control. At high frequency, we still observe a plateau regime in the optimal region for $\sigma_b$. At low frequency, however, we find that the upper bound of the region approaches a different plateau, while the lower bound decays with decreasing frequency. As a result, the optimal region for $\sigma_b$ in fact expands as the frequency decreases, in contrast to the limited zone of validity in the low-frequency region under strain control.

A substantial consequence of the distinct trends noted above for strain- and stress-controlled experiments is that if one wishes to conduct a frequency sweep spanning both the low- and high-frequency regimes of a material in strain control, one must first conduct experiments at each frequency to determine the appropriate strain amplitude to impose, in order that the result is neither dominated by bias nor variance. This adds significant experimentation time to a medium-amplitude frequency sweep, and this further decreases the data-throughput. In stress control, however, it is possible to choose a single input stress amplitude that is within the optimal region at \emph{all} frequencies, as indicated by the gray shaded region between the dashed black lines in Figure \ref{fig:bounds_white}b. Such an amplitude may be found by conducting a single stress-amplitude sweep at a frequency $\omega\tau > 1$, and selecting an amplitude on the lower side of the optimal region. This is a significant advantage for stress-controlled experimental design, as an appropriate input stress amplitude need only be determined once, and the same amplitude can then be applied at all frequencies without substantially affecting the total error. For this reason, the experiments discussed in Section \ref{experiments} are all performed in the stress-controlled mode.

\subsubsection{Estimating Uncertainty in MAPS Measurements}
\label{sec:uncertainty}

In the experimental design developed in the previous sections, a cubic fit to two three-tone experiments will be exact, as we provide two degrees of freedom to describe two data points on each frequency channel. The fit itself provides no estimate of the uncertainty of the measured third order MAPS point on a given channel, thus any estimate of the uncertainty of measured quantities must come from consideration of the bias and variance presented in Section \ref{sec:vandermonde}. This requires knowledge of the fifth order material response and the instrument response on that channel, which are in general not precisely known. Thus any precise statement of the uncertainty of measured MAPS quantities are difficult to make in the present framework.

Another methodology for obtaining uncertainty estimates on MAPS quantities is to add a third measurement to the experimental design by implementing the same three-tone input signal with amplitude $\gamma_c$. The first and third order elements of the material response at a channel $\omega^*$ can now be determined by the least-squares solution to the following system:
\begin{equation}
\left(\begin{array}{c} \hat \sigma_a( \omega^* ) \\ \hat \sigma_b( \omega^* ) \\ \hat \sigma_c(\omega^*) \end{array}\right) = \left(\begin{array}{ccc}  r_a & r_a^3 \\ r_b & r_b^3 \\ 1 & 1 \end{array}\right) \left(\begin{array}{c} \gamma_c \hat \sigma^{(1)}( \omega^* ) \\ \gamma_c^3 \hat \sigma^{(3)}( \omega^* ) \end{array}\right) + \boldsymbol{\Delta\sigma}(\omega^*) \label{eq:vandermonde_lstsq}
\end{equation}
with $r_a = \gamma_a / \gamma_c$, $r_b = \gamma_b / \gamma_c$, and $0 < r_a \leq r_b < 1$. A reasonable estimate for the error, due to bias and variance, is the residual root-mean-squared error: $\Delta\sigma(\omega^*) = \frac{1}{\sqrt{3}}|\boldsymbol{\Delta\sigma}(\omega^*)|_2$. The uncertainty in the measured third order element of the response on the channel is:
\begin{equation}
    (\Delta \sigma^{(3)}(\omega^*))^2 = \textbf{e}^T_2(\textbf{V}^T\textbf{V})^{-1}\textbf{e}_2,
\end{equation}
with $\textbf{e}_2 = \Delta\sigma \times (0,1)$. The design variables $r_a$ and $r_b$ can be used to optimally condition $\textbf{V}^T\textbf{V}$ to minimize this uncertainty. The minimal error introduced by the experimental design in this case occurs for $r_a = r_b = 0.5$, for which $(\Delta \sigma^{(3)}(\omega^*))^2 = 16 (\Delta \sigma(\omega^*))^2 / 3$. Using this design, the uncertainty in the measured third order element of the material response can be determined, and subsequently propagated to the measured values of the relevant MAPS material function.

\subsubsection{Selecting the Input Phases}

The last design variables for the three-tone MAPS experiment to address are the phases of the input tones, $\alpha_j$. The previous analysis has revealed that the specific values of $G^*_3(\omega_1,\omega_2,\omega_3)$ probed by a three-tone MAPS experiment are independent of the values of $\alpha_j$. So too is the magnitude of the stress response on each output channel, though the relative weight of the real and imaginary components of the response on each channel does depend on the phases of the input tones (Table \ref{tab:channels}). The input phases therefore serve a distinct purpose in the experimental design, as a tool to control the peak value of the strain, strain rate, or torque attained by the oscillating rheometer fixture. This control is especially important for multi-tone signals with moderate amplitudes, which might operate near the safe limits of strain rate or torque of the rheometer. For example, if all phases in equation \ref{eq:thresholds} were left at zero, then the peak strain rate of such a signal would be $\gamma_0\sum_j\omega_j$, which may be quite large when considering high frequency tones. Even worse, if the phases were each set to $\pi/2$, then the peak acceleration would be $\gamma_0\sum_j\omega_j^2$. When large $\omega_j$ and large $\gamma_0$ are employed, these peak values can quickly exceed the instrumental limits. The same problem exists for stress-controlled experiments, as rheometers can only implement rapid changes in the stress with finite accuracy, and large imposed stresses can also result in strain rates or torques outside of the instrumental limits. Careful adjustments of the input phases, however, can dramatically reduce the peak value of the input signal or its derivatives, allowing one to operate within the instrumental limits even at large amplitudes and high frequencies.

The dependencies of the peak value of multi-tone signal and its derivatives on the input phases are all highly nonlinear. This problem has received much attention in fields such as electronics, where multi-tone signals are widely employed \cite{boyd-1986,friese-1997}. The problem of choosing phases is often formulated in terms of minimizing the peak value of the signal relative to the root-mean-squared value of the signal, a measure known as the \emph{crest factor}. Formally, the crest factor is defined as follows:
\begin{equation}
    CF(u) \equiv \frac{||u||_{\infty}}{||u||_2},
\end{equation}
with the $L^{\infty}$ norm:
\begin{equation}
    ||u||_{\infty} \equiv \sup_t |u(t)|,
\end{equation}
and the $L^2$ norm:
\begin{equation}
    ||u||_2 \equiv \left(\frac{1}{T}\int_0^T(u(t))^2dt\right)^{1/2},
\end{equation}
where $T$ is the period of the periodic input signal $u(t)$.

Crest factor minimization for general multi-tone signals is quite difficult, so specific guidelines for selecting input phases to minimize the crest factor are often heuristic and are valid only for certain sets of input tones. Fortunately, we have found that for three-tone signals that probe asymptotic nonlinearities in viscoelastic materials, the required stresses and strain rates are often much lower than the instrumental upper bounds. For the results in Section \ref{experiments}, even at the maximum crest factor for a given set of input tones, the signal always operates within the rheometer's operating limits. Therefore, we will set $\alpha_j = 0$ for the remainder of this work. For the set of input tones $\{1,4,16\}$, this results in a crest factor of $CF(u) = 2.30$ for the input signal, and a crest factor of $CF(\dot u) = 1.76$ for its time derivative. For the set $\{5,6,9\}$, the resulting crest factor of the input signal is $CF(u) = 2.27$ and a crest factor of $CF(\dot u) = 2.21$ for its derivative. With only three input tones, therefore, the crest factor of the signal is not dramatically higher than the crest factor for a single tone, which is equal to $\sqrt{2}$. However, because this na\"ive choice would result in the peak value of the input signal and its derivative growing quickly with the number of input tones, the problem of crest factor minimization will be an important topic in future explorations of multi-tone signals.

\section{MAPS Rheology of a Model Complex Fluid}
\label{experiments}

In designing an experimental protocol to measure MAPS functions in a high-dimensional, high-throughput manner, careful consideration was given to implementing procedures that are accessible in commercial rheometers, and to standardizing the procedure for data collection and analysis. In this section, we demonstrate the experimental design proposed in this work with measurements on a model complex fluid: a surfactant solution of entangled wormlike micelles \cite{rehage-1991,hoffmann-1992}. These experiments were performed in stress control, due to the ease in implementing a frequency sweep in this mode at fixed stress amplitude (as discussed in Section \ref{sec:amplitude_region}). All measurements were taken using a DHR-3 Discovery Hybrid Rheometer from TA Instruments with inputs signals generated in TRIOS v5.0.0.

\begin{figure*}[t]
    \centering
    \includegraphics[width = 0.8\textwidth]{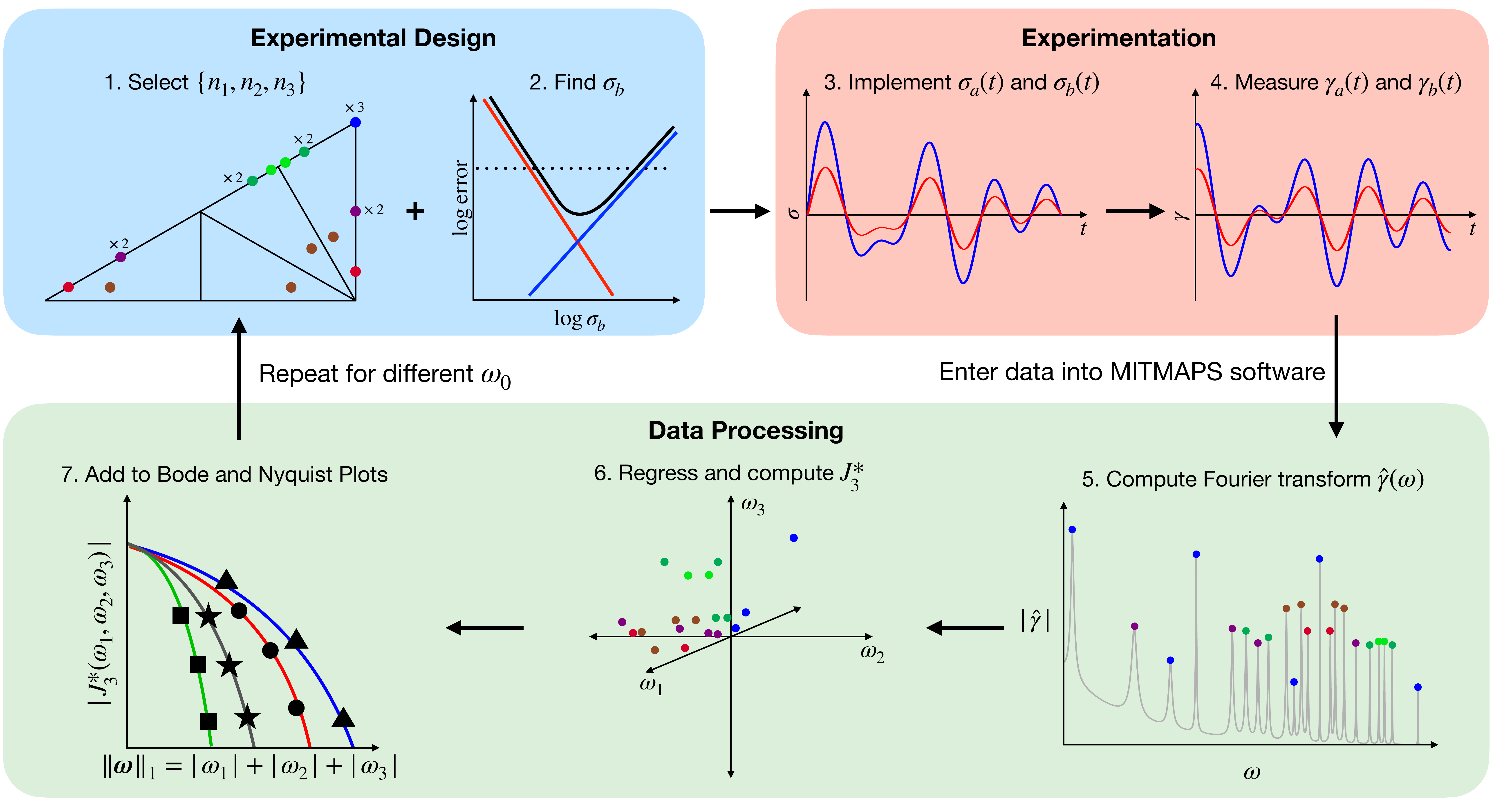}
    \caption{Flowchart for the full three-tone experimental protocol for MAPS rheology in stress control. Experimental design includes selection of the input tones $\{n_1,n_2,n_3\}$ (Section \ref{sec:input_tones}) and determining the optimal $\sigma_b$ (Section \ref{sec:vandermonde}). These design variables are used to construct input signals according to $\sigma(t) = \sigma_0 \sum_j \sin(n_j \omega_0 t)$ with $\sigma_0 = \sigma_b$ and $\sigma_0 = \sigma_a = \sigma_b/2$, for some selected $\omega_0$, and these signals are implemented in a rheometer through its own arbitrary wave of multiwave control software. The time-series stress and strain data are recorded, and fed into a stand-alone software package for data processing. Data processing includes taking the discrete Fourier transform, regressing third order components of the response (Sections \ref{sec:vandermonde} and \ref{sec:uncertainty}), and associating each output channel to a discrete measurement of the third order complex compliance (Table \ref{tab:channels}). The results of this MAPS experiment can be added to Bode and Nyquist plots within the software, and the procedure can then be repeated for different values of $\omega_0$ to sweep through a range of different values of the $L^1$-norm: $|\boldsymbol{\omega}|_1 = |\omega_1| + |\omega_2| + |\omega_3|$.}
    \label{fig:flowchart}
\end{figure*}

Before we present the measured MAPS data, let us review the procedure for experimental design, implementation, and data analysis developed in the preceding sections. A flowchart depicting the full MAPS experimental protocol for a three-tone input signal is presented in Figure \ref{fig:flowchart}, and the steps in this protocol are enumerated below. 

\begin{enumerate}
    \item With the stress-controlled mode selected, the experimental design consists principally of two steps. The first is to select the input tones $\{n_1,n_2,n_3\}$. Here, we select $\{5,6,9\}$ as an example of a set that will study the regions in the vicinity of the MAOS vertices, and $\{1,4,16\}$ as an example of a set that will broadly study the entire MAPS domain.
    
    \item The second step in the experimental design is to determine an appropriate input stress amplitude $\sigma_b$ that is suitable for experiments at all fundamental frequencies. This requires some experimentation -- specifically, an amplitude sweep at a frequency for which $\omega\tau > 1$. The characteristic relaxation time of the material, $\tau$, is easily determined from a SAOS sweep, after which an appropriate frequency is selected for an amplitude sweep and $\sigma_b$ determined from the analysis set forth in Section \ref{sec:vandermonde}.
    
    \item Next, the fundamental frequency $\omega_0$ for the MAPS experiment is selected, and the three-tone waveforms entered into the rheometer's control software with amplitudes $\sigma_b$ and at least one replicate of $\sigma_a = \sigma_b/2$ (one replicate is sufficient to regress the third order response, and a second replicate with $\sigma_c = \sigma_a$ will produce an estimate of its uncertainty).
    
    \item The raw time-series data for both the input stress and output strain are then measured by the rheometer over a duration of time sufficient to reach a steady oscillatory state. By default, the software package included with this work assumes that the sample is allowed to complete at least five periods with respect to the fundamental frequency ($5\times (2\pi/\omega_0)$) once the steady oscillatory state is reached.
    
    \item Finally, the raw time-series data is input into a software package that we have developed to process the data. The first step in data processing is to compute the discrete Fourier transform of both the input and output time-series data, and identify the magnitude and phase of the signal on each input and output channel.
    
    \item The software package then fits the data to a cubic polynomial according to the procedure of either Section \ref{sec:vandermonde} or \ref{sec:uncertainty}, and uses Table \ref{tab:channels} to associate each channel to a discrete measurement of the third order complex compliance.
    
    \item The software package outputs the value of the third order complex compliance in both tabular and graphical forms. If linear response data is available, the software converts the measured values of the third order complex compliance to corresponding values of the third order complex modulus and third order complex viscosity using the known MAPS inversion relationships, and outputs these values as well.
\end{enumerate}
The final three of these steps have all been automated by the MITMAPS software package, so that the user need only input the time-series data from the rheometer, the experimental design parameters: $\omega_0$ and $\{n_1,n_2,n_3\}$, and the desired format for the output.

Once completed, the experimental protocol can be repeated for either different input tones $\{n_1,n_2,n_3\}$ or a different fundamental frequency, $\omega_0$. For a particular set of input tones, the software package allows the user to merge data sets collected with different $\omega_0$ to create a MAPS frequency sweep, which is conveniently depicted through Bode or Nyquist plots. Frequency sweeps obtained with different $\{n_1,n_2,n_3\}$ can also be merged by the software, and these frequency sweeps can be depicted on the same or separate figures.

Below, each step in the experimental protocol is demonstrated in a study of a surfactant solution of wormlike micelles. The resulting MAPS data is presented in two forms: Bode plots of the third order complex compliance as a function of the frequency $L^1$-norm, obtained directly from three-tone MAPS experiments, and Bode plots of the third order complex viscosity, obtained using the data for the third order complex compliance in conjunction with a fit to SAOS data and the inversion formulas presented in Part 1 of this work. We choose to show both forms of data to demonstrate the versatility of the MAPS representation, in that it is capable of simultaneously describing both stress and strain control, and because nonlinear rheological data and solutions to viscoelastic constitutive models are very often obtained in a strain-controlled fashion, thus the complex viscosity representation may look more familiar to some readers. We then compare the experimental MAPS data to the predictions of the corotational Maxwell model and the Giesekus model as a simple exercise in evaluating the suitability of constitutive models for describing MAPS data. Finally, this section concludes with a comparison of the MAPS data to data obtained with MAOS amplitude sweeps, as a means to validate that our three-tone MAPS experiments have indeed measured an intrinsic material response.

\begin{figure*}[t]
    \centering
    \includegraphics[width = 0.8\textwidth]{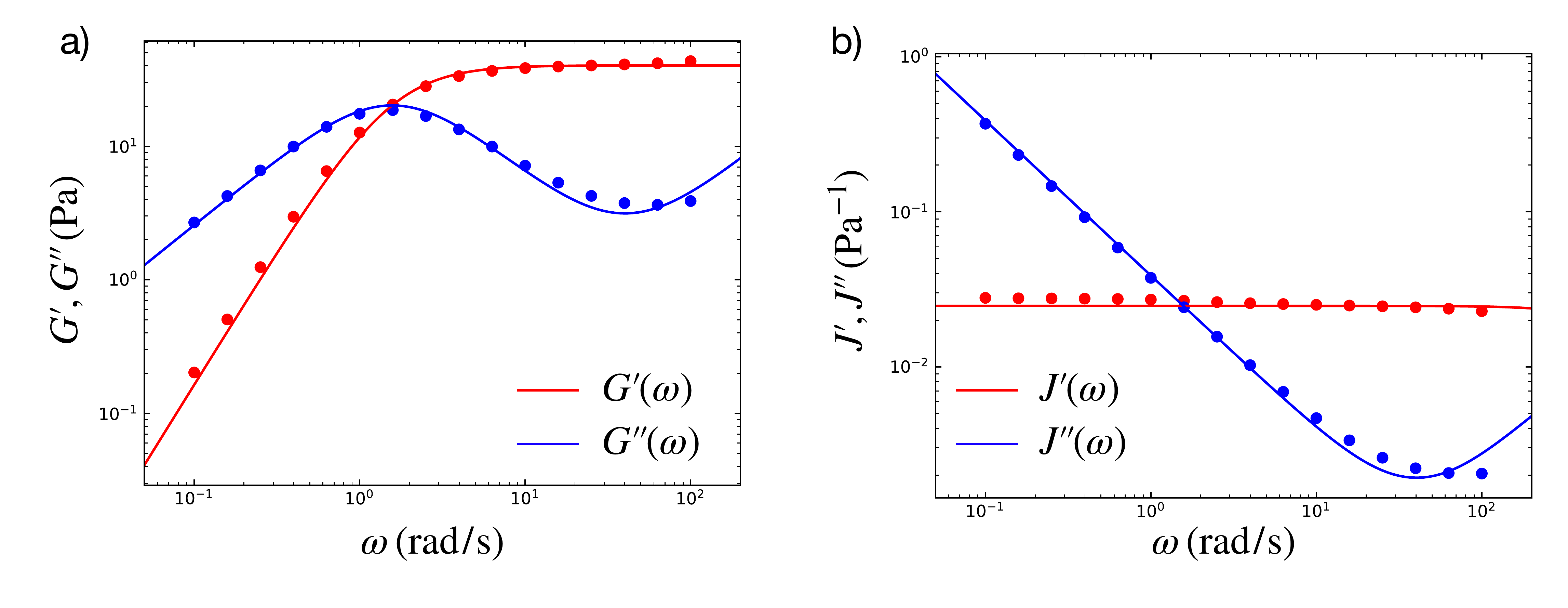}
    \caption{SAOS frequency sweep of the wormlike micellar solution in controlled stress, with $\sigma_0 = 0.1$ Pa. Circles show experimental values, lines represent the fit to a single Maxwell element plus an effective Newtonian solvent contribution (equation \ref{eq:maxwell_LR}) with $\eta_0 = 25.7$ Pa$\cdot$s, $\tau = 0.64$ s, and $\eta_{\infty} = 0.039$ Pa$\cdot$s. a) The complex modulus, $G^*(\omega) = 1/J^*(\omega) = G'(\omega) + iG''(\omega)$. b) The complex compliance, $J^*(\omega) = J'(\omega) - iJ''(\omega)$.}
    \label{fig:saos}
\end{figure*}

\subsection{Experimental Details}

The surfactant solution of wormlike micelles used in this study was composed of cetylpyridinium chloride (CPyCl), sodium salicylate (NaSal), and sodium chloride (NaCl) in deionized water at concentrations of 100 mM, 60 mM, and 33 mM, respectively (CPyCl and NaSal supplied by Alfa Aesar, reagent grade NaCl purchased from Sigma Aldrich). All experiments were conducted with a 60 mm, 2\degree aluminum cone with truncation gap of 58 $\mu$m and a bottom Peltier plate maintained at 25\degree C. To reduce solvent evaporation over the duration of each experiment, the sample was sealed using a thin layer of hexadecane oil.

\subsection{SAOS Frequency Sweep}

A stress-controlled SAOS frequency sweep was conducted to measure the linear viscoelastic spectrum of the solution of wormlike micelles. The sweep was conducted over the range $\omega \in [0.1,100]$ rad/s with five points per decade, at a constant stress amplitude of $\sigma_0 = 0.1$ Pa. The resulting measurements of the storage and loss compliances, $J'(\omega)$ and $J''(\omega)$ (with $J^*_1(\omega) = J'(\omega) - iJ''(\omega)$) are shown in Figure \ref{fig:saos}, along with the corresponding values of the storage and loss moduli, $G'(\omega)$ and $G''(\omega)$ (with $G^*_1(\omega) = G'(\omega) + iG''(\omega) = 1/J^*_1(\omega)$). At low and moderate frequencies, the linear viscoelastic behavior of this wormlike micellar solution is known to be close to that of an ideal Maxwell model with a single relaxation time \cite{cates-1990,yesilata-2006}. At high frequencies, the observed behavior deviates from the predictions of the model due to the presence of faster-relaxing Rouse modes. Over the range for which SAOS data is available, the linear viscoelastic response can be compactly represented by a single-mode linear Maxwell element plus an effective Newtonian solvent contribution:
\begin{equation}
    G^*_1(\omega) = \frac{1}{J^*_1(\omega)} = \frac{\eta_0i\omega}{1 + i\tau\omega} + \eta_{\infty}i\omega,
    \label{eq:maxwell_LR}
\end{equation}
with the parameters: $\eta_0 = 25.7$ Pa$\cdot$s, $\tau = 0.64$ s, and $\eta_{\infty} = 0.039$ Pa$\cdot$s.

\subsection{Oscillatory Amplitude Sweep}

\begin{figure}
    \centering
    \includegraphics[width = \columnwidth]{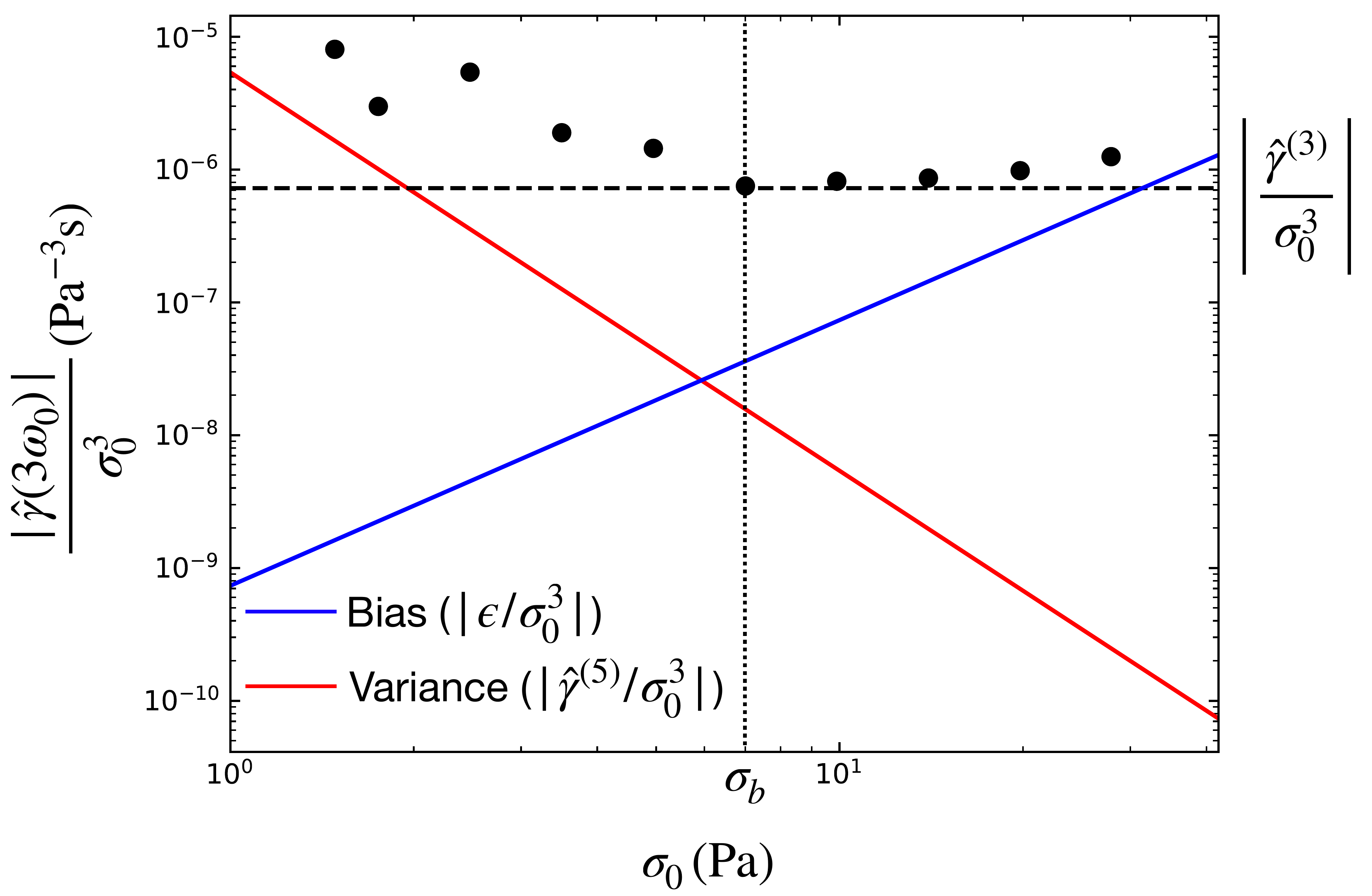}
    \caption{The magnitude of the third harmonic response, $|\hat{\gamma}(3\omega_0)|$, to a single tone oscillatory stress input at a frequency $\omega_0 = 5.12$ rad/s with varying amplitude $\sigma_0$. Circles indicate experimental results, with the components of a fifth-order polynomial fit shown by lines: the bias (linear) term in blue, the variance (fifth-order) term in red, and the weakly nonlinear (third order) term as a horizontal dashed line. The ordinate is scaled by the cube of the imposed stress amplitude to reflect the form of Figure \ref{fig:biasvar}. An input stress amplitude in the optimal range is indicated with a vertical dotted line at $\sigma_b = 7$ Pa.}
    \label{fig:amp_sweep}
\end{figure}

Next, an oscillatory stress amplitude sweep was performed at a frequency of $\omega_0 = 5.12$ rad/s, for which $\tau\omega = 3.27 > 1$. The stress amplitude was swept over the range $\sigma_0 \in [1,30]$ Pa. At each imposed stress amplitude, the oscillatory strain output from the rheometer was recorded as a function of time, and from the Fourier transform of this strain signal, the magnitude of the third harmonic strain response was computed. To assess the magnitude of the error due to bias and variance, the third harmonic signal was normalized by the cube of the imposed stress amplitude and plotted against the imposed amplitude. The resulting data is shown in Figure \ref{fig:amp_sweep}, which closely reflects the form of the theoretical bias-variance plot shown in Figure \ref{fig:biasvar}. Figure \ref{fig:amp_sweep} also shows the linear (red), and fifth-order (blue) contributions of a fit to the third harmonic strain data, which represent approximately the variance and bias, respectively. The horizontal dashed black line shows the regressed value of the third order response, $|\hat\gamma^{(3)}/\sigma_0^3|$.

Figure \ref{fig:amp_sweep} demonstrates that the SAOS frequency sweep ($\sigma_0 = 0.1$ Pa) was indeed performed at an amplitude that was sufficiently small to excite predominantly linear behavior. The data also reveals a range of intermediate amplitudes over which the measured value of the third harmonic signal is nearly equal to the regressed third order response.
The analysis of Section \ref{sec:amplitude_region} indicated that $\sigma_b$ should be selected towards the lower end of this optimal region, so that the same stress amplitude can be used across a range of frequencies in a MAPS frequency sweep. Therefore, the value of $\sigma_b = 7$ Pa is selected for the forthcoming MAPS experiments, as indicated with a vertical dotted line.

\subsection{MAPS Frequency Sweeps}

Two MAPS frequency sweeps were performed with input tone sets $\{n_1,n_2,n_3\} = \{5,6,9\}$ and $\{n_1,n_2,n_3\} = \{1,4,16\}$, sweeping over the fundamental frequencies $\omega_0 = 0.16$, $0.32$, $0.64$, and $1.28$ rad/s. These input tone sets were selected to demonstrate the diversity of information that can be obtained from MAPS experiments. Specifically, the input tone set $\{5,6,9\}$ is able to illustrate the sensitivity of MAPS data in a small region near the MAOS vertices, while the $\{1,4,16\}$ set probes how MAPS data varies over the entire domain of the third order response function. The fundamental frequencies were selected to probe a window of time-scales surrounding the characteristic time-scale of the material determined from the linear response ($\tau = 0.64$ s). It is especially important to capture data at high frequencies ($\omega\tau > 1$) because in this regime, the behavior of different constitutive models has been shown to vary widely, while in the low-frequency regime the behavior of different models is much more regular \cite{ewoldt-2014,lennon-2020}. High-frequency data is therefore critical in assessing the capabilities of a specific constitutive model to accurately capture measured experimental data, as will be shown in the following section.

At each imposed $\omega_0$, values of the third order complex compliance were determined using the procedure developed in Section \ref{sec:uncertainty}. Specifically, input signals of the form:
\begin{equation}
    \sigma(t) = \sigma_0 \sum_j \sin(n_j\omega_0 t)
\end{equation}
were implemented using the Arbitrary Wave mode in the TRIOS software provided by TA Instruments, with one trial using a stress amplitude $\sigma_b = 7$ Pa, and two trials using a stress amplitude $\sigma_a = 0.5\sigma_b = 3.5$ Pa. These three trials were implemented for each $\omega_0$ with both input tone sets, resulting in a total of 24 trials. Each trial was run for a time equal to $10\times(2\pi/\omega_0)$, or 10 periods with respect to the fundamental frequency, to allow the sample to reach its steady oscillatory state. The sampling rate was set such that at least 2000 data points were collected per period with respect to the fundamental frequency, to ensure that high-frequency channels would be distinguishable in the response. In general, the sampling rate should be at least an order of magnitude larger than the highest frequency expected in the weakly nonlinear response. Setting the sampling rate much higher, however, will result in file sizes that become large and computationally expensive to process. For more details on how a MAPS frequency sweep should be implemented in the TRIOS software for compatibility with the MITMAPS software package, see the Supplementary Material. Note that while we have used the Arbitrary Wave mode in this work to generate the three-tone input signals, it may be possible to generate the same signals using a built-in multiwave functionality in the control software for the rheometer by activating only the frequencies of interest. However, any tests run in multiwave mode must still export the raw shear stress and shear strain time-series data for subsequent data processing.

After all trials were completed, the measured stress and strain time series were exported to a text file for processing by the software. Data processing consists simply of windowing the data to exclude the initial transient response (done by removing the first five periods of the signal), taking the Fourier transform of the time series data, computing the third order components of the response by least-squares regression, and associating each peak with a value of the third order complex compliance. Because each combination of $\omega_0$ and $\{n_1,n_2,n_3\}$ results in 19 measured values of $J^*_3(\omega_1,\omega_2,\omega_3)$, these two MAPS frequency sweeps over four frequencies each produce a total of $2 \times 4 \times 19 = 152$ measurements of $J^*_3(\omega_1,\omega_2,\omega_3)$ while requiring less than 80 minutes of experimentation. Compared to similar rheological techniques such as MAOS, this data throughput is remarkably high.

\begin{figure*}[t!]
    \centering
    \includegraphics[width = \textwidth]{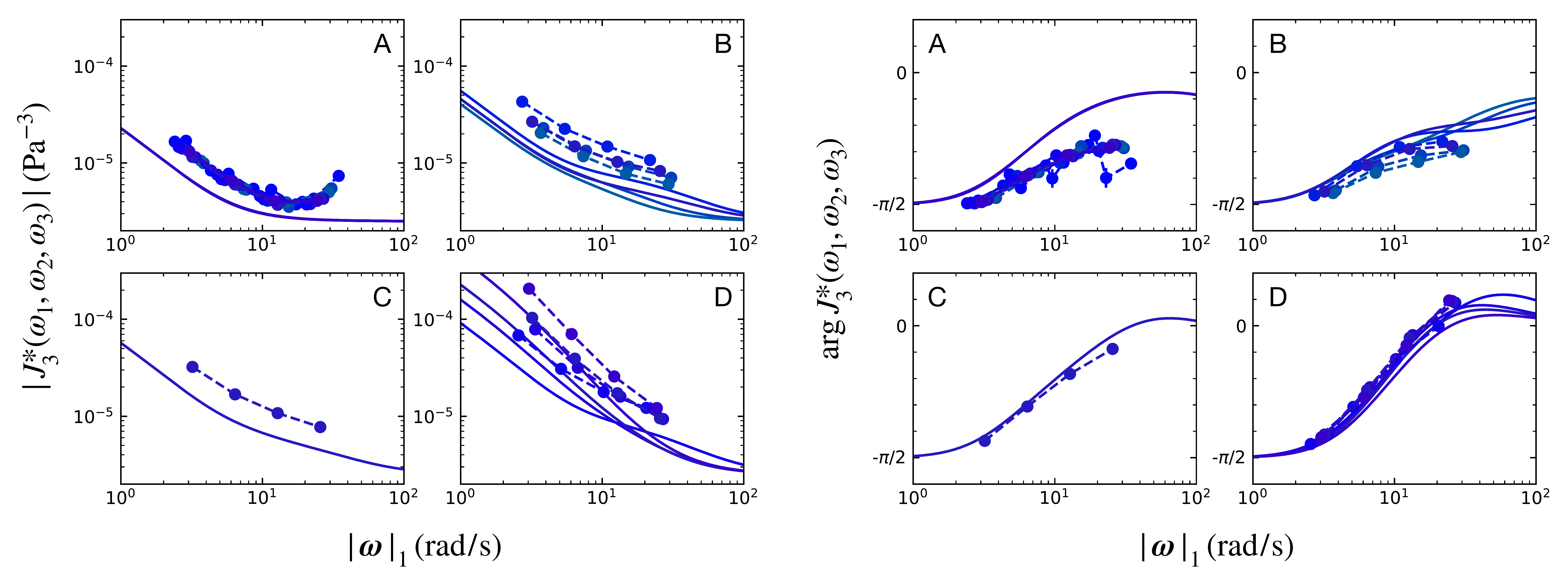}
    \caption{Values of the third order complex compliance, $J^*_3(\omega_1,\omega_2,\omega_3)$, obtained from a MAPS frequency sweep with the integer triplet $\{n_1,n_2,n_3\} = \{5,6,9\}$, at frequencies of $\omega_0 =$ 0.16, 0.32, 0.64, and 1.28 rad/s. The data is visualized according to the procedure set forth in Section \ref{sec:volterra}, wherein the measured points are divided into four subspaces, and within each subspace points are colored on the basis of their barycentric coordinates. The magnitude of the third order complex compliance, $|J^*_3(\omega_1,\omega_2,\omega_3)|$ plotted against the frequency $L^1$-norm, $|\boldsymbol{\omega}|_1 = \omega_1 + \omega_2 + \omega_3$, within each subspace is shown to the left, and the argument (phase) of the third order complex compliance, $\mathrm{arg} \, J^*_3(\omega_1,\omega_2,\omega_3)$, plotted against the frequency $L^1$-norm within each subspace is shown to the right. Circles indicate experimental results, with dashed lines connecting points associated with the same barycentric coordinates within a subspace. Solid lines indicate predictions from the two parameter corotational Maxwell model plus a Newtonian solvent mode, with $\eta_0 = 25.7$ Pa$\cdot$s, $\tau = 0.64$ s, and $\eta_{\infty} = 0.039$ Pa$\cdot$s.}
    \label{fig:569_J3}
\end{figure*}

\begin{figure*}[t!]
    \centering
    \includegraphics[width = \textwidth]{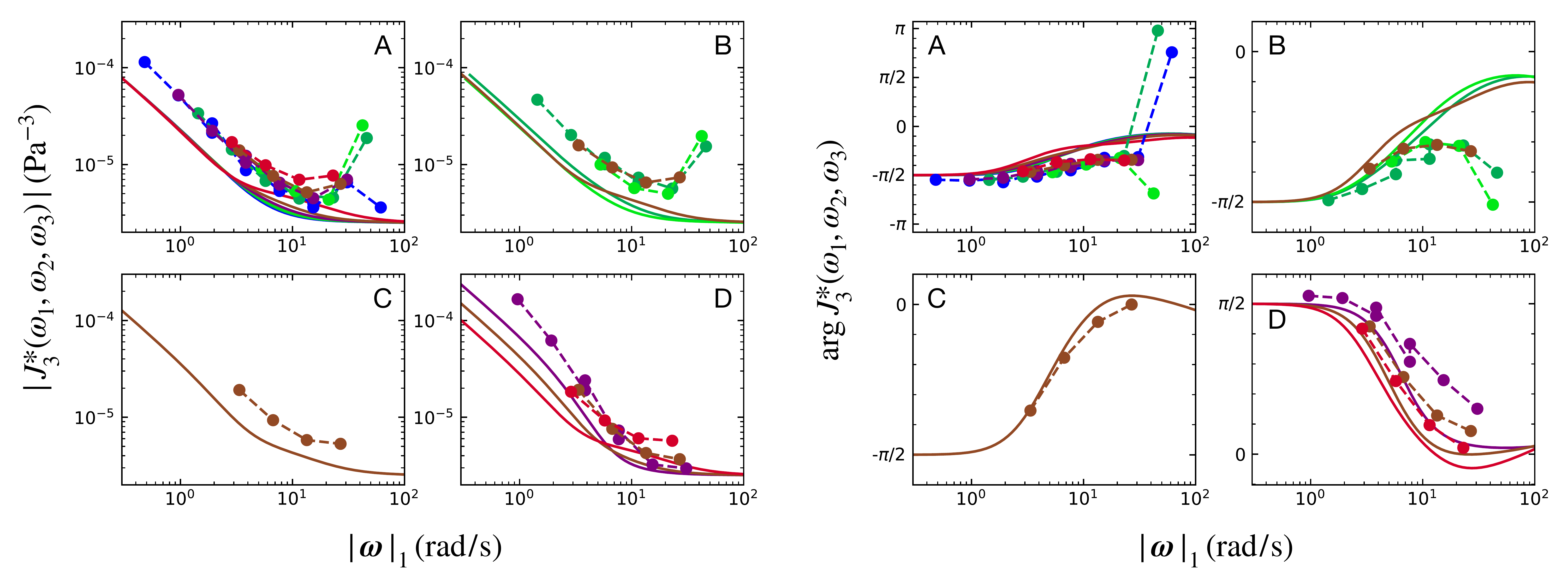}
    \caption{Values of the third order complex compliance, $J^*_3(\omega_1,\omega_2,\omega_3)$, obtained from a MAPS frequency sweep with the integer triplet $\{n_1,n_2,n_3\} = \{1,4,16\}$, at frequencies of $\omega_0 =$ 0.16, 0.32, 0.64, and 1.28 rad/s. The data is visualized according to the procedure set forth in Section \ref{sec:volterra}, wherein the measured points are divided into four subspaces, and within each subspace points are colored on the basis of their barycentric coordinates. The magnitude of the third order complex compliance, $|J^*_3(\omega_1,\omega_2,\omega_3)|$ plotted against the frequency $L^1$-norm, $|\boldsymbol{\omega}|_1 = \omega_1 + \omega_2 + \omega_3$, within each subspace is shown to the left, and the argument (phase) of the third order complex compliance, $\mathrm{arg} \, J^*_3(\omega_1,\omega_2,\omega_3)$, plotted against the frequency $L^1$-norm within each subspace is shown to the right. Circles indicate experimental results, with dashed lines connecting points associated with the same barycentric coordinates within a subspace. Solid lines indicate predictions from the two parameter corotational Maxwell model plus a Newtonian solvent mode, with $\eta_0 = 25.7$ Pa$\cdot$s, $\tau = 0.64$ s, and $\eta_{\infty} = 0.039$ Pa$\cdot$s.}
    \label{fig:1416_J3}
\end{figure*}

After data processing, the MITMAPS software was used to generate Bode plots of the frequency dependence in the measured third order complex compliance, according to the plotting scheme discussed in Section \ref{sec:volterra}. These plots are shown in Figure \ref{fig:569_J3} for the input tone set $\{5,6,9\}$, and in Figure \ref{fig:1416_J3} for the input tone set $\{1,4,16\}$. Though points are colored by their barycentric coordinates in the respective MAPS subspace (cf. Figures \ref{fig:barycentric} and \ref{fig:frequencies}), dashed lines have been added to connect points of the same barycentric coordinates to help in identifying trends with increasing $|\boldsymbol{\omega}|_1$, and trends with changing barycentric coordinates. We do not include Nyquist plots of the data here because the real and imaginary parts of $J^*_3(\omega_1,\omega_2,\omega_3)$ measured by the MAPS frequency sweeps vary over orders of magnitude, thus much of the data is indistinguishable on a linear-scaled Nyquist plot.

While this new rheological data may be challenging to interpret on first inspection, it is possible to make a few important general observations. First, we can clearly see the diversity and density of data obtained by the MAPS experimental protocol. In Figure \ref{fig:569_J3}, subspaces B and D demonstrate that, although points are obtained at barycentric coordinates in a small range near the first harmonic MAOS vertex, curves at different barycentric coordinates are clearly distinguishable both in magnitude and phase. This is in contrast to subspace A, where points obtained near the third harmonic MAOS vertex show almost no variation with barycentric coordinates. Thus, the weakly nonlinear rheological response of this micellar solution across subspaces is quite different as well. Figure \ref{fig:1416_J3} emphasizes these observations. In particular, it is clear that the behavior of the phase angle in subspace D at points sampled near the odd PS vertex (those closer to red in color) is distinct from points sampled near the MAOS vertex (those closer to blue in color) in the same subspace. It is the evident richness of these data sets that distinguishes MAPS rheology from other weakly nonlinear experimental protocols, which can probe points along only a very limited subset of the curves shown in Figures \ref{fig:569_J3} and \ref{fig:1416_J3}.

Besides the richness and diversity of the data, Figures \ref{fig:569_J3} and \ref{fig:1416_J3} demonstrate two key features that should instill confidence in the experimental results. Firstly, the fundamental frequencies have been chosen to be exactly factors of two apart, such that the frequency sweep with the input tone set $\{1,4,16\}$ produces multiple measurements at the same point in the MAPS domain. For example, the third order response coordinate $(\omega_0,\omega_0,\omega_0)$ with $\omega_0 = 0.64$ rad/s and the coordinate $(4\omega_0,4\omega_0,4\omega_0)$ with $\omega_0 = 0.16$ rad/s are one and the same. These points serve as internal consistency checks for the experiment, in that the two measurements on the same point should be equal, or at least very nearly so. From Figure \ref{fig:1416_J3}, we see that this is indeed the case. In subspace A, consistency checks exist along the blue, purple, and darker green curves. Any variability between two points at the same value of $|\boldsymbol{\omega}|_1 = \omega_1 + \omega_2 + \omega_3$ in this subspace is small compared to the variations in the data across the range of $|\boldsymbol{\omega}|_1$. In subspaces B and D, consistency checks exist along the green and purple curves, respectively. In the plots of the magnitude of the third order complex compliance, differences between points at the same $|\boldsymbol{\omega}|_1$ are again very small. In the plots of the phase angle, experimental variations are evident, but remain small. Given the sensitivity of the measurements of the phase angle, discrepancies that are so small compared to the range of possible phase angles ($\Delta \mathrm{arg}J^*_3(\omega_1,\omega_2,\omega_3)/2\pi < 0.03$) are still a good indicator of the precision of the experiments.

\begin{figure*}[t!]
    \centering
    \includegraphics[width = \textwidth]{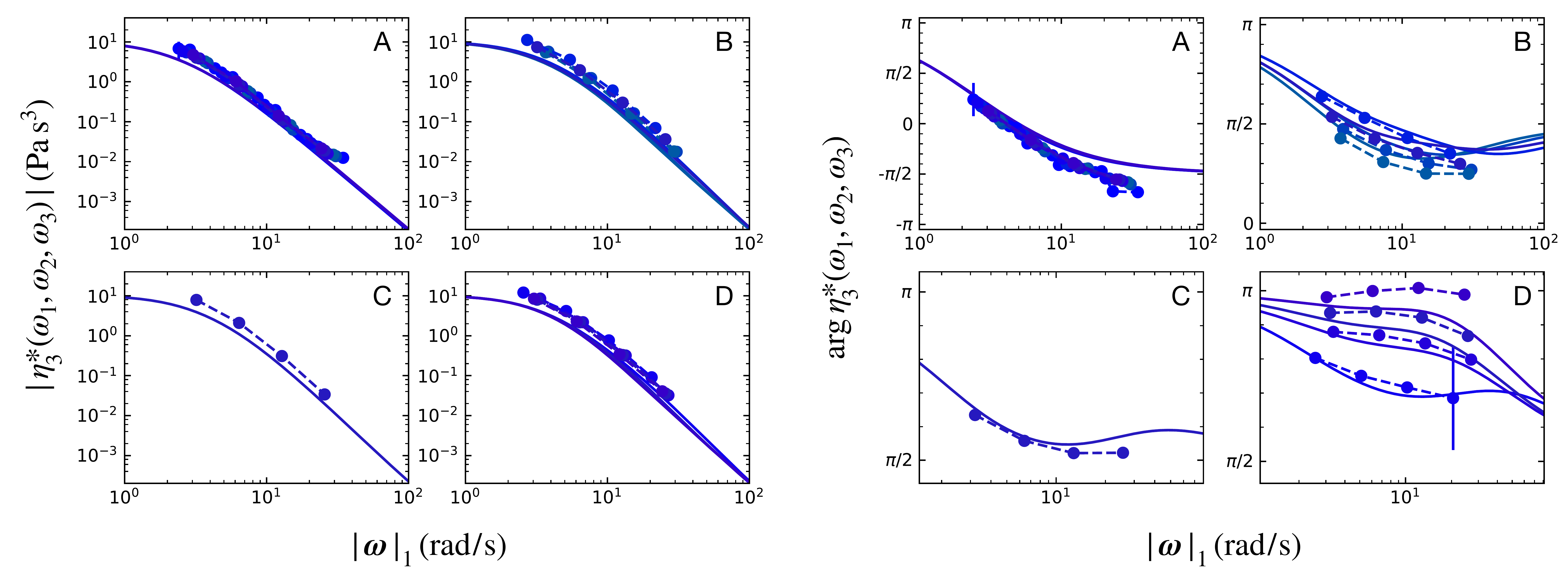}
    \caption{Values of the third order complex viscosity, $\eta^*_3(\omega_1,\omega_2,\omega_3)$, obtained from applying equations \ref{eq:J_to_G} and \ref{eq:G_to_eta} to the data in Figure \ref{fig:569_J3}, using a single mode Maxwell model with Newtonian solvent (equation \ref{eq:maxwell_LR}) with $\eta_0 = 25.7$ Pa$\cdot$s, $\tau = 0.64$ s, and $\eta_{\infty} = 0.039$ Pa$\cdot$s for the linear response. The magnitude of the third order complex viscosity, $|\eta^*_3(\omega_1,\omega_2,\omega_3)|$ plotted against the frequency $L^1$-norm, $|\boldsymbol{\omega}|_1 = \omega_1 + \omega_2 + \omega_3$, within each subspace is shown to the left, and the argument (phase) of the third order complex viscosity, $\mathrm{arg} \, \eta^*_3(\omega_1,\omega_2,\omega_3)$, plotted against the frequency $L^1$-norm within each subspace is shown to the right. Circles indicate experimental results, with dashed lines connecting points associated with the same barycentric coordinates within a subspace. Solid lines indicate predictions from the two parameter corotational Maxwell model, with $\eta_0 = 25.7$ Pa$\cdot$s and $\tau = 0.64$ s.}
    \label{fig:569_eta3}
\end{figure*}

\begin{figure*}[t!]
    \centering
    \includegraphics[width = \textwidth]{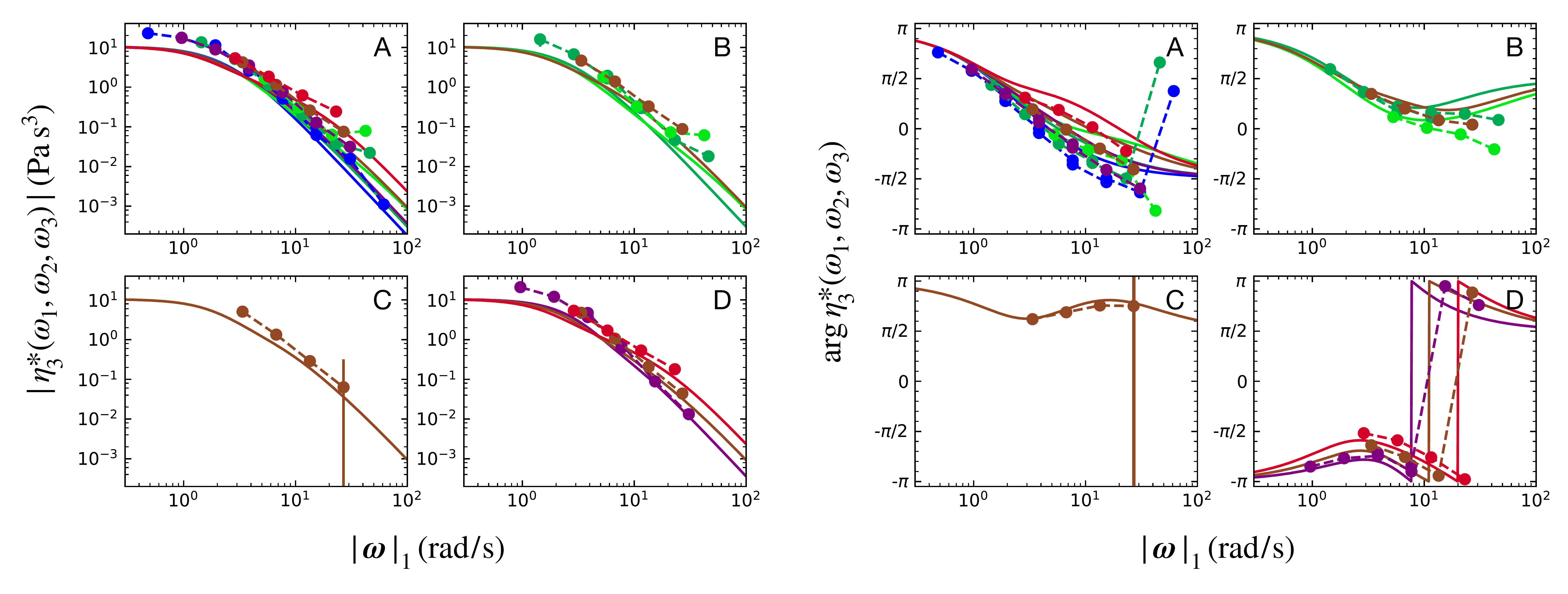}
    \caption{Values of the third order complex viscosity, $\eta^*_3(\omega_1,\omega_2,\omega_3)$, obtained from applying equations \ref{eq:J_to_G} and \ref{eq:G_to_eta} to the data in Figure \ref{fig:1416_J3}, using a single mode Maxwell model with Newtonian solvent (equation \ref{eq:maxwell_LR}) with $\eta_0 = 25.7$ Pa$\cdot$s, $\tau = 0.64$ s, and $\eta_{\infty} = 0.039$ Pa$\cdot$s for the linear response. The magnitude of the third order complex viscosity, $|\eta^*_3(\omega_1,\omega_2,\omega_3)|$ plotted against the frequency $L^1$-norm, $|\boldsymbol{\omega}|_1 = \omega_1 + \omega_2 + \omega_3$, within each subspace is shown to the left, and the argument (phase) of the third order complex viscosity, $\mathrm{arg} \, \eta^*_3(\omega_1,\omega_2,\omega_3)$, plotted against the frequency $L^1$-norm within each subspace is shown to the right. Circles indicate experimental results, with dashed lines connecting points associated with the same barycentric coordinates within a subspace. Solid lines indicate predictions from the two parameter corotational Maxwell model, with $\eta_0 = 25.7$ Pa$\cdot$s and $\tau = 0.64$ s.}
    \label{fig:1416_eta3}
\end{figure*}

The second key observation about the data that reinforces the validity of the experimental results pertains to the low-frequency behavior of the measured material function. It has been shown that, in strain-controlled experiments, the intrinsic nonlinearities measured by MAOS follow a distinct low-frequency expansion \cite{ewoldt-2014}. This result has been extended to the third order complex viscosity, $\eta^*_3(\omega_1,\omega_2,\omega_3)$ by the present authors \cite{lennon-2020}. In particular, the third order complex viscosity for a simple fluid, as defined by Coleman and Noll \cite{coleman-1961,noll-1958}, possesses the low-frequency expansion:
\begin{equation}
    \eta^*_3(\omega_1,\omega_2,\omega_3) = a + ib\sum_j\omega_j + \frac{c}{2}\sum_{j \neq k}\omega_j\omega_k + d\sum_j\omega_j^2.
    \label{eq:lowfreq}
\end{equation}
Therefore, in the limit that $\tau|\boldsymbol{\omega}|_1 \rightarrow 0$, $\eta^*_3(\omega_1,\omega_2,\omega_3) \rightarrow a$. In other words, for the third order complex viscosity, curves at every barycentric coordinate in all four subspaces should approach the same constant, purely real value, indicative of a purely viscous third order contribution to the steady shear viscosity. This plateau should occur at a characteristic scale set by the zero-shear viscosity and characteristic relaxation time of the material: $a \sim \eta_0\tau^2$. The same result could be extended to the third order complex compliance and used to check the data in Figures \ref{fig:569_J3} and \ref{fig:1416_J3} for consistency with the expected low-frequency behavior. However, one principal advantage of the MAPS rheology representation is that it allows the direct comparison of weakly nonlinear data obtained in stress control to data obtained in strain control, through the inversion formula:
\begin{align}
    & G^*_3(\omega_1,\omega_2,\omega_3) = \label{eq:J_to_G} \\ 
    & \quad\quad - \frac{J^*_3(\omega_1,\omega_2,\omega_3)}{J^*_1(\omega_1)J^*_1(\omega_2)J^*_1(\omega_3)J^*_1(\sum_j \omega_j)} \nonumber.
\end{align}
To convert from the third order complex compliance to the third order complex modulus only requires data about the linear response at the frequencies $\omega_1$, $\omega_2$, $\omega_3$, and $\omega_1+\omega_2+\omega_3$. Though SAOS data was not explicitly obtained at all of these frequencies, we can employ the model given by equation \ref{eq:maxwell_LR} with the fit parameters $\eta_0 = 25.7$ Pa$\cdot$s and $\tau = 0.64$ s to interpolate at the required points. When equation \ref{eq:J_to_G} is combined with the relationship between the third order complex modulus and third order complex viscosity:
\begin{equation}
    \eta^*_3(\omega_1,\omega_2,\omega_3) = i\left(\prod_{j}\omega_j\right)^{-1}G^*_3(\omega_1,\omega_2,\omega_3),
    \label{eq:G_to_eta}
\end{equation}
the data in Figures \ref{fig:569_J3} and \ref{fig:1416_J3} can be reformulated in terms of the third order complex viscosity, and directly compared to equation \ref{eq:lowfreq}. To demonstrate this salient feature of MAPS rheology, we perform this interconversion, with the resulting Bode plots of $\eta^*_3(\omega_1,\omega_2,\omega_3)$ shown in Figures \ref{fig:569_eta3} and \ref{fig:1416_eta3}.

From Figures \ref{fig:569_eta3} and \ref{fig:1416_eta3}, we indeed observe that many of the curves at constant barycentric coordinates approach a constant value of approximately 20 Pa$\cdot$s\textsuperscript{3} at low $|\boldsymbol{\omega}|_1$, though in some cases data is not available at low enough frequencies to accurately determine a plateau value. Indeed, we see that this plateau occurs at a value consistent with the expected characteristic scale of $a \sim \eta_0\tau^2$. In the cases where a low-frequency plateau is evident, the phase angle does indeed tend to a value of $\pi$, indicating a purely viscous asymptotic nonlinearity. Both of these observations are consistent with the expected low frequency expansion in equation \ref{eq:lowfreq}. We also briefly note that, because the range of the phase angle is constrained to $(-\pi,\pi]$, the phase angle may appear to jump discontinuously as in subsection D in Figure \ref{fig:1416_eta3}. These jumps are purely reflective of the circular nature of the phase angle, and do not represent any discontinuity in the response functions. In fact, the phase angle passing through the value of either $\pi$ or $-\pi$ signifies that the value of the imaginary part of the MAPS response function, in this case $\eta''_3(\omega_1,\omega_2,\omega_3)$, is passing through zero and changing sign. The same is true for the imaginary part at a phase angle of zero, and sign changes in the real part of the response function occur at phase angles of $\pm \pi/2$.

The data presented in Figures \ref{fig:569_J3} and Figures \ref{fig:1416_J3}, or alternatively in Figures \ref{fig:569_eta3} and \ref{fig:1416_eta3}, reveals the potential richness of MAPS rheology. The MAPS experimental protocol outlined in the previous sections are able to obtain large data sets such as these in a relatively short time. The data sets can be quite diverse internally (depending on the values of $\{n_1,n_2,n_3\}$ selected), and adjusting experimental parameters such as the input tone sets can probe features of the material response that are evidently quite distinct. Moreover, some simple analysis of the data reveals that the data sets are both internally consistent and also consistent with the expected low-frequency behavior of a simple fluid. However, this high-dimensional data is difficult to analyze on its own. Fully understanding the variations in the magnitude and phase angle of the MAPS response functions as a function of $|\boldsymbol{\omega}|_1$ and of the barycentric coordinates within each MAPS subspace requires accompanying constitutive modelling. A detailed study of the nonlinear constitutive modelling for this surfactant solution of wormlike micelles is outside of the scope of this work, and will be a topic of future study. However, it is still instructive to consider a comparison of the data to two simple constitutive models that are appropriate for describing the weakly nonlinear rheology of wormlike micellar solutions, in particular to validate that the trends observed in the data are indeed consistent with typical viscoelastic behaviors. In the next section, we consider the comparison of the data to both the corotational Maxwell model and the Giesekus model.

\subsection{Comparison to the Corotational Maxwell Model}

Perhaps one of the simplest nonlinear viscoelastic constitutive models possible that is consistent with the linear response function given by equation \ref{eq:maxwell_LR} is one in which the total stress consists of the linear combination of a solvent stress ($\boldsymbol{\sigma}_S$) and a viscoelastic extra stress ($\boldsymbol{\sigma}_E$):
\begin{equation}
    \boldsymbol{\sigma} = \boldsymbol{\sigma}_S + \boldsymbol{\sigma}_E,
    \label{eq:stress_components}
\end{equation}
where the solvent stress is Newtonian ($\boldsymbol{\sigma}_S = \eta_{\infty}\boldsymbol{\dot{\gamma}}$) and the extra stress is given by the corotational Maxwell model \cite{dewitt-1955}:
\begin{equation}
    \boldsymbol{\sigma}_E + \tau\frac{\mathcal{D}\boldsymbol{\sigma}_E}{\mathcal{D}t} = \eta_0\boldsymbol{\dot{\gamma}}.
\end{equation}
In the corotational Maxwell model, $\mathcal{D\boldsymbol{\sigma}}/\mathcal{D}t$ represents the corotational derivative of the stress tensor:
\begin{equation}
    \frac{\mathcal{D}\boldsymbol{\sigma}_E}{\mathcal{D}t} \equiv \frac{D\boldsymbol{\sigma}_E}{Dt} + \frac{1}{2}\{\boldsymbol{\omega}\cdot\boldsymbol{\sigma}_E - \boldsymbol{\sigma}_E\cdot\boldsymbol{\omega}\},
\end{equation}
with deformation rate and vorticity tensors:
\begin{equation}
    \boldsymbol{\dot{\gamma}} \equiv \nabla \textbf{u} + (\nabla \textbf{u})^T,
\end{equation}
\begin{equation}
    \boldsymbol{\omega} \equiv \nabla \textbf{u} - (\nabla \textbf{u})^T
\end{equation}
that depend on the velocity profile $\textbf{u}$. In this model, the solvent mode produces only a linear and additive contribution to the total stress. The nonlinear strain-controlled response is therefore entirely due to the nonlinearity of the corotational Maxwell model. In Part 1 of this work, we showed that the third order complex viscosity for the corotational Maxwell model is:
\begin{align}
    &\eta^*_3(\omega_1,\omega_2,\omega_3) = -\frac{\eta\tau^2}{6}\left(\frac{1}{1 + i\tau\sum_j\omega_j}\right) \label{eq:crm_eta3_body} \\
    &\times \sum_j\left[\left(\frac{1}{1 + i\tau\sum_{k\neq j}\omega_k}\right)\sum_{k\neq j}\left(\frac{1}{1 + i\tau\omega_k}\right)\right]. \nonumber
\end{align}
The third order complex compliance for the corotational Maxwell model can be found by combining equation \ref{eq:crm_eta3_body} with equations \ref{eq:G_to_eta} and \ref{eq:J_to_G}, using equation \ref{eq:maxwell_LR} as the linear response function. Note that, because the linear response includes the solvent mode, the third order complex compliance is affected by the inclusion of the solvent, unlike the third order complex viscosity. However, because the solvent viscosity is much less than the zero-shear viscosity from the polymeric contribution, this contribution to the third order complex compliance does not become evident until frequencies higher than those studied experimentally in this work. 

The corotational Maxwell model is especially simple because it contains only two adjustable material parameters, $\eta_0$ and $\tau$, which are directly determined from the linear viscoelastic response data alone. With these parameters, and the effective solvent viscosity $\eta_{\infty}$, determined from a SAOS sweep as in the previous section, there are no remaining adjustable parameters to fit the nonlinear response. Therefore, the data obtained by MAPS frequency sweeps can be directly compared to the predictions of the corotational Maxwell model without the need to supply any additional information. These comparisons are shown in Figures \ref{fig:569_J3} and \ref{fig:1416_J3} in terms of the third order complex compliance, and in Figures \ref{fig:569_eta3} and \ref{fig:1416_eta3} in terms of the third order complex viscosity, with solid lines representing the predictions of the corotational Maxwell model.

The corotational Maxwell model evidently predicts the general features of the weakly nonlinear response quite well. This is especially remarkable given that the model has no adjustable parameter to fit the MAPS data; the predicted behavior is determined completely by the linear viscoelastic response. In plots of the phase angle, the predictions of the corotational Maxwell model are in near-quantitative agreement with the data, capturing the trends with increasing $|\boldsymbol{\omega}|_1$, with varying barycentric coordinates, and across the different subspaces. Moreover, the model captures the same trends in the magnitude of either the third order complex modulus or third order complex compliance. The largest discrepancy between the model predictions and the data is that the model underestimates the magnitude of both $J^*_3(\omega_1,\omega_2,\omega_3)$ and of $\eta^*_3(\omega_1,\omega_2,\omega_3)$ by a factor of approximately two, which appears to be constant across all barycentric coordinates and values of $|\boldsymbol{\omega}|_1$. Another discrepancy between the model and data occurs for some very high values of $|\boldsymbol{\omega}|_1$, for example at the highest values measured in the A and B subspaces in the $\{1,4,16\}$ frequency sweep. At those $|\boldsymbol{\omega}|_1$, the magnitude and phase of $J^*_3(\omega_1,\omega_2,\omega_3)$ appear to deviate sharply from the trend of the lower-frequency data. Though the exact origin of these deviations is unknown, it is possible that they are due to the rotational inertia of the cone-and-plate fixture that becomes important at high frequencies in controlled-stress devices, or due to the inertia of the fluid preventing a homogeneous state of shear from fully developing in the sample gap on the time-scales of these observations. This latter effect is examined more closely in Appendix \ref{app:gap_loading}, in the context of the gap-loading limit.

\begin{figure*}[t!]
    \centering
    \includegraphics[width = \textwidth]{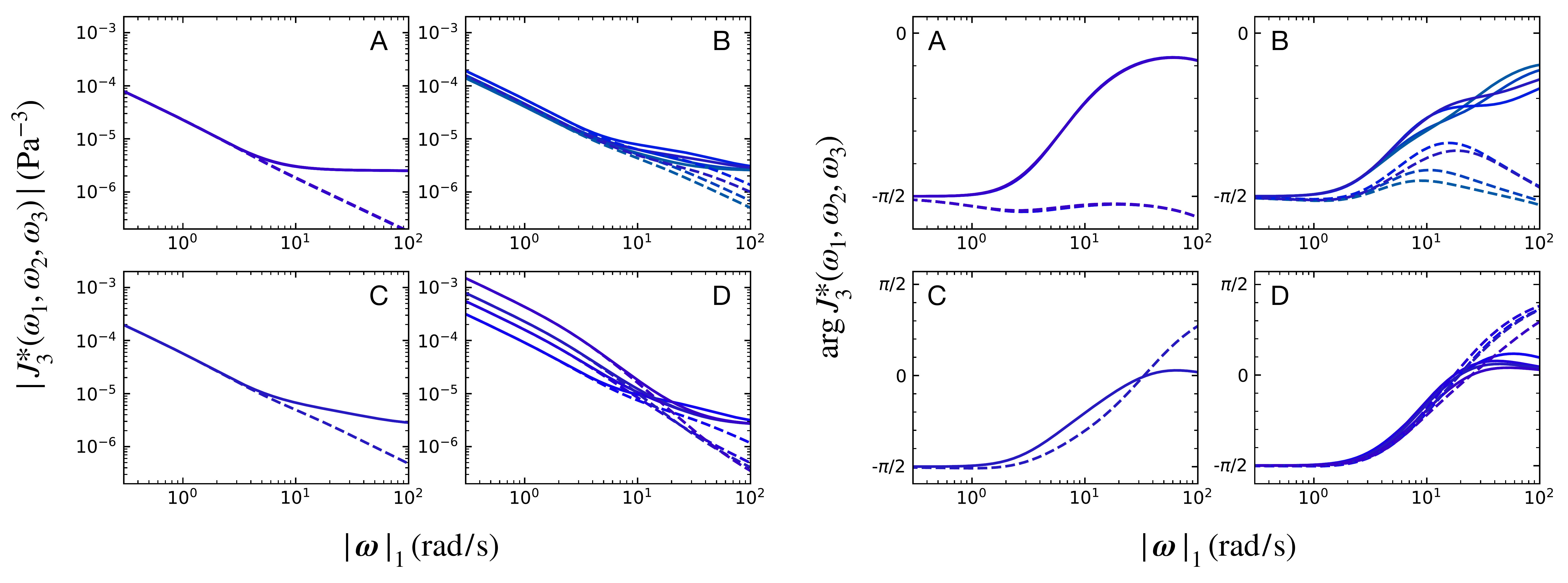}
    \caption{Bode plots of the third order complex compliance $J^*_3(\omega_1,\omega_2,\omega_3)$ at barycentric coordinates probed by a MAPS experiment with the input tone set $\{5,6,9\}$, computed from the analytical solution to equation \ref{eq:stress_components} with a Newtonian solvent stress $\boldsymbol{\sigma}_S$ with viscosity $\eta_{\infty} = 0.039$ Pa$\cdot$s, and an extra stress $\boldsymbol{\sigma}_E$ given by either the corotational Maxwell model (solid lines) with $\eta_0 = 25.7$ Pa$\cdot$s and $\tau = 0.64$ s, or the Giesekus model (dashed lines) with $\eta_0 = 25.7$ Pa$\cdot$s, $\tau = 0.64$ s, and $\alpha = 0.5$.}
    \label{fig:giesekus_vs_crm_569}
\end{figure*}

\begin{figure*}[t!]
    \centering
    \includegraphics[width = \textwidth]{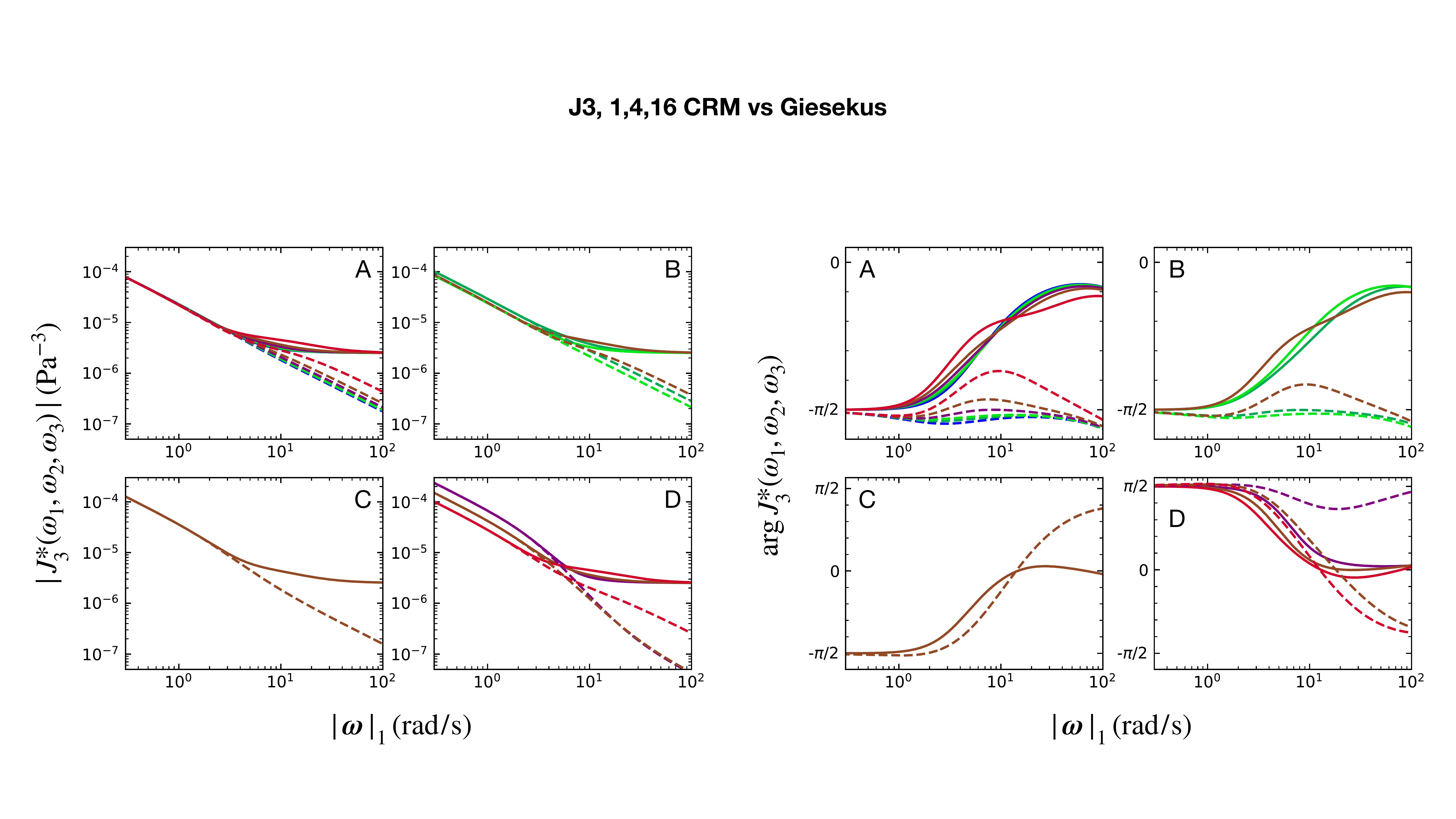}
    \caption{Bode plots of the third order complex compliance $J^*_3(\omega_1,\omega_2,\omega_3)$ at barycentric coordinates probed by a MAPS experiment with the input tone set $\{1,4,16\}$, computed from the analytical solution to equation \ref{eq:stress_components} with a Newtonian solvent stress $\boldsymbol{\sigma}_S$ with viscosity $\eta_{\infty} = 0.039$ Pa$\cdot$s, and an extra stress $\boldsymbol{\sigma}_E$ given by either the corotational Maxwell model (solid lines) with $\eta_0 = 25.7$ Pa$\cdot$s and $\tau = 0.64$ s, or the Giesekus model (dashed lines) with $\eta_0 = 25.7$ Pa$\cdot$s, $\tau = 0.64$ s, and $\alpha = 0.5$.}
    \label{fig:giesekus_vs_crm_1416}
\end{figure*}

At low values of $|\boldsymbol{\omega}|_1$, it is perhaps unsurprising that the qualitative behavior of the model and the data are similar. Such an agreement should be expected for a constitutive model and viscoelastic material that are both Coleman-Noll simple fluids, as the same low-frequency expansion (equation \ref{eq:lowfreq}) will apply to both. The more interesting observation is that the high-frequency behavior of the model and data also agree quite well. For instance, the corotational Maxwell model predicts both the low-frequency and high-frequency limits of the phase angle given by the MAPS response. It also predicts the correct high-frequency scaling of the data with $|\boldsymbol{\omega}|_1$, namely that $|J^*_3(\omega_1,\omega_2,\omega_3)|$ plateaus and that $|\eta^*_3(\omega_1,\omega_2,\omega_3)| \propto |\boldsymbol{\omega}|_1^{-3}$ at high frequencies. One significant consequence of this observation is that many other models, which possess different high-frequency characteristics, are inconsistent with the data (see for example the recent extensive study by Hyun and co-workers \cite{hyun-2020}). In Part 1 of this work we explored the weak nonlinearities of multiple constitutive models, and of those only time-strain separable models with a Maxwellian linear response, of which the corotational Maxwell model is one example, display high-frequency characteristics consistent with these observations. 

Notably, the Giesekus model is a constitutive model that produces high-frequency behavior that is apparently inconsistent with the observed trends in the experimental data. The Giesekus constitutive model is similar in form to the corotational Maxwell model \cite{giesekus-1982}:
\begin{equation}
    \boldsymbol{\sigma}_E + \tau\boldsymbol{\sigma}_{E(1)} + \frac{\alpha\tau}{\eta_0}\boldsymbol{\sigma}_E\cdot\boldsymbol{\sigma}_E = \eta_0\boldsymbol{\dot{\gamma}},
    \label{eq:giesekus}
\end{equation}
with a single nonlinear mobility parameter $\alpha$ and the upper convected derivative:
\begin{equation}
    \boldsymbol{\sigma}_{E(1)} = \frac{D\boldsymbol{\sigma}_E}{Dt} - (\nabla\textbf{u})^T\cdot\boldsymbol{\sigma}_E - \boldsymbol{\sigma}_E\cdot(\nabla\textbf{u}).
\end{equation}
Other authors have reported that the Giesekus model is suitable for describing the behavior of solutions of wormlike micelles under LAOS deformation \cite{gurnon-2012}, or under PS and orthogonal superposition deformation \cite{kim-2013}, in part because (for suitable choices in the value of $\alpha$) the Giesekus model can also predict the formation of shear bands that have been observed in solutions of wormlike micelles \cite{yoo-1989,salmon-2003,moorcroft-2014,fardin-2012}. However, it becomes clear from inspection of the MAPS signatures of the Giesekus model that it fails to capture key features of the measured MAPS response of this wormlike micellar solution. Figures \ref{fig:giesekus_vs_crm_569} and \ref{fig:giesekus_vs_crm_1416} show how the predicted weakly nonlinear response of the Giesekus model, which we have obtained in Part 1 of this work, compares to that of the corotational Maxwell model for MAPS experiments with input tone sets $\{5,6,9\}$ and $\{1,4,16\}$, respectively. While the corotational Maxwell model predicts a high-frequency plateau in the magnitude of the third order complex compliance, consistent with the experimental data, the Giesekus model predicts a continued decay at high frequencies. The Giesekus model also predicts that, in subspace A of the $\{5,6,9\}$ experiment (Figure \ref{fig:giesekus_vs_crm_569}), the phase of the third order complex compliance should be approximately constant with variation in $|\boldsymbol{\omega}|_1$ near a value of $-\frac{\pi}{2}$, as opposed to the steady increase in phase from $-\frac{\pi}{2}$ to 0 predicted by the corotational Maxwell model. This invariant behavior predicted by the Giesekus model is again inconsistent with the experimental data, which is measured to steadily increase in phase along these barycentric coordinates. Although the Giesekus model uses a different frame-invariant derivative than does the corotational Maxwell model, its frequency-dependent MAPS signatures arise directly from the inclusion of the $\boldsymbol{\sigma}\cdot\boldsymbol{\sigma}$ term in equation \ref{eq:giesekus}. In fact, excluding this term ($\alpha = 0$) reduces the Giesekus model to the upper-convected Maxwell model, which produces a purely linear shear stress response in simple shear at all strain amplitudes and frequencies.

Figures \ref{fig:giesekus_vs_crm_569} and \ref{fig:giesekus_vs_crm_1416} also reveal the importance of high-frequency data ($|\boldsymbol{\omega}|_1 \tau > 1$) in assessing the quantitative ability of different constitutive models to describe MAPS data. At low frequencies, the predictions of the Giesekus and corotational Maxwell models are nearly indistinguishable. Both models in this regime possess the same polynomial expansion, given in equation \ref{eq:lowfreq} for the third order complex viscosity, thus their MAPS responses differ only in the numerical values of the coefficients in this expansion. At high frequencies, however, the predictions diverge substantially, to the point where they predict nonlinearities that are completely out of phase with one another, and scale differently with $|\boldsymbol{\omega}|_1$. These differences allow us to distinguish the corotational Maxwell model as superior for describing the weakly nonlinear response of this wormlike micellar solution to simple shear.

\begin{figure*}[t]
    \centering
    \includegraphics[width = \textwidth]{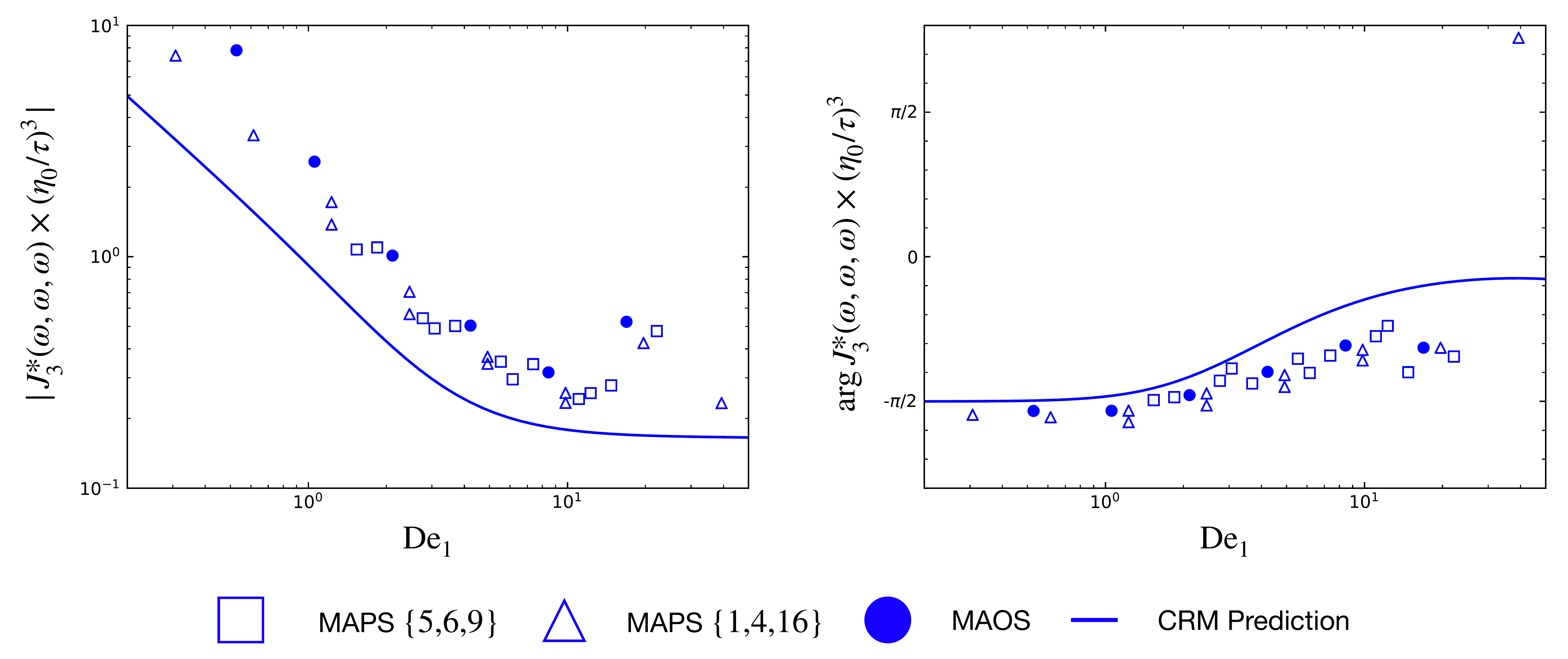}
    \caption{Bode plots of the magnitude (left) and phase (right) of the dimensionless third order complex viscosity, plotted against the Deborah number with respect to the frequency $L^1$-norm, on the third harmonic MAOS vertex for which $\omega_1 = \omega_2 = \omega_3$. Unfilled squares represent points obtained in the MAPS frequency sweep with the input tone set $\{5,6,9\}$. Unfilled triangles represent points obtained in the MAPS frequency sweep with the input tones set $\{1,4,16\}$. Filled circles represent points obtained from the MAOS amplitude sweeps. The prediction of the corotational Maxwell model (CRM) is indicated with a solid blue line.}
    \label{fig:maos_J3_comparison}
\end{figure*}

This section serves as a preliminary illustrative exercise in evaluating the fit of a constitutive model to MAPS data. Though the corotational Maxwell model is quite simple and does not require fitting any model parameters to the MAPS data, it predicts the trends of the data remarkably well. However, this nonlinear continuum mechanics representation provides no insight into the underlying microscopic physics governing the rheology of the wormlike micelle solution, which is one of the central purposes of constitutive modeling. Furthermore, we have not explored here whether the predictions of the corotational Maxwell model similarly match the response of this wormlike micellar solution to different deformation protocols, such as startup of steady shear. Moreover, it is not clear whether the corotational Maxwell model is able to accurately describe strongly nonlinear features of the simple shear rheology of solutions of wormlike micelles, such as the shear banding instabilities observed for certain solutions including some CPyCl-based wormlike micellar solutions \cite{salmon-2003,moorcroft-2014,fardin-2012}. In future work, we will address these concerns through a more detailed study of CPyCl-based wormlike micelles, including an examination of a microstructural-based constitutive model specifically derived for such materials \cite{cates-1990}.

\subsection{Validation with MAOS Experiments}

The close resemblance of the measured MAPS data to a canonical nonlinear viscoelastic constitutive model, as well as the internal consistency of the data sets and adherence to the expected low-frequency expansion, illustrate that the experimental protocol proposed in this work enables us to systematically measure the intrinsic nonlinearities of viscoelastic materials. One potential concern is, of course, that the measured nonlinearities might instead reflect nonlinearities in the electromechanical transfer function of the experimental equipment. Recognizing that the measurements closely reflect expected viscoelastic behavior should assuage these concerns, as it is unlikely that the transfer function of the instrument produces the same type of nonlinearity. However, with any new experimental protocol, it is beneficial to validate results against a more established protocol for which it is known that the instrument response is of little concern. To validate MAPS data, MAOS is an obvious candidate. The procedure for obtaining MAOS data is well-established, and MAOS data can be easily compared with MAPS data because MAOS tests are, in fact, a subset of MAPS.

MAOS data was obtained by running stress-controlled amplitude sweeps with \emph{single tone} oscillations at frequencies $\omega_0 =$ 0.32, 0.64, 1.28, 2.56, 5.12, and 10.24 rad/s. The amplitude of oscillation was varied in the range $\sigma_0 \in [1,30]$ Pa. The third harmonic strain response $\hat{\gamma}(3\omega_0)$ for each combination of $\omega_0$ and $\sigma_0$ was measured, and regressed to a fifth-order polynomial:
\begin{equation}
    \hat{\gamma}(3\omega_0) = \epsilon(3\omega_0) + \sigma_0^3\hat{\gamma}^{(3)}(3\omega_0) + \sigma_0^5\hat{\gamma}^{(5)}(3\omega_0),
\end{equation}
which assumes that there is negligible spectral leakage from the first harmonic response at the third harmonic peak. Note that the same assumption should not be made for MAPS data, as many of the output channels reside very close to at least one of the distinct first harmonic peaks corresponding to each input tone, and therefore warrant a linear spectral leakage term in their fit. Finally, the value of $\hat{\gamma}^{(3)}(3\omega_0)$ was used to compute the value of $J^*_3(\omega_0,\omega_0,\omega_0) = 4\hat\gamma^{(3)}(3\omega_0)/i\pi$.

Both the MAPS frequency sweep with the input tone set $\{5,6,9\}$ and with the input tone set $\{1,4,16\}$ measure points along the third harmonic MAOS vertex, i.e. for coordinates where $\omega_1 = \omega_2 = \omega_3$. During experimentation, it was observed that the combined duration of the MAPS and MAOS experiments was long enough that sample evaporation effects became evident, as detected by small changes in the SAOS data before and after the set of medium amplitude experiments. Therefore, it was necessary to replace the sample between the MAOS and MAPS sweeps to fully eliminate the effects of sample evaporation. While the duration of each of these sweeps separately was short enough that evaporation effects became negligible, as confirmed by comparing SAOS data taken before and after the sweeps, the linear response of the sample studied in the MAOS experiments differed slightly from the sample studied during the MAPS experiments, due to sample-to-sample variability. For the sample studied by the MAOS experiments, SAOS data was fit to a single-mode Maxwell model plus effective Newtonian solvent with $\eta_0 = 20.5$ Pa$\cdot$s, $\tau = 0.55$ s, and $\eta_{\infty} = 0.044$ Pa$\cdot$s. To account for these small variations in the linear viscoelastic response of the samples, the third order complex compliance is made dimensionless by $(\eta_0/\tau)^3$, and an appropriate Deborah number defined as:
\begin{equation}
    \mathrm{De}_{1} \equiv 3\omega_0\tau
    \label{eq:dimensionless}
\end{equation}
for MAOS experiments and
\begin{equation}
    \mathrm{De}_{1} \equiv |\boldsymbol{\omega}|_1\tau
\end{equation}
for MAPS experiments, with $\eta_0$ and $\tau$ determined from linear response data taken immediately before the respective experiment (i.e 25.7 Pa$\cdot$s and 0.64 s for the MAPS experiments, and 20.5 Pa$\cdot$s and 0.55 s for the MAOS experiment).

In Figure \ref{fig:maos_J3_comparison}, the points on the third harmonic MAOS vertex from each MAPS frequency sweep have been extracted, scaled with $(\eta_0/\tau)^3$, and the resulting magnitude and phase plotted against $\mathrm{De}_1$. The MAOS data obtained by the amplitude sweeps have been made dimensionless and plotted as well. From this figure, it is clear not only that the two MAPS frequency sweeps produce consistent measurements of the third order complex modulus along the MAOS vertex, but also that the MAOS experiment agrees quite well with the MAPS experiments. This is an especially encouraging sign that the MAPS experimental protocol measures true intrinsic material nonlinearities. The relatively narrow spread of data in Figure \ref{fig:maos_J3_comparison} is also indicative of the precision of our proposed MAPS protocol.

\begin{figure}[t]
    \centering
    \includegraphics[width = \columnwidth]{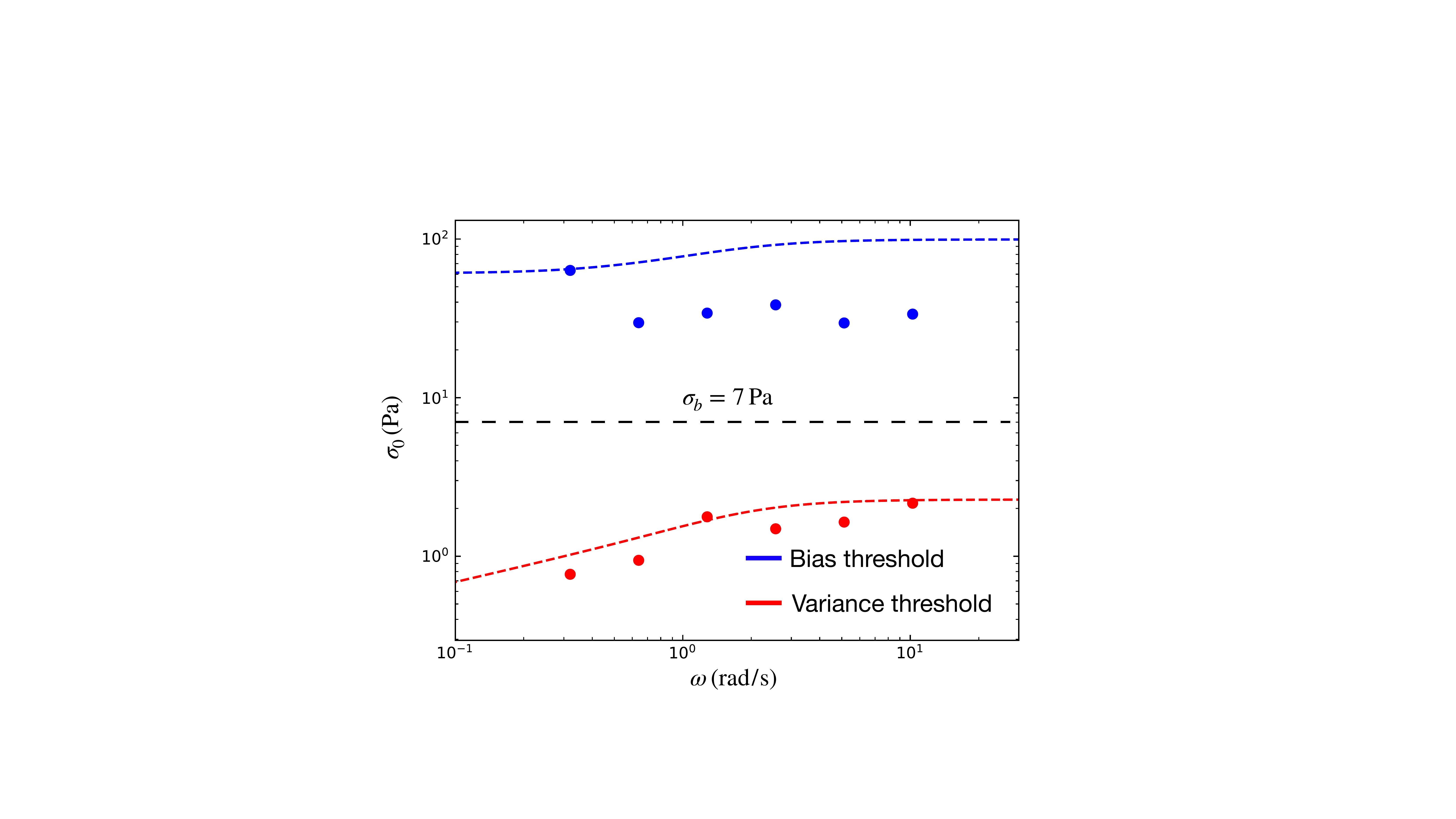}
    \caption{The upper (high-bias) and lower (high-variance) bounds for the imposed amplitudes of a MAOS experiment, as measured by the magnitude of the third harmonic response. Circles show values obtained from experimental MAOS sweeps. Dashed lines show predictions from the corotational Maxwell model with $\eta_0 = 20.5$ Pa$\cdot$s and $\tau = 0.55$ s, with white instrumental noise at $\epsilon(\omega) = 10^{-5}$. The imposed-stress amplitude ($\sigma_b$) selected for the MAPS experiments in previous sections is indicated with a horizontal dashed black line.}
    \label{fig:amp_bounds_maos}
\end{figure}

The MAOS amplitude sweeps discussed in this section also validate the conclusions of Section \ref{sec:amplitude_region}, in which it was determined that by using stress-controlled test inputs, a single stress amplitude was suitable to minimize bias and variance at all frequencies. To show this, we construct from each amplitude sweep a plot identical to Figure \ref{fig:amp_sweep}. The threshold error is set equal to the magnitude of the third order component of the response. Thus, lower and upper bounds of the optimal amplitude region can be approximated by the intersection of the red lines with the dashed black line, and of the blue lines with the dashed black line, respectively. Figure \ref{fig:amp_bounds_maos} plots these upper and lower bounds at each imposed frequency. The results show qualitative agreement with the prediction of the upper and lower bounds of the same threshold for the corotational Maxwell model, with $\eta_0 = 20.5$ Pa$\cdot$s and $\tau = 0.55$ s, and with an instrument noise level: $\epsilon(\omega) = 10^{-5}$ s. The model overestimates the upper bound for the optimal amplitude range, probably because the corotational Maxwell model underestimates the magnitude of the fifth-order response that is excited in the actual micellar test fluid. However, most importantly, the observed bounds indeed reveal a range of possible imposed-stress amplitudes that are optimal for all frequencies under consideration. Though these bounds will change slightly with small sample-to-sample variations in $\eta_0$ and $\tau$, amplitudes near the center of the optimal region, such as the value $\sigma_b = 7$ Pa selected for our MAPS frequency sweeps, should remain comfortably between the upper and lower bounds.

\section{Discussion}

The mathematical framework of MAPS rheology reveals the fundamental richness of weakly nonlinear measurements of complex fluids that is captured in the form of the three-dimensional, complex-valued MAPS response functions such as $G^*_3(\omega_1,\omega_2,\omega_3)$. However, there have been no attempts until now to directly measure the MAPS response functions or their time-domain analogs across their entire domain \cite{bierwirth-2019}. The experimental design presented in this work therefore represents the first attempt to fully exploit the richness of weakly nonlinear shear rheometry, taking a major step forward both in the dimensionality of the nonlinear rheological data obtained and the data throughput. In the first part of this work we noted that a MAOS experiment measures two complex data points (corresponding to $[e_1](\omega) + i\omega[v_1](\omega)$ and $[e_3](\omega) - i\omega[v_3](\omega)$) embedded on a one-dimensional manifold defined by the imposed oscillation frequency, $\omega$. In this same sense, the experimental protocol for MAPS rheology presented herein measures 19 complex data points embedded on a three-dimensional manifold defined by the coordinates $(\omega_1,\omega_2,\omega_3)$. The difference in dimensionality and data-throughput is striking, and it is even more notable when we realize that by carefully separating the fundamental frequency $\omega_0$ and input tones $\{n_1,n_2,n_3\}$ in the experimental design, we can control the acquisition time of a MAPS frequency sweep to be comparable to that of a MAOS frequency sweep. Furthermore, the data obtained from a single three-tone MAPS experiment parallels the data throughput and dimensionality of even highly nonlinear rheometric test protocols such as LAOS.

Though the increased complexity of the multi-tone deformation protocols discussed in this work affords high-dimensional and high-throughput measurements, it does at the same time introduce the possibility of additional complications. Two complications emanate from the fact that multi-tone signals can require rapid changes in strain and strain rate. In the implementation of such signals, a rheometer must impose large angular accelerations, and therefore large torques. Even for modest strain amplitudes, these large torques might exceed the capabilities of the instrument. As we have briefly mentioned, one strategy for minimizing the peak acceleration required in such a signal is to adjust the phase of the input tones. This problem is often formulated in terms of minimizing the crest factor of the signal \cite{boyd-1986,friese-1997}. Though fortunately the risk of large accelerations is minimal in three-tone signals with modest amplitudes, it will warrant closer attention when considering more complex signals composed of more than three tones. 

In addition to the complication of implementing signals with rapid changes in strain rate, such signals also present a difficulty in ensuring the linearity of the imposed flow profile in the gap, a critical kinematic assumption in defining simple shear flow. This problem amounts to ensuring that the transport of linear momentum across the gap occurs on a time scale much less than the observation time scale. This effect can be controlled, to an extent, by adjusting the size of the sample gap; however, the gap separation cannot be decreased indefinitely due to alignment limits or the microstructure of the test material. This places limitations on the range of deformation protocols that will still ensure homogeneous kinematics. It is possible to solve analytically for the limiting gap size for which the assumption of a linear flow profile holds given an arbitrary shear deformation protocol, which is often called the ``gap loading limit'' \cite{schrag-1977}. This solution is presented in Appendix \ref{app:gap_loading}. For the large majority of the observations in this work, the assumption of homogeneous kinematics clearly holds. For the highest-frequency measurements, however, the gap at the rim of the cone-and-plate fixture approaches the same order of magnitude as the gap loading limit. This is one possible explanation for the deviation of these measurements from the lower-frequency trends. Because assessing whether MAPS measurements are made within the gap loading limit is an important step in all measurements, we have included a feature in the MITMAPS software package to compute an estimate of the gap loading limit for a MAPS experiment based on a material's density and linear viscoelastic response.

At several points in our discussion we have alluded to multi-tone experiments with more than three input tones. The above discussion reveals that such protocols require some additional considerations, such as crest factor minimization and close attention to the gap loading limit. Additionally, the constraint that the input tone set has only unique triplet sums becomes more prohibitive as the number of input tones is increased. Therefore, such experiments may indeed require a reformulation of the experimental design. Given the already high data throughput of a three-tone experiment, and the complications introduced by adding more tones, one might question whether such experiments are worthwhile. Though the ultimate feasibility of carrying out such experiments is still unknown, let us briefly consider the potential rewards. While the three-tone experiments developed in this work yield 19 complex-valued data points after measurements at two amplitudes, a four-tone experiment would produce as many as 60 distinct complex-valued data points after measurements at five amplitudes, and a five-tone experiment would produce as many as 110 complex data points after measurements at six amplitudes. Thus despite the additional complexity, the promise of such rich data sets warrants future study.

\section{Conclusions}

Part 1 of this work laid the mathematical foundations for MAPS rheology, revealing a common language for weakly nonlinear shear rheology in terms of a new material function: the third order complex modulus. Through existing rheometric techniques such as MAOS and PS, rheologists have studied some projections of this material function for years now, although only on the periphery of its domain (as discussed in Section \ref{sec:input_tones}). In this second part of the article series, an experimental protocol was developed to access the interior of this domain, allowing rheologists to study the entire continuum of weakly nonlinear shear responses spanning between those projections captured by MAOS and PS for the first time. The many levers that are available within this new experimental design, such as the selected set of input tones $\{n_1,n_2,n_3\}$ and the range of fundamental frequencies $\omega_0$ imposed, allow rheologists the flexibility to explore the nonlinear shear response space of viscoelastic materials in regimes that suit their specific research goals. Furthermore, such experiments allow rheologists to obtain rich data sets from commercial rheometers in a high-throughput manner. With only a modest number of MAPS experiments it is possible, therefore, to begin constructing a more complete picture of a material's weakly nonlinear shear response space.

Though we hope that this accessible experimental protocol will facilitate the implementation of MAPS rheology, there are still a number of questions about both the mathematical and experimental framework yet to be addressed. For instance, while the guidelines of Section \ref{sec:input_tones} demonstrate how different regions of the MAPS domain can be studied by choosing different input tone sets, it is not clear which regions are most important to study for particular purposes. For example, certain regions of the MAPS domain may be more strongly correlated with certain aspects of a given material's molecular physics. Moreover, which data sets are most useful for data-driven problem formulations, such as automated material classification, is not yet known. Whether MAPS data alone provides enough information to reliably distinguish materials whose rheological signatures differ in other experiments is also an open question, as is the limit to which weakly nonlinear data can accurately predict a material's response to even stronger deformations or other flow types (e.g. planar elongation). Finally, in Part 1 of this work we discussed briefly the possibility that not all materials may possess a convergent Volterra series approximation for infinite periodic inputs. It may be the case that those classes of materials (which potentially include thixotropic and yield stress materials whose memory of past deformations does not necessarily fade over time) are not amenable to the present experimental protocol. Determining the context in which MAPS rheology applies to these materials, if it does at all, is an interesting avenue for future exploration.

Each of the above challenges and open questions, along with those discussed in Part 1 of this work, represents a potentially rich area of study. We hope that future work in this field will address many of these questions. Although MAPS rheology is in its infancy, our illustrative experiments with a wormlike micellar solution show that it is already a powerful tool for studying and understanding the weakly nonlinear shear response space of viscoelastic materials. By directly applying the same experimental protocol developed in this work to study a broad range of complex fluids and soft solids, rheologists can begin to construct the same sort of large, rich data sets that now drive innovation in many other fields.

\section*{Supplementary Material}

A version of the MITMAPS software tool is include in the Supplementary Material along with documentation, links to the data files used to construct Figure \ref{fig:saos} and Figures \ref{fig:569_J3} through \ref{fig:1416_eta3}, and a tutorial on how to construct these figures as well as Figures \ref{fig:giesekus_vs_crm_569} and \ref{fig:giesekus_vs_crm_1416}.

\section*{Acknowledgements}

K.R.L. was supported by the US Department of Energy  Computational  Science  Graduate  Fellowship  program  under  grant  DE-SC0020347.

\bibliography{biblio}

% Appendix

% Appendices

\appendix

\section{Fifth-order Vandermonde Interpolation for MAPS}
\label{app:fifth_order}

The experimental design in this paper focused on isolating the third-order elements of a material's response using a cubic fit. This fit leaves the possibility of bias due to fifth-order effects, which can be limited by careful selection of the input amplitude. It is possible to further correct for this bias by performing a fifth-order fit to the material response along a given channel, $\omega^*$, thereby pushing the bias to higher-order effects, which are weaker in the weakly nonlinear regime. To do so, three strain amplitudes are selected, $ \gamma_a $, $ \gamma_b $, and $ \gamma_c $ and the same three tone MAPS experiment is performed with each of these amplitudes.  The Fourier transformation of the stress in the experiment with strain amplitude $ \gamma_a $ on a channel of interest, $ \omega^* $, can now be expressed as a fifth order polynomial:
\begin{equation*}
\hat \sigma_a( \omega^* ) = \gamma_a \sigma^{(1)} + \gamma_a^3 \hat \sigma^{(3)}( \omega^* ) + \gamma_a^5 \hat \sigma^{(5)}( \omega^* ),
\end{equation*}
and likewise for the stress resulting from the experiments with strain amplitude $ \gamma_b $ and $ \gamma_c $.  By convention, we choose that: $ \gamma_a = r_a \gamma_c $ and $ \gamma_b = r_b \gamma_c $, with $ 0 < r_a < r_b < 1 $. The linear system representing these sets of experiments is:
\begin{equation}
\left(\begin{array}{c} \hat \sigma_a( \omega^* ) \\ \hat \sigma_b( \omega^* ) \\
\sigma_c( \omega^* )\end{array}\right) = \left(\begin{array}{ccc}  r_a & r_a^3 & r_a^5 \\ r_b & r_b^3 & r_b^5 \\ 1 & 1 & 1 \end{array}\right) \left(\begin{array}{c} \gamma_c \hat \sigma^{(1)}( \omega^* ) \\ \gamma_c^3 \hat \sigma^{(3)}( \omega^* ) \\ \gamma_c^5 \hat \sigma^{(5)}( \omega^* ) \end{array}\right).  \label{eq:vandermonde_appendix}
\end{equation}
The $L^\infty$-norm condition number of the fifth-order Vandermonde matrix is:
\begin{equation}
    \| \mathbf{V} \|_\infty \| \mathbf{V}^{-1} \|_\infty = \frac{1+r_a - r_b}{r_a r_b (1 + r_a ) ( 1 - r_b ) ( r_b - r_a ) }.
\end{equation}
The values of $ r_a $ and $ r_b $ that minimize the condition number of the Vandermonde matrix are: $ r_a = ( \sqrt{5} - 1 ) / 4 \approx 0.31 $ and $ r_b = ( \sqrt{5} + 1 ) / 4 \approx 0.81 $, with minimal condition number: 16.

This fifth order fit to the stress on the channel $ \omega^* $ accomplishes two tasks.  First, it extracts and quantifies the cubic nonlinearity, which is associated with the quantity we want to measure.  Second, at finite strain amplitude, the channel may carry information about the next highest nonlinearity in the stress, which is fifth order.  Although we do not investigate how this stress relates to a fifth order modulus (and it may not relate uniquely), determining this fifth order component by regression prevents the next highest order nonlinearity from influencing the quantity of interest: $ \hat \sigma^{(3)}( \omega^* ) $.  The polynomial regression reduces bias in the inferred value of $ \hat \sigma^{(3)}( \omega^* ) $ but increases slightly the variance in the regressed result.  The bias in this fit is: $ (13/16) \, \gamma_c^4 | \hat \sigma^{(7)}( \omega^* ) | $, where $ \hat \sigma^{(7)}( \omega^* ) $ is the coefficient of the seventh order response that is now neglected in the expansion of the measured stress. The variance in the inferred value of $ \hat \sigma^{(3)}( \omega^* )$ with this fit is set by the value of $ \sqrt{\mathbf{e}_3^T ( \mathbf{V}^T \mathbf{V} )^{-1} \mathbf{e}_3} $, with $ \mathbf{e}_3 = ( 0, 1, 0 ) $, and is  $ ( 12\sqrt{29} / 5 )  \, \epsilon( \omega^* ) \gamma_c^{-3} $.

If we repeat the bias-variance trade off analysis performed for the cubic fit, we find that the optimal strain amplitude is:
\begin{equation}
    \gamma_c \approx 1.4 \left( \frac{\epsilon( \omega^* )}{ |\hat \sigma^{(7)}( \omega^* ) |} \right)^{1/7}. \label{eq:biasvar_appendix}
\end{equation}
The minimal error in the inferred value of the third order response function resulting from equal bias and variance  is approximately:
\begin{equation}
    7.8 \, \epsilon( \omega^* )^{4/7} | \sigma^{(7)}( \omega^* ) |^{3/7}.
\end{equation}  
The minimal error is slightly more sensitive to the instrumental noise response than the material response captured with the fifth-order fit, in contrast to the cubic fit (equation \ref{eq:minimum_error}) where the power law exponents are $2/5$ and $3/5$, respectively.

\section{Fifth-order Solution to the Corotational Maxwell Model}
\label{app:crm}

The corotational Maxwell model can be expressed as the following differential equation:
\begin{equation}
    \boldsymbol{\sigma} + \tau\frac{\mathcal{D}\boldsymbol{\sigma}}{\mathcal{D}t} = \eta_0\boldsymbol{\dot{\gamma}},
\end{equation}
where $\boldsymbol{\dot{\gamma}} = \nabla \textbf{u} + (\nabla \textbf{u})^T$ is the rate-of-strain tensor, $\textbf{u}$ is the velocity field, and the corotational derivative is given by:
\begin{equation*}
    \frac{\mathcal{D}\boldsymbol{\sigma}}{\mathcal{D}t} \equiv \frac{D\boldsymbol{\sigma}}{Dt} + \frac{1}{2}\{\boldsymbol{\omega}\cdot\boldsymbol{\sigma} - \boldsymbol{\sigma}\cdot\boldsymbol{\omega}\},
\end{equation*}
with the vorticity tensor $\boldsymbol{\omega} = \nabla \textbf{u} - (\nabla\textbf{u})^T$. The plateau modulus of the corotational Maxwell model is given by $G_0 = \eta_0 / \tau$.

In Part 1 of this work, we obtained the solution for the first and third order complex viscosities in the corotational Maxwell model:
\begin{equation}
    \frac{\eta^*_1(\omega)}{\eta_0} = \frac{1}{1 + i\tau\omega},
\end{equation}
\begin{align}
    & \frac{\eta^*_3(\omega_1,\omega_2,\omega_3)}{\eta_0\tau^2} = -\frac{1}{6}\left(\frac{1}{1 + i\tau_0\sum_j\omega_j}\right) \label{eq:crm_eta3} \\
    & \times \sum_j \left[\left(\frac{1}{1 + i\tau_0\sum_{k\neq j}\omega_k}\right)\sum_{k\neq j}\left(\frac{1}{1 + i\tau_0\omega_k}\right)\right]. \nonumber
\end{align}
The solution for the fifth order complex viscosity can be similarly obtained by asymptotic analysis. This solution is:
\begin{widetext}
\begin{align}
    \frac{\eta^*_5(\omega_1,\omega_2,\omega_3,\omega_4,\omega_5)}{\eta_0\tau^4} =& -\frac{1}{5!}\left(\frac{1}{1 + i\tau_0\sum_{j=1}^5 \omega_j}\right) \times \sum_{j=1}^5 \left(\frac{1}{1 + i\tau_0\sum_{k\neq j}\omega_k}\right)\sum_{k\neq j}\left(\frac{1}{1 + i\tau_0\sum_{l\neq k,j}\omega_l}\right) \label{eq:crm_eta5} \\
    & \times \sum_{l \neq k,j}\left(\frac{1}{1 + i\tau_0\sum_{m\neq k,j,l}\omega_l}\right)\sum_{m\neq k,j,l}\left(\frac{1}{1 + i\tau_0\omega_m}\right). \nonumber
\end{align}
\end{widetext}
These solutions for the first, third, and fifth order complex viscosities can be easily converted to those for the complex moduli via the relationship:
\begin{equation}
    G^*_n(\omega_1,...,\omega_n) = \left(\prod_{m=1}^n i\omega_m\right)\eta^*_n(\omega_1,...,\omega_n).
\end{equation}
The first, third, and fifth order elements of the stress response at the fundamental frequency in a LAOS experiment ($\gamma(t) = \gamma_0\sin(\omega_0 t)$) are:
\begin{equation}
    \hat{\sigma}^{(1)}(\omega_0) = -i\pi G^*_1(\omega_0),
\end{equation}
\begin{equation}
    \hat{\sigma}^{(3)}(\omega_0) = -\frac{3i\pi}{4}G^*_3(\omega_0,\omega_0,-\omega_0),
\end{equation}
\begin{equation}
    \hat{\sigma}^{(5)}(\omega_0) = -\frac{10i\pi}{16}G^*_5(\omega_0,\omega_0,\omega_0,-\omega_0,-\omega_0).
\end{equation}
With some algebra, it can be verified that these solutions agree with those previously derived for the corotational Maxwell model in strain-controlled LAOS \cite{giacomin-2011}. Note that in the expression for $\hat{\sigma}^{(5)}(\omega_0)$, we have left the fraction $10/16$ in non-reduced form. This was done to reflect that the denominator arises directly from the $2^{n-1}$ term in the frequency-domain Volterra expansion (see equation 4 of \cite{lennon-2020}) for $n = 5$, and that the numerator is a permutation symmetry factor (${5 \choose 2} = 10$) related to the arguments of the fifth order complex modulus.

The forms of the first, third, and fifth order complex compliances can be found using the inversion relationships:
\begin{equation}
    J^*_1(\omega) = \frac{1}{G^*_1(\omega)},
\end{equation}
\begin{align}
    & J^*_3(\omega_1,\omega_2,\omega_3) = \\ 
    & \quad\quad - \frac{G^*_3(\omega_1,\omega_2,\omega_3)}{G^*_1(\omega_1)G^*_1(\omega_2)G^*_1(\omega_3)G^*_1(\sum_{j=1}^3\omega_j)} \nonumber,
\end{align}
\begin{widetext}
\begin{align}
    J^*_5(\omega_1,\omega_2,\omega_3,\omega_4,\omega_5) =& -\frac{1}{G^*_1(\omega_1)G^*_1(\omega_2)G^*_1(\omega_3)G^*_1(\omega_4)G^*_1(\omega_4)G^*_1(\sum_{j=1}^5\omega_j)}\Bigg[G^*_5(\omega_1,\omega_2,\omega_3,\omega_4,\omega_5) \\
    & \left. - \frac{1}{20}\sum_{i,j,k,l,m \in \{1,2,3,4,5\}}\frac{G^*_3(\omega_i,\omega_j,\omega_k+\omega_l+\omega_m)G^*_3(\omega_k,\omega_l,\omega_m)}{G^*_1(\omega_k+\omega_l+\omega_m)}\right]. \nonumber
\end{align}
\end{widetext}
The resulting first, third, and fifth order elements of the strain response at the fundamental frequency in a stress-controlled LAOS experiment ($\sigma(t) = \sigma_0\cos(\omega_0 t)$) are therefore:
\begin{equation}
    \hat{\gamma}^{(1)}(\omega_0) = \pi J^*_1(\omega_0),
\end{equation}
\begin{equation}
    \hat{\gamma}^{(3)}(\omega_0) = \frac{3\pi}{4}J^*_3(\omega_0,\omega_0,-\omega_0),
\end{equation}
\begin{equation}
    \hat{\gamma}^{(5)}(\omega_0) = \frac{10\pi}{16}J^*_5(\omega_0,\omega_0,\omega_0,-\omega_0,-\omega_0).
\end{equation}

% Gap loading limit
\section{The Gap Loading Limit for MAPS Experiments}
\label{app:gap_loading}

One of the critical assumptions made in cone-and-plate or parallel-plate rheometry is that shear deformation is homogeneous in the direction of the velocity gradient across the sample gap. That is, the velocity profile in the direction of the velocity gradient is linear throughout the sample. Because measured properties, such as the third order MAPS response functions, associate a measured stress with a measured strain or strain rate, accurate measurements require that the shear rate and shear stress transduced to the cone or plate must be indicative of the state of deformation in the material throughout the gap.

The assumption of a linear velocity profile in the gap, or equivalently of a uniform shear rate, is equivalent to the assumption that dynamic equilibrium is established in the velocity profile on a time-scale much faster than the shortest observation time-scale. For smaller gap sizes, equilibrium is established quicker as linear momentum needs to be transported over a shorter distance. Thus, for any material and any time scale of observation, it is possible to decrease the gap to a scale at which the assumption of a uniform shear rate is valid (neglecting, for now, other sources of flow inhomogeneity within the fluid, such as shear banding). This limiting value of the gap size is called the \emph{gap loading limit}.

The gap loading limit can be derived by considering the equation for momentum transport across the gap in simple shear flow between parallel plates:
\begin{equation}
    \rho \frac{\partial u}{\partial t} = \frac{\partial \sigma}{\partial y},
    \label{eq:cauchy}
\end{equation}
with boundary conditions $u(H,t) = \gamma_0 U(t)$ and $u(0,t) = 0$, which is obtained directly from the Cauchy momentum equation with the assumption of incompressibility and unidirectional flow in the $x$-direction that is independent of $x$. In equation \ref{eq:cauchy}, $\sigma \equiv \sigma_{xy}$ denotes the shear stress within the material, $u$ denotes the $x$-component of the flow velocity, $\rho$ denotes the material's density, $H$ is the size of the gap, $U(t)$ is some time-dependent function that is $\mathcal{O}(1)$, and $\gamma_0$ is an amplitude parameter for the deformation. We will examine the Fourier transform of this equation:
\begin{equation}
    i\omega\rho \hat{u}(\omega) = \frac{\partial}{\partial y}\hat{\sigma}(\omega),
    \label{eq:cauchy_freq}
\end{equation}
with boundary conditions $\hat{u}(H,\omega) = \gamma_0 \hat{U}(\omega)$ and $\hat{u}(0,\omega) = 0$. In simple shear, for which the rate-of-strain tensor is $\boldsymbol{\dot{\gamma}} = \frac{\partial u}{\partial y}$, the shear stress can be written as a Volterra series in the velocity gradient:
\begin{align}
    &\hat{\sigma}(\omega) \approx \eta^*_1(\omega) \frac{\partial \hat{u}(\omega)}{\partial y} \\
    & \quad + \iiint_{-\infty}^{\infty}\eta^*_3(\omega_1,\omega_2,\omega_3)\delta(\omega - \sum_{j=1}^3\omega_j)\prod_{j=1}^3\frac{\partial \hat{u}(\omega_j)}{dy}d\omega_j. \nonumber
\end{align}

With the Volterra series relating the shear stress to the velocity profile, equation \ref{eq:cauchy_freq} can be solved asymptotically. The velocity profile can be written as a perturbation series in $\gamma_0$:
\begin{equation}
    \hat{u}(\omega) = \gamma_0 \hat{u}^{(1)}(\omega) + \gamma_0^2 \hat{u}^{(2)}(\omega) + \gamma_0^3\hat{u}^{(3)}(\omega) + \mathcal{O}(\gamma_0^4).
\end{equation}
At $O(\gamma_0)$, equation \ref{eq:cauchy_freq} reads:
\begin{equation}
    i\omega\rho \hat{u}^{(1)}(\omega) = \eta^*_1(\omega) \frac{\partial^2 \hat{u}^{(1)}(\omega)}{\partial y^2}.
\end{equation}
The solution to this differential equation can be written in the form:
\begin{equation}
    \hat{u}^{(1)}(\omega) = A \sinh{ky} + B \cosh{ky}
    \label{eq:linear_profile}
\end{equation}
with the complex wavenumber:
\begin{equation}
    k = \sqrt{\frac{i\omega\rho}{\eta^*_1(\omega)}}.
\end{equation}
Given the boundary conditions at the plates, it is straightforward to show that $B = 0$ and $A = \hat{U}(\omega)/\sinh(kH)$. The important quantity, however, is the wavenumber $k$, which determines the length scales for variations in the velocity profile. In particular, the wavenumber can be expressed in terms of a wavelength $\lambda_s$ and penetration depth $d_s$ \cite{vermant-2020}:
\begin{equation}
    k = \frac{2\pi}{\lambda_s} - \frac{i}{d_s},
\end{equation}
with:
\begin{equation}
    \lambda_s = \frac{2\pi}{\sqrt{\frac{\omega\rho}{|\eta^*_1(\omega)|}}\cos\frac{\delta}{2}}, \quad d_s = \frac{1}{\sqrt{\frac{\omega\rho}{|\eta^*_1(\omega)|}}\sin\frac{\delta}{2}},
\end{equation}
where $\delta \equiv \tan^{-1}(\eta'_1(\omega)/\eta''_1(\omega))$ and $|\eta^*_1(\omega)| = \sqrt{\eta'^2_1(\omega) + \eta''^2_1(\omega)}$ are the phase and magnitude of the complex viscosity, respectively.

In a small-amplitude oscillatory experiment, terms of $\mathcal{O}(\gamma_0^2)$ and above are negligible compared to the linear terms. Therefore, $\hat{u}(\omega) \approx \gamma_0\hat{u}^{(1)}(\omega)$, thus we need only be concerned that the first-order velocity profile is linear. Based on the above analysis, we see that a linear approximation to equation \ref{eq:linear_profile} on the interval $[0,H]$ is valid if both $H/\lambda_s \ll 1$ and $H/d_s \ll 1$.

In a medium amplitude experiment such as the one outlined in this work, we must consider effects up to $\mathcal{O}(\gamma_0^3)$. Continuing the above analysis, we find that the second-order component to the velocity profile obeys the same homogeneous linear differential equation as the first-order component, thus its solution is also of the form of \ref{eq:linear_profile} with boundary conditions $\hat{u}(0,\omega) = \hat{u}(H,\omega) = 0$. In general, these boundary conditions require that the second-order velocity profile is zero everywhere, unless $kH = in\pi$ for some integer $n$.

At third order, the velocity profile satisfies the inhomogeneous differential equation:
\begin{align}
    &i\omega\rho\hat{u}^{(3)}(\omega) = \eta^*_1(\omega)\frac{\partial^2 \hat{u}^{(3)}(\omega)}{\partial y^2} \label{eq:cubic_profile} \\
    &+ \frac{\partial}{\partial y}\iiint \eta^*_3(\omega_1,\omega_2,\omega_3)\delta(\omega - \sum_{j=1}^3\omega_j)\prod_{j=1}^3\frac{\partial \hat{u}^{(1)}(\omega_j)}{\partial y}d\omega_j \nonumber
\end{align}
again with the boundary conditions $\hat{u}^{(3)}(0,\omega) = \hat{u}^{(3)}(H,\omega) = 0$. We will not solve analytically for the full velocity profile here, but it is evident from equation \ref{eq:cubic_profile} what length scales govern the third-order solution. The homogeneous portion of the solution obeys the same linear differential equation as did the first-order profile, thus is governed by the length scales $\lambda_s$ and $d_s$. The inhomogeneity includes the product of three factors of the first order solution. Because this solution depends exponentially on $y/\lambda_s$ and $y/d_s$, we know that this product will be governed by smaller length scales $l_j$ of the form:
\begin{equation}
    \frac{1}{l_1} = \frac{3}{\lambda_s}, \quad \frac{1}{l_2} = \frac{3}{d_s}, \quad \frac{1}{l_3} = \frac{2}{\lambda_s} + \frac{1}{d_s}, ...
\end{equation}
To ensure that the third-order profile is approximately linear over the gap, the gap separation must be much smaller than the smallest of these length scales, which is set either by $\lambda_s$ or $d_s$:
\begin{equation}
    l_\mathrm{max} = \min\left(\frac{\lambda_s}{3}, \frac{d_s}{3}\right).
\end{equation}
Therefore, to be in the gap loading limit for a MAPS experiment:
\begin{equation}
    H \ll l_\mathrm{max}.
\end{equation}

One observation from this analysis is that the gap loading limit $l_\mathrm{max}$ depends only on a material's properties through its density and linear viscoelasticity, but not on weakly nonlinear viscoelastic effects. For a fluid with a Maxwellian linear response, we know that:
\begin{equation}
    \eta'_1(\omega) = \frac{\eta_0}{1 + \tau^2\omega^2} + \eta_{\infty}, \quad \eta''_1(\omega) = \frac{\eta_0 \tau\omega}{1 + \tau^2 \omega^2}.
\end{equation}
Using these expressions, the maximum acceptable gap height can be calculated for an observation timescale set by $\omega$ once $\eta_0$, $\tau$, and $\rho$ are specified. For the wormlike micelle solution studied in this work, measurements show that $\eta_0 = 25.7$ Pa$\cdot$s, $\tau = 0.64$ s, and $\eta_{\infty} = 0.039$ Pa$\cdot$s, and the density is assumed to be approximately that of water ($\rho \approx 1000$ kg/m\textsuperscript{3}). The range of frequencies under study in the MAPS experiments are $\omega \in [0.16, 61.44]$ rad/s. In this range, both $\lambda_s$ and $d_s$ are monotonically decreasing functions of $\omega$, therefore the minimum values occur for the shortest timescale under study: $\omega = 61.44$ rad/s. At this point, we find that $\lambda_s = 0.021$ m, and $d_s = 0.077$ m. Therefore, $l_\mathrm{max} = \lambda_s/3 = 0.0069$ m. Because the truncation gap size in the experiments was set to $H = 58$ $\mu$m, we indeed have that:
\begin{equation}
    H = 5.8\times10^{-5} \, \mathrm{m} \ll l_\mathrm{max} = 6.9\times10^{-3} \,  \mathrm{m}.
\end{equation}
This indicates that the gap loading limit is valid for the truncation gap of the cone-and-plate geometry. However, due to the conical geometry, the gap increases radially outwards along the radial coordinate of the cone. For the 2\degree, 60 mm diameter cone, the maximum gap separation at the edge of the sample is:
\begin{align}
    H_\mathrm{max} &= 5.8\times 10^{-5} \, \mathrm{m} + (30 \times 10^{-3} \, \mathrm{m})\tan(2\degree) \nonumber \\
    &= 1.1\times10^{-3} \, \mathrm{m} < l_\mathrm{max},
\end{align}
The gap size in this case is still less than the gap loading limit by a factor of six. However, for these highest-frequency measurements when $l_{\mathrm{max}}$ is smallest, it is possible that the flow profile is not fully homogeneous at the rim of the cone. Because this region has the most influence on the measured torque, high-frequency rheological measurements for which $H_\mathrm{max}$ is on, or near, the same order of $l_\mathrm{max}$ may be influenced by these inertial effects. This is one possible explanation for the abrupt deviations of the very highest frequency measurements in Figures \ref{fig:1416_J3} and \ref{fig:1416_eta3} from the lower frequency trends. For lower frequency measurements, the gap loading constraint is less severe, and it is likely that the effects of fluid inertia in these measurements are negligible.

\end{document}